\documentclass[njp,10pt]{iopart}
\usepackage{graphicx}
\expandafter\let\csname equation*\endcsname\relax
\expandafter\let\csname endequation*\endcsname\relax
\usepackage{amsmath}
\usepackage{amsfonts}
\usepackage{amssymb}
\usepackage{soul}
\usepackage{dsfont}
\usepackage{hyperref}
\usepackage{amstext}
\usepackage{braket}
\usepackage[caption=false]{subfig}
\usepackage{braket}
\usepackage{color,soul}

\date{\today}

\hypersetup{
    colorlinks=true,
    linkcolor=blue,
    filecolor=magenta,      
    urlcolor=cyan,
    pdftitle={Sharelatex Example},
    bookmarks=true,
    pdfpagemode=FullScreen,
}

\begin{document}

\title[Time-evolution of nonlinear optomechanical systems: Interplay of arbitrary mech...]{Time-evolution of nonlinear optomechanical systems: Interplay of  mechanical squeezing and non-Gaussianity}

\author{Sofia Qvarfort$^{1}$, Alessio Serafini$^{1}$, 
  Andr\'e Xuereb$^{2}$, Daniel Braun$^{3}$, Dennis R\"atzel$^{4}$, 
  David Edward Bruschi$^{5,6,7}$\footnote{'Theoretical Physics, Universit\"at des Saarlandes, 66123 Saarbr\"ucken, Germany' }}
\address{$^{1}$ Department of Physics and Astronomy, University College London, Gower Street, WC1E 6BT London, United Kingdom}
\address{$^{2}$ Department of Physics, University of Malta, Msida MSD 2080, Malta}
\address{$^{3}$ Institut f\"{u}r Theoretische Physik, Eberhard-Karls-Universit\"{a}t T\"{u}bingen, D-72076 T\"{u}bingen, Germany} 
\address{$^{4}$ Institut f\"{u}r Physik, Humboldt-Universit\"{a}t zu Berlin, 12489 Berlin, Germany}
\address{$^{5}$ Faculty of Physics, University of Vienna, 1090 Vienna, Austria}
\address{$^{6}$ Institute for Quantum Optics and Quantum Information - IQOQI Vienna, Austrian Academy of Sciences,  1090 Vienna, Austria}
\address{$^{7}$ Central European Institute of Technology (CEITEC), Brno University of Technology, 621 00 Brno, Czech Republic}
\ead{sofiaqvarfort@gmail.com, david.edward.bruschi@gmail.com}

\begin{abstract} 
We solve the time evolution of a nonlinear optomechanical Hamiltonian with arbitrary time-dependent mechanical displacement, mechanical single-mode squeezing and a time-dependent optomechanical coupling up to the solution of two second-order differential equations. The solution is based on identifying a minimal and finite Lie algebra that generates the time-evolution of the system. This reduces the problem to considering a finite set of coupled ordinary differential equations of real functions. To demonstrate the applicability of our method, we compute the degree of non-Gaussianity of the time-evolved state of the system by means of a measure based on the relative entropy of the non-Gaussian state and its closest Gaussian reference state. We find that the addition of a constant mechanical squeezing term to the standard optomechanical Hamiltonian generally decreases the overall non-Gaussian character of the state. For sinusoidally modulated squeezing, the two second-order differential equations mentioned above take the form of the Mathieu equation. We derive perturbative solutions for a small squeezing amplitude at parametric resonance and show that they correspond to the rotating-wave approximation at times larger than the scale set by the mechanical frequency. We find that the non-Gaussianity of the state increases with both time and the squeezing parameter in this specific regime. 
\end{abstract}

\maketitle

\section{Introduction}\label{intro}

The mathematical understanding of optomechanical systems operating in the nonlinear quantum regime is a major topic of current interest. While most experiments effectively undergo linear dynamics, governed by quadratic Hamiltonians that emerge following
 a `linearisation' procedure~\cite{aspelmeyer2014cavity,bowen2015quantum, serafini2017quantum}, many experiments now operate in the fully nonlinear regime~\cite{sankey2010strong,leijssen2017nonlinear, fogliano2019cavity} where this procedure fails.  It is therefore highly desirable to provide a complete and analytic characterisation of the fully nonlinear system dynamics. Analytic solutions have previously been found for a constant light--matter coupling~\cite{bose1997preparation, mancini1997ponderomotive} and,  more recently, the time-dependent case was solved~\cite{qvarfort2019enhanced}. 

The inherently nonlinear interaction between the optical field and the mechanical element in an optomechanical system allows for the generation of non-Gaussian states. Starting from a broad class of initial states, including coherent states, the vacuum, and thermal states, this is only possible in the nonlinear regime;  in contrast, quadratic Hamiltonians take input Gaussian states to output Gaussian states.  As such, investigating the non-Gaussianity of optomechanical states can only be performed once the time-evolution in the nonlinear regime has been solved, which is the primary aim of this work. Interestingly, a number of non-classical and non-Gaussian states have been found to constitute an important resource for sensing. Schr\"{o}dinger cat states~\cite{mancini1997ponderomotive, bose1997preparation}, compass states~\cite{zurek2001sub, toscano2006sub} and hypercube states~\cite{howard2018hypercube} -- which are all non-Gaussian  and highly non-classical states -- have all been found to have applications for sensing. 
 More generally, the detection and generation of non-Gaussianity in optomechanical systems has been extensively studied in theoretical proposals~\cite{lemonde2016enhanced,latmiral2016probing, yin2017nonlinear} as well as in experiments~\cite{sankey2010strong, doolin2014nonlinear, leijssen2017nonlinear}. Beyond optomechanics, the presence of a nonlinear element is also key to a number of quantum information tasks, such as obtaining a universal gate set for quantum computing~\cite{lloyd1999quantum, menicucci2006universal}, teleportation~\cite{dell2010teleportation}, distillation of entanglement~\cite{fiuravsek2002gaussian, giedke2002characterization}, error correction~\cite{niset2009no}, and non-Gaussianity has been explored as the basis of an operational resource theory~\cite{zhuang2018resource, takagi2018convex, PhysRevA.98.052350}. 
 
Optomechanical systems offer a natural nonlinear coupling which, if strong enough, may lead to substantial non-Gaussianity in the evolved state. It is therefore essential to better understand the dynamics of such systems, with special emphasis on
 the interplay between nonlinearities and other Hamiltonian terms in this dynamics. An important question to be answered is thus \textit{how do the different aspects of an optomechanical system affect the non-Gaussianity of the state at a given time?} A preliminary study of non-Gaussianity in standard optomechanical systems  provided the first tools to approach this question~\cite{qvarfort2019enhanced}, however, optomechanical systems can exhibit additional, potentially more interesting, effects. An important  non-classical effect that can be included into optomechanical systems is \textit{squeezing} of the optical or mechanical modes.  The addition of squeezing has been found to be beneficial for sensing since it increases the sensitivity in a specific field quadrature. For example, it has been shown that squeezed light injected into LIGO can be used to enhance the detection of gravitational waves~\cite{aasi2013enhanced}. Similarly, mechanical squeezing 
can aid the amplification and measurement of weak mechanical signals~\cite{clerk2010introduction}. 

In this work, we study the non-Gaussianity of a quantum system of two bosonic modes characterised by an optomechanical Hamiltonian with the standard cubic light--matter interaction term, and with the addition of a mechanical displacement term and a mechanical squeezing term. We extend a recently developed solution of the time evolution operator induced by a plain optomechanical Hamiltonian~\cite{qvarfort2019enhanced} to include the additional terms of interest here. Interestingly, for time-dependent squeezing modulated at resonance, we find that the dynamics are governed by the well-studied Mathieu equation. We subsequently derive perturbaive solutions and show that these coincide with the physically intuitive rotating-wave approximation for large times. The decoupling methods used in this work have a long tradition in quantum theory~\cite{wei1963lie,wilcox1967exponential, puri2001mathematical} and were recently applied to problems such as the one at hand~\cite{bruschi2018mechano, qvarfort2019enhanced, bruschi2018time}. We use the resulting analytic solutions to compute the amount of non-Gaussianity of the state using a measure of relative entropy~\cite{genoni2008quantifying,marian2013relative} for both a constant and a time-dependent mechanical squeezing parameter. 

Our results indicate that the non-Gaussian character of an initially coherent state \emph{decreases} in general with an increasing squeezing parameter. However, when the squeezing is applied periodically at mechanical resonance, the non-Gaussianity increases approximately linearly with time and the amplitude of the squeezing.  The competition between the amount of squeezing and the strength of the nonlinear term is difficult to compute explicitly; instead, we provide asymptotic expressions in terms of upper and lower bounds to the non-Gaussianity in different regimes. A conclusive answer requires further investigation, potentially providing a concise expression where such competition can be easily understood. 

The paper is structured as follows. In Section~\ref{dynamics}, we introduce the nonlinear Hamiltonian with mechanical squeezing. This is followed by Section~\ref{tools} where we provide a short introduction to the methods used to solve the dynamics. The full derivation can be found in~\ref{appendix:Hamiltonian:nonlinear:decoupling}. Following this, we review the measure of non-Gaussianity and derive expressions for an asymptotic expression and a reduced measure in Section~\ref{sec:measure}. In Section~\ref{sec:applications}, we then specialise to two specific cases and compute the amount of non-Guassianity for constant squeezing (Section~\ref{sec:constant:squeezing}), and for modulated squeezing (Section~\ref{sec:modulated:squeezing}). Finally, we conclude with a discussion in Section~\ref{discussion} and some final remarks in Section~\ref{conclusions}. 

\section{Dynamics}\label{dynamics}
In this section we present the optomechanical Hamiltonian of interest to this work and explain the origin of the various terms.  An extensive introduction to optomechanics can be found in the literature~\cite{aspelmeyer2014cavity}.

\subsection{Hamiltonian}\label{optomech}
In this work we consider the two-mode Hamiltonian
\begin{align}\label{main:time:independent:Hamiltonian:to:decouple:dimensionful}
	\hat {H} &= \hat{H}_0+ \hbar \,\mathcal{D}_1(t)\,\left(\hat b+ \hat b{}^\dagger\right) +  \hbar \, \mathcal{D}_2(t)\left(\hat b+ \hat b{}^\dagger \right)^2 - \hbar\,\mathcal{G}(t) \hat a^\dagger\hat a \, \left(\hat b+ \hat b{}^\dagger\right),
\end{align}
where $\hat{H}_0:=\hbar\,\omega_\mathrm{c} \hat a^\dagger \hat a + \hbar\,\omega_\mathrm{m} \,\hat b{}^\dagger \hat b$ is the free Hamiltonian, while $\omega_\mathrm{c}$ and $\omega_\mathrm{m}$ are the frequencies of the cavity mode and the mechanical resonator respectively.

\begin{figure}[ht!]
\centering
{\includegraphics[width=.5\linewidth, trim = -10mm 0mm 0mm -10mm]{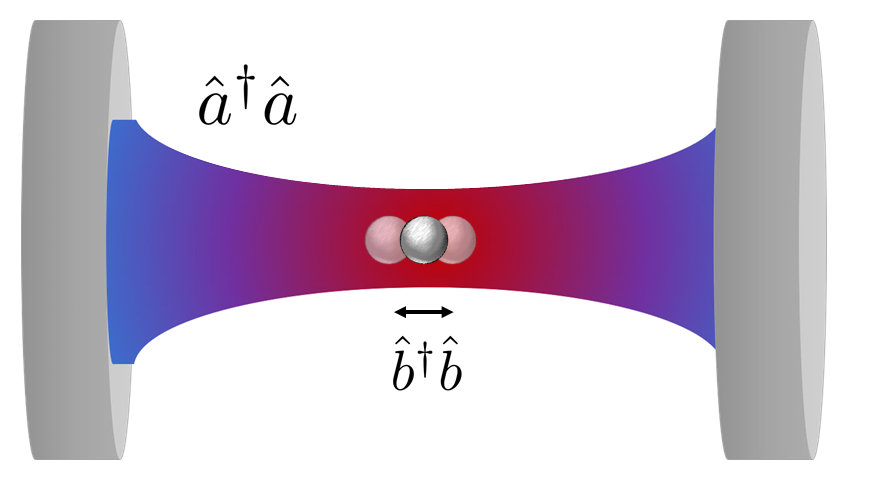}}\hfill
\caption{A levitated nanosphere in a cavity. The optical field is described by annihilation and creation operators $\hat{a}$ and $\hat{a}^\dagger$, while the mechanics -- in this case the mechanical motion of the nanosphere -- 
  is described by annihilation and creation operators $\hat{b}$ and $\hat{b}^\dagger$. The system evolves under the optomechanical Hamiltonian~\eqref{main:time:independent:Hamiltonian:to:decouple}.}\label{the:system}
\end{figure}

The Hamiltonian~\eqref{main:time:independent:Hamiltonian:to:decouple:dimensionful} describes the dynamics of a number of different systems. For example, $\mathcal{G}(t)$ appears in optomechanical systems as a standard coupling term due to radiation pressure obtained for Fabry-P\'{e}rot cavities, where one end of the cavity is a mirror that can move freely~\cite{favero2009optomechanics}. Such
coupling appears also within systems with a central translucent membrane in a rigid optical cavity~\cite{jayich2008dispersive},
levitated nanodiamonds~\cite{yin2013large} or optomechanical crystals~\cite{eichenfield2009picogram, safavi2014two}. A depiction of a levitated nanosphere interacting with cavity modes can be found in Figure~\ref{the:system}. 

The Hamiltonian~\eqref{main:time:independent:Hamiltonian:to:decouple:dimensionful} reduces to the standard optomechanical Hamiltonian when $\mathcal{D}_1=\mathcal{D}_2=0$.\footnote{We note here that non-Gaussianity for the case $\mathcal{D}_1=\mathcal{D}_2=0$ has been already studied~\cite{qvarfort2019enhanced}.}  The term weighted by the coupling $\mathcal{D}_1$ corresponds to an externally imposed displacement of the mechanical part, which can be induced by a piezoelectric element connected to its support or by an external acceleration, such as that caused by the gravitational force acting on the mechanical element~\cite{qvarfort2018gravimetry, armata2017quantum}. 
The term governed by $\mathcal{D}_2$ can be thought of as a modulation of the trap frequency and leads to squeezing of the mechanics., which can be externally imposed employing another strong optical field or an electrostatic force~\cite{blencowe2004quantum}.  
 
\subsection{Dimensionless dynamics}\label{optomech}
 
To understand which dimensionless parameters are relevant to the dynamics of the system, we start by introducing dimensionless quantities and rescaling the Hamiltonian.  Such a rescaling also serves to simplify the notation and any graphical representation of the system dynamics. We achieve this by dividing the functions in the Hamiltonian by the mechanical frequency  $\omega_{\mathrm{m}}$. The action corresponds to switching from the laboratory time $t$ to $\tau=\omega_{\mathrm{m}}\,t$, where $\tau$ is the new, dimensionless time. The optical frequency becomes $\Omega_{\mathrm{c}}:=\omega_{\mathrm{c}}/\omega_{\mathrm{m}}$. In addition, the couplings in the Hamiltonian become $\tilde{\mathcal{G}}(\tau) = \mathcal{G}(\omega_{\mathrm{m}}\,t) / \omega_{\mathrm{m}}$, $\tilde{\mathcal{D}}_1(\tau) = \mathcal{D}_1(\omega_{\mathrm{m}}\,t)/\omega_{\mathrm{m}}$ and $\tilde{\mathcal{D}}_2(\tau) = \mathcal{D}_2(\omega_{\mathrm{m}}\,t)/\omega_{\mathrm{m}}$. We also rescale the Hamiltonian by $\hbar$, meaning that $\hat {H}$ becomes
\begin{align}\label{main:time:independent:Hamiltonian:to:decouple}
	\frac{\hat {H}}{\hbar \,  \omega_{\mathrm{m}}} = \, & \Omega_{\mathrm{c}}\,\hat{a}^\dag\hat{a}+\hat{b}^\dag\hat{b}+ \tilde{\mathcal{D}}_1(\tau)\,\left(\hat b+ \hat b^\dagger\right)+  \tilde{\mathcal{D}}_2(\tau)\left(\hat b+ \hat b^\dagger \right)^2 - \tilde{\mathcal{G}}(\tau)\,\hat a^\dagger\hat a \,\left(\hat b+ \hat b^\dagger\right),
\end{align}
which is the Hamiltonian that we will be working with.

\section{Solving the dynamics}\label{tools}
Our aim is to provide the techniques to be used to understand the interplay of mechanical squeezing and non-Gaussianity in an optomechanical system, for which an analytic expression for the state evolution~\cite{qvarfort2019enhanced} is central. In this section we introduce the tools needed to solve the dynamics generated by~\eqref{main:time:independent:Hamiltonian:to:decouple}. See~\ref{appendix:decoupling} for a more in-depth introduction to the underlying concepts, and~\ref{appendix:Hamiltonian:nonlinear:decoupling} for the full calculations.

\subsection{Continuous variables and covariance-matrix formalism} 
\label{tools:cmf}
When solving the dynamics, we employ methods from the continuous
variable formalism~\cite{adesso2014continuous,
  serafini2017quantum}.  Specifically, the methods are used to solve the time-evolution of the quadratic part of the system, and to describe its action on the nonlinear light--matter interaction term. We briefly review the continuous variable formalism
here.  

In recent years, thanks to progress in the mathematical framework provided by the covariance matrix formalism~\cite{adesso2014continuous,
  serafini2017quantum}, it has become clear that Gaussian states constitute
an extremely valuable toolkit to investigate quantum information
processing in quantum setups, and in relativistic ones as well
\cite{alsing2012observer}. The main advantage is that the covariance
matrix formalism provides a powerful set of mathematical tools to
treat Gaussian states of bosonic fields that undergo linear 
transformations of the creation and annihilation 
operators fully
analytically~\cite{adesso2014continuous}. Ultimately, Gaussian states
are the paramount resource for continuous variables quantum
information processing and computation~\cite{lloyd1999quantum} and
have become a standard feature in most quantum optics laboratories. However, it should also be pointed out that these methods can be used to describe the evolution of operators in the Heisenberg picture, even when the states considered are not Gaussian.

In quantum mechanics, the initial state $\hat{\rho}_i$ of a system of
$N$ bosonic modes with operators $\{\hat{a}_n,\hat{a}^{\dag}_n\}$
evolves to a final state $\hat{\rho}_f$ through the standard
Schr\"{o}dinger equation 
$\hat{\rho}_f=\hat{U}\,\hat{\rho}_i\,\hat{U}^{\dag}$, where $\hat U$
implements the transformation of interest, such as time evolution. If
the state $\hat{\rho}$ is Gaussian and the Hamiltonian $\hat H$ is
quadratic in the operators, it is convenient to introduce the vector
$\hat{\mathbb{X}}=(\hat{a}_1,\ldots,\hat{a}_N,\hat{a}^{\dag}_1,\ldots,\hat{a}^{\dag}_N)^{\rm{T}}$,
where $\rm{T}$ denotes the transpose of the vector. 
Similarly, the vector of first moments $d:=\langle\hat{\mathbb{X}}\rangle$ and
the covariance matrix $\boldsymbol{\sigma}$ are defined by 
$\sigma_{nm}:=\langle\{\hat{X}_n,\hat{X}^{\dag}_m\}\rangle-2\langle\hat{X}_n\rangle\langle\hat{X}_m^{\dag}\rangle$,
where $\{\cdot,\cdot\}$ stands for anticommutator and all expectation
values of an operator $\hat{\mathcal{A}}$ are defined by $\langle
\hat{\mathcal{A}}\rangle:=\text{Tr}(\hat{\mathcal{A}}\,\hat{\rho})$. 

In
this language, the canonical commutation relations read
$[\hat{X}_n,\hat{X}_m^{\dag}]=i\,\Omega_{nm}$, where the $2N\times2N$
matrix $\boldsymbol{\Omega}$ is known as the symplectic form
\cite{adesso2014continuous}. We then notice that, while arbitrary
states of bosonic modes are, in general, characterised by infinite
real parameters, a Gaussian state is uniquely determined
by its first and second moments, $d_n$ and $\sigma_{nm}$ respectively
\cite{adesso2014continuous}. Furthermore, unitary transformations
quadratic in the annihilation and creation operators, 
such as Bogoliubov transformations~\cite{birrell1984quantum}, preserve
the Gaussian character of a Gaussian state and can always be
represented by a $2N\times2N$ symplectic matrix $\boldsymbol{S}$ that
preserves the symplectic form, i.e.,
$\boldsymbol{S}^{\dag}\,\boldsymbol{\Omega}\,\boldsymbol{S}=\boldsymbol{S}\,\boldsymbol{\Omega}\,\boldsymbol{S}^{\dag}=\boldsymbol{\Omega}$.\footnote{Note that $\boldsymbol{\sigma}$ is complex in our choice of basis, which implies taking the Hermitian conjugate  of $\boldsymbol{S}$.} 
In a similar manner, the symplectic matrix $\boldsymbol{S}$ that encodes the evolution of a state is generated by the Hamiltonian matrix $\boldsymbol{H}$, which is defined by $\hat H = \hat {\mathbb{X}}^\dag \boldsymbol{H} \hat{\mathbb{X}}/2  + \hat{\mathbb{X}}^\dag d$. The symplectic matrix becomes $\boldsymbol{S} = \mathrm{exp} \left[ \boldsymbol{\Omega} \boldsymbol{H}\right]$.
 
The Schr\"{o}dinger  equation can be translated in this language to the simple equation $\boldsymbol{\sigma}_f=\boldsymbol{S}\,\boldsymbol{\sigma}_i\,\boldsymbol{S}^{\dag}$ for the second moments, and $\boldsymbol{r}_f = \boldsymbol{S} \,
\boldsymbol{r}_i $ for the first moments, which shifts the problem of usually untreatable operator algebra to simple $2N\times2N$ matrix multiplication. Here, the indices $i$ and $f$ denote the initial and final state, respectively.
Finally, Williamson's theorem guarantees that any $2N\times2N$ Hermitian matrix, such as the covariance matrix $\boldsymbol{\sigma}$, can be decomposed as $\boldsymbol{\sigma}=\boldsymbol{S}^{\dag}\,\boldsymbol{\nu}_{\oplus}\,\boldsymbol{S}$, where $\boldsymbol{S}$ is an appropriate symplectic matrix. 
The diagonal matrix $\boldsymbol{\nu}_{\oplus}=\textrm{diag}(\nu_1,\dots,\nu_N,\nu_1,\dots,\nu_N)$ is known as the Williamson form of the state and $\nu_n:=\coth(\frac{\hbar\,\omega_n}{2\,k_B\,T})\geq1$ 
(where we have introduced normal frequencies $\omega_n$ and a nominal temperature $T$)
are the symplectic eigenvalues of the state~\cite{williamson1936algebraic}. 

Williamson's form $\boldsymbol{\nu}_{\oplus}$ contains information about the local and global mixedness of the state of the system
\cite{adesso2014continuous}. The state is pure if $\nu_n = 1$ for all $n $ and is mixed otherwise. As an example, the thermal state
$\boldsymbol{\sigma}_{th}$ of a $N$-mode bosonic system is simply given by its Williamson form, i.e., $\boldsymbol{\sigma}_{th}=\boldsymbol{\nu}_{\oplus}$. 

\subsection{Decoupling of a time evolution operator}\label{tools:time:evolution}
The time evolution of a system with time-dependent Hamiltonian $\hat{H}(t)$ is 
\begin{equation}\label{general:time:evolution:operator}
\hat{U}(t)=\overset{\leftarrow}{\mathcal{T}}\,\exp\left[-\frac{i}{\hbar}\int_0^{t} dt'\,\hat{H}(t')\right],
\end{equation}
where $\overset{\leftarrow}{\mathcal{T}}$ is the time ordering operator. This expression simplifies dramatically when the Hamiltonian $\hat{H}$ is time independent, in which case one  obtains $\hat U(t)=\exp[-\frac{i}{\hbar}\,\hat{H}\,t]$ as a solution to the time-dependent Schr\"{o}dinger equation. We are, however, interested in Hamiltonians with time-dependent parameters. 
Any Hamiltonian can be cast in the form $\hat{H}=\sum_n \hbar\, g_n(t)\,\hat{G}_n$, where the $\hat{G}_n$ are time independent, Hermitian operators and the $g_n(t)$ are time-dependent real functions. The choice of $\hat{G}_n$ need not be unique, and if this is the case, a specific choice is motivated by the specific aims of the problem.

We say that the time evolution operator~\eqref{general:time:evolution:operator} has been \textit{decoupled} if it can be written as~\cite{wei1963lie, wilcox1967exponential}
\begin{equation}\label{decoupled:time:evolution:operator}
\hat{U}(t)=\prod_n\hat{U}_n(t)= \prod_n \exp[-i\,F_n(t)\,\hat{G}_n],
\end{equation}
where the real functions $F_n(t)$ are in general time-dependent. It has been shown that these functions can be found as solutions to a set of differential equations and are determined solely by the parameters $g_n(t)$ of the Hamiltonian~\cite{wei1963lie}. The order of the operators in~\eqref{decoupled:time:evolution:operator} is not unique; a different order  changes the form of the functions $F_n(t)$, but the not the expectation value of physical quantities. A more detailed outline of these decoupling techniques may be found in~\ref{appendix:decoupling}. 

It is possible to obtain an even more explicit decoupling~\eqref{decoupled:time:evolution:operator} in the context of Gaussian states and linear (i.e., quadratic in the operators) interactions. Given a set of $N$ bosonic modes, there are $N\,(2\,N+1)$ independent quadratic Hermitian operators, which we can denote $\hat{G}_n$, that can be formed by arbitrary quadratic combinations of the creation and annihilation operators~\cite{bruschi2013time}.\footnote{For example, $\hat{G}_1=\hat a_1^{\dag}\hat a_1+\hat a_1 \hat a_1^{\dag}$ or $\hat{G}_{8}= \hat a_2^{\dag} \hat a_5^{\dag}+\hat a_5 \hat a_2$, where the numbering and ordering of the generators $\hat{G}_n$ is a matter of convenience. Work in this direction has also been done in~\cite{PhysRevD.87.084062}} We also recall that any unitary transformation induced by a quadratic operator, including the quadratic time evolution operator (\ref{general:time:evolution:operator}), can be represented by a $2N\times2N$ symplectic matrix $\boldsymbol{S}$. Combining all of this together, it can be shown~\cite{bruschi2013time} that the symplectic matrix $\boldsymbol{S}$ that represents the time evolution operator (\ref{general:time:evolution:operator}) takes the form
\begin{equation}\label{general:symplectic:decomposition}
\boldsymbol{S}=\prod_{n=1}^{N\,(2N+1)}\boldsymbol{S}_n,
\end{equation}
where the symplectic matrices $\boldsymbol{S}_n$ are given by $\boldsymbol{S}_n:=\exp[F_n(t)\,\boldsymbol{\Omega}\,\boldsymbol{G}_n]$ and the matrices $\boldsymbol{G}_n$ can be obtained through $\hat G_n=\frac{1}{2}\,\mathbb{X}^{\dag}\,\boldsymbol{G}_n\,\mathbb{X}$, with the restriction that the generator matrix $\boldsymbol{G}_n$ must be Hermitian. The techniques to obtain the real, time-dependent functions $F_n(t)$ are the same as in the more general case described above. More details can be found in~\ref{appendix:decoupling}.

\subsection{Decoupling algebra of the nonlinear Hamiltonian}
Decoupling of the Hamiltonian~\eqref{main:time:independent:Hamiltonian:to:decouple:dimensionful} can be done using different choices of the Hermitian operators $\hat{G}_n$. Here, we find it convenient to consider the \emph{closed} finite $9$-dimensional Lie algebra generated by the following set of Hermitian basis operators
\begin{align}\label{basis:operator:Lie:algebra}
	 	\hat{N}^2_a &:= (\hat a^\dagger \hat a)^2 
	& \hat{N}_a &:= \hat a^\dagger \hat a &
	\hat{N}_b &:= \hat b^\dagger \hat b \nonumber\\
	\hat{B}_+ &:=  \hat b^\dagger +\hat b &
	\hat{B}_- &:= i\,(\hat b^\dagger -\hat b) &
	 & \nonumber\\
	\hat{B}^{(2)}_+ &:= \hat b^{\dagger2}+\hat b^2 &
	\hat{B}^{(2)}_- &:= i\,(\hat b^{\dagger2}-\hat b^2) &
	 &  \nonumber\\
	\hat{N}_a\,\hat{B}_+ &:= \hat{N}_a\,(\hat b^{\dagger}+\hat b) &
	\hat{N}_a\,\hat{B}_- &:= \hat{N}_a\,i\,(\hat b^{\dagger}-\hat b), &
	 & 
\end{align}
which form the smallest set of operators in the Lie algebra that generate the Hamiltonian~\eqref{main:time:independent:Hamiltonian:to:decouple}.\footnote{In particular, the Hamiltonian~\eqref{main:time:independent:Hamiltonian:to:decouple} is generated by a linear combination of the Hermitian operators $\hat{N}_a$, $\hat{N}_b$, $\hat{B}_+$, $\hat{B}^{(2)}_+$ and $\hat{N}_a\,\hat{B}_+$.}

A generic time evolution operator $\hat{U}(\tau)$ induced by an arbitrary Hamiltonian cannot in general be written in the form~\eqref{decoupled:time:evolution:operator} for finite number of operators $\hat{G}_n$. A finite decoupling~\eqref{decoupled:time:evolution:operator} is however possible when the operators forms a finite Lie algebra that is closed under commutation. This is the case for the Hamiltonian in~\eqref{main:time:independent:Hamiltonian:to:decouple:dimensionful}, since the commutator of any two elements in the algebra~\eqref{basis:operator:Lie:algebra} yields a linear combination of the elements of the algebra. 
This allows us to make an informed ansatz for the evolution operator as we will see below. 

\subsection{Decoupling of a nonlinear time-dependent optomechanical Hamiltonian}\label{sub:algebra:techniques}
In order to achieve the main aim of this work, we need an analytical expression of the decoupling~\eqref{decoupled:time:evolution:operator} given our Hamiltonian
\eqref{main:time:independent:Hamiltonian:to:decouple:dimensionful}. While we will show that we can always obtain a formal expression for the evolution, the coefficients that make up the evolution cannot always be computed analytically, as will be clear for certain choices of the mechanical squeezing function $\tilde{D}_2(\tau)$. 

We find it convenient to proceed by collecting all quadratic terms -- including the squeezing term with $\tilde{\mathcal{D}}_2$ in~\eqref{main:time:independent:Hamiltonian:to:decouple} -- as a separate operator which we call $\hat{U}_{\mathrm{sq}}(\tau)$. This choice allows us to study the action of the quadratic and nonlinear parts separately, which can be solved through different means. Since we are interested in computing the first and second moments of the system for the purpose of computing the non-Gaussianity, it is straight-forward to apply $\hat U_{\mathrm{sq}}(\tau)$ to the operator $\hat b$ as a symplectic transformation. 

We now make an ansatz for the time-evolution operator $\hat U(\tau)$ as a finite product of the operators in the algebra:
\begin{align}\label{U}
\hat U(\tau):=&e^{-i\,\Omega_\textrm{c} \hat{N}_a\,\tau}\,\hat{U}_{\mathrm{sq}}\,e^{-i\,F_{\hat{N}_a}\,\hat{N}_a}\,e^{-i\,F_{\hat{N}^2_a}\,\hat{N}^2_a}\,e^{-i\,F_{\hat{B}_+}\,\hat{B}_+}\,e^{-i\,F_{\hat{N}_a\,\hat{B}_+}\,\hat{N}_a\,\hat{B}_+}\,e^{-i\,F_{\hat{B}_-}\,\hat{B}_-}\nonumber\\
&\times\,e^{-i\,F_{\hat{N}_a\,\hat{B}_-}\,\hat{N}_a\,\hat{B}_-},
\end{align}
where we have defined an evolution operator $\hat U_{\rm{sq}} $ as a
quadratic evolution operator of the 
mechanical degree of freedom:
\begin{align}\label{decoupling:form:to:be:used}
\hat{U}_{\mathrm{sq}} &= \overleftarrow{T} \exp\biggl[ - i \int^\tau_0 d\,\tau' \, 2\,\left(\frac{1}{2} \,+\tilde{\mathcal{D}}_2(\tau')\right)\,\hat{N}_b+ \tilde{\mathcal{D}}_2(\tau')\,\hat{B}^{(2)}_+ \biggr].
\end{align}
Here, we have effectively divided the Hamiltonian into a quadratic  contribution $\hat U_{\rm{sq}}(\tau)$ and a remaining nonlinear contribution with the addition of linear term proportional to $\hat B_\pm$.

The coefficients in the decoupling above can now be obtained in terms of definite integrals. The full calculations can be found in~\ref{appendix:Hamiltonian:nonlinear:decoupling}. We obtain
\begin{align}\label{sub:algebra:decoupling:solution}
F_{\hat{N}_a}&= -2\,\int_0^\tau\,\mathrm{d}\tau'\,\tilde{\mathcal{D}}_1(\tau')\,\Im\xi(\tau')\int_0^{\tau'}\mathrm{d}\tau''\,\tilde{\mathcal{G}}(\tau'')\,\Re\xi(\tau'')\, \nonumber \\
&\quad -2 \,\int^\tau_0\,\mathrm{d}\tau' \,\tilde{\mathcal{G}}(\tau')\, \Im \xi \, \int^{\tau'}_0 \,\mathrm{d}\tau''\, \tilde{\mathcal{D}}_1(\tau'') \, \Re \xi(\tau'') \, ,  \nonumber\\
F_{\hat{N}^2_a}&= 2\,\int_0^\tau\,\mathrm{d}\tau'\,\tilde{\mathcal{G}}(\tau')\,\Im\xi(\tau')\int_0^{\tau'}\mathrm{d}\tau''\,\tilde{\mathcal{G}}(\tau'')\,\Re\xi(\tau'') \, , \nonumber\\
F_{\hat{B}_+}&=\int_0^\tau\,\mathrm{d}\tau'\,\tilde{\mathcal{D}}_1(\tau')\,\Re\xi(\tau') \, , \nonumber\\
F_{\hat{B}_-}&=- \int_0^\tau\,\mathrm{d}\tau'\,\tilde{\mathcal{D}}_1(\tau')\,\Im\xi(\tau') \, , \nonumber\\
F_{\hat{N}_a\,\hat{B}_+}&=- \int_0^\tau\,\mathrm{d}\tau'\,\tilde{\mathcal{G}}(\tau')\,\Re\xi(\tau') \, , \nonumber\\
F_{\hat{N}_a\,\hat{B}_-}&=\int_0^\tau\,\mathrm{d}\tau'\,\tilde{\mathcal{G}}(\tau')\,\Im\xi(\tau'),
\end{align}
where we have introduced the function
\begin{equation} \label{eq:def:xi}
\xi(\tau) = P_{11}(\tau)-i\,\int_0^\tau\,\mathrm{d}\tau'\,P_{22}(\tau),
\end{equation}
and where $P_{11}(\tau)$ and $P_{22}(\tau)$ are defined below. 

The only problem that we encounter is computing a decoupled form of $\hat U_{\rm{sq}}$ in~\eqref{decoupling:form:to:be:used}. In fact, it has been shown that decoupling of the evolution operator does not yield analytical results except in very specific cases~\cite{moore2016tuneable}. For our purposes, this is not problematic, because we can calculate the action of $\hat U_{\rm{sq}}$ on the first and second moments analytically using the covariance matrix formalism. 

\subsection{Action of the single-mode squeezing component}
Although it is not possible to obtain an analytical decoupling of~\eqref{decoupling:form:to:be:used}, it is possible to obtain an expression for its action on the operators $\hat{b}$ and $\hat{b}^\dag$. First of all, we note that a Bogoliubov transformation of a single mode operator always has  the general expression $\hat{U}_{\mathrm{sq}}^\dag\,\hat{b}\,\hat{U}_{\mathrm{sq}}=\alpha(\tau)\,\hat{b}+\beta(\tau)\,\hat{b}^\dag$, see~\cite{moore2016tuneable}. The challenge is to find an explicit expression for the Bogoliubov coefficients $\alpha(\tau)$ and $\beta(\tau)$, which satisfy the only nontrivial Bogoliubov identity $|\alpha(\tau)|^2-|\beta(\tau)|^2=1$. 
In~\ref{appendix:Hamiltonian:nonlinear:decoupling} we show that the Bogoliubov coefficients $\alpha(\tau)$ and $\beta(\tau)$ can be obtained through 
\begin{align} \label{eq:bogoliubov:coefficients}
\alpha(\tau)=&\frac{1}{2}\biggl[P_{11}(\tau)+P_{22}(\tau)-i \int_0^\tau \mathrm{d}\tau' P_{22}(\tau')  -i \int_0^\tau \mathrm{d}\tau' (1+4\,\tilde{\mathcal{D}}_2(\tau'))\,P_{11}(\tau')\biggr] \, , \nonumber\\
\beta(\tau) =&\frac{1}{2}\biggl[P_{11}(\tau)-P_{22}(\tau)+i \int_0^\tau \mathrm{d}\tau' P_{22}(\tau')  -i \int_0^\tau \mathrm{d}\tau' (1+4\,\tilde{\mathcal{D}}_2(\tau'))\,P_{11}(\tau')\biggr] \,  ,
\end{align}
whose explicit form can be obtained once an explicit expression of the functions $P_{11}(\tau)$ and $P_{22}(\tau)$ is found. Given the previously defined function $\xi(\tau)$ in~\eqref{eq:def:xi}, we also find  $\alpha(\tau) = \frac{1}{2}(\xi(\tau) + i \dot{\xi}(\tau))$ and $\beta(\tau) = \frac{1}{2}(\xi^*(\tau) + i \dot{\xi}^*(\tau))$, where dotted functions imply differentiation with respect to $\tau$. 

The two functions $P_{11}$ and $P_{22}$ are determined by the two following uncoupled differential equations:
\begin{align}\label{differential:equation:written:in:paper}
\ddot{P}_{11}+(1+4\,\tilde{\mathcal{D}}_2(\tau))\,P_{11}&=0\nonumber\\
\ddot{P}_{22}-4\frac{\dot{\tilde{\mathcal{D}}}_2(\tau)}{1+4\,\tilde{\mathcal{D}}_2(\tau)}\,\dot{P}_{22}+(1+4\,\tilde{\mathcal{D}}_2(\tau))\,P_{22}&=0,
\end{align}
where the dot stands for a derivative with respect to $\tau$ and the initial conditions are $P_{11}(0)=P_{22}(0)=1$ and $\dot{P}_{11}(0)=\dot{P}_{22}(0)=0$. Furthermore, the second equation in~\eqref{differential:equation:written:in:paper} can be written as
\begin{equation} \label{eq:IP22}
\ddot{I}_{P_{22}}  + ( 1 + 4 \, \tilde{\mathcal{D}}_2 (\tau) ) I_{P_{22}} = 0 \, ,
\end{equation}
which now has boundary conditions $I_{P_{22}}(0) = 0$ and $\dot{I}_{P_{22}} = 1$, and where
\begin{equation}
I_{P_{22}} = \int^\tau_0 \mathrm{d}\tau' P_{22} ( \tau') \, .
\end{equation}
The solutions to $P_{11}$ and $P_{22}$ (or $I_{P_{22}}$) can then be used in the expressions~\eqref{sub:algebra:decoupling:solution},~\eqref{eq:def:xi} and~\eqref{eq:bogoliubov:coefficients} to find the full dynamics of the state. While the solutions must in general be obtained numerically, we anticipate that there are scenarios, such as constant $\tilde{\mathcal{D}}_2$, where the equations above  admit analytical solutions. 

\subsection{Initial state}
In this work, we assume that both the optical and mechanical modes are initially in a coherent states, namely $\ket{\mu_{\mathrm{c}}}$ and $\ket{\mu_{\mathrm{m}}}$ respectively, defined as the eigenstates of the annihilation operators, i.e., $\hat{a} \ket{\mu_{\mathrm{c}}} = \mu_{\mathrm{c}} \ket{\mu_{\mathrm{c}}}$ and $\hat{b} \ket{\mu_{\mathrm{m}}} = \mu_{\mathrm{m}} \ket{\mu_{\mathrm{m}}}$. 

For optical fields, this is generally a good assumption. On the other hand, within optomechanical systems the mechanical element is typically found initially in a thermal state. Our choice of initial coherent state can be generalised to that of a thermal state in a straight-forward manner, that is, by integrating over the coherent state parameter with an appropriate kernel 
(as any thermal state may be written as Gaussian average of coherent states, as per its P-representation). 
Restricting ourselves hence to a single coherent state also for the mechanical oscillator, the initial state $|\Psi(0)\rangle$  reads
\begin{equation}\label{initial:state:two}
|\Psi(0)\rangle = \ket{\mu_{\mathrm{c}}}\otimes \ket{\mu_{\mathrm{m}}}.
\end{equation}
We now proceed to apply~\eqref{U} to this state. 

\smallskip
\subsection{Full state evolution for general dynamics}
For completeness, we present here the full state derived under the evolution with $\hat{U}(\tau)$ for two initially coherent states~\eqref{initial:state:two}:
\begin{align}\label{non:linear:state:evolution}
\ket{\Psi(\tau)} =& \,    e^{ - i  \,\left( F_{\hat B_+}\,  F_{\hat B_-}+\Im\left(K  \, \mu_{\rm{m}} \right)\right) } \,e^{- \frac{1}{2}|\mu_{\rm{c}}|^2}\,\sum_n \biggl[ \frac{\mu_{\rm{c}}^n}{\sqrt{n!}} \,   e^{- i \,\left(F_{\hat{N}^2_a}+  F_{\hat N_a \, \hat B_+} \, F_{\hat N_a \, \hat B_-}  \right)\, n^2}\nonumber \\
&e^{ - i   \, \left( \Omega_\textrm{c}\,\tau+F_{\hat{N}_a}+F_{\hat N_a \, \hat B_+} \, F_{\hat B_-} + F_{\hat N_a \, \hat B_-} \, F_{\hat B_+}+\Im\left(K_{\hat N_a}\, \mu_{\rm{m}} \right)\right)\,  n} \, 
\ket{n} \otimes  \ket{\phi_n, \tilde{\mathcal{D}}_2(\tau)} \biggr] \, ,
\end{align}
where we have defined $K:=F_{\hat{B}_-} + i F_{\hat{B}_+}$ and
$K_{\hat{N}_a}:=F_{\hat{N}_a \, \hat{B}_- }+ i F_{\hat{N}_a \, \hat{B}_+}$, where $|\phi_n(\tau), \tilde{\mathcal{D}}_2(\tau) \rangle$ is a mechanical coherent squeezed state where 
$|\phi_n(\tau), \tilde{\mathcal{D}}_2(\tau) \rangle = \hat{U}_{\mathrm{sq}}(\tau) |\phi_n(\tau)\rangle $,  and where $|\phi_n(\tau)\rangle$ is a coherent state with $\phi_n(\tau):= K^*+ n \, K_{\hat N_a}^*+\mu_{\mathrm{m}}$. Note that, in the above, we have kept the dependence on $\tau$ implicit 
but, in general, all exponentials will oscillate in time. We also note that the state~\eqref{non:linear:state:evolution} contains all terms that have been considered in the literature before, including the contributions from a constant nonlinear light--matter term~\cite{bose1997preparation}, a time-dependent light--matter term~\cite{qvarfort2019enhanced}, and a linear, mechanical displacement term~\cite{qvarfort2018gravimetry}. The main addition here is $\hat{U}_{\mathrm{sq}}$, which takes on the trivial form $\hat U_{\rm{sq}} = e^{- i \tau \hat b^\dag \hat b}$ only if $\tilde{\mathcal{D}}_2 \neq 0$. 

We note here that the expression of~\eqref{non:linear:state:evolution} allows us to compute the reduced state of the mechanics $\hat{\rho}_{\text{Mech.}}(\tau)$ at any time $\tau$, which reads
\begin{align}\label{non:linear:reduced:mechanical:state:evolution}
\hat{\rho}_{\text{Mech.}}(\tau) &=   e^{- |\mu_{\rm{c}}|^2}\,\sum_n \frac{|\mu_{\rm{c}}|^{2\,n}}{n!} 
\ket{\phi_n, \tilde{\mathcal{D}}_2(\tau)} \bra{\phi_n, \tilde{\mathcal{D}}_2(\tau)}.
\end{align}
We are now ready to consider the non-Gaussianity of the evolved state. 

\section{Measures of deviation from Gaussianity} \label{sec:measure}
The time evolution~\eqref{U} is not linear. Therefore, an initial Gaussian state will evolve, in general, to a non-Gaussian state. Here we ask: \emph{is it possible to quantify the deviation from Gaussianity of the state evolving from an initial Gaussian state}?

To answer this question we need to find one or more suitable measures of deviation from Gaussianity. In this work we choose to employ a measure based on the comparison between the entropy of the final state and that of a reference Gaussian state~\cite{genoni2008quantifying}. This measure can be understood simply as follows: Let us assume that our initial state $\hat{\rho}(0)$ evolves into the state $\hat{\rho}(\tau)$ at time $\tau$. We can analytically compute the first and second moments of $\hat{\rho}(\tau)$. We then consider a Gaussian state, 
which we call $\hat{\rho}_{\mathrm{G}} (\tau)$, with the same first and second moments as $\hat{\rho}(\tau)$. It has been shown that this reference state $\hat \rho_{\rm{G}}(\tau)$ is indeed the Gaussian state that is closest to $\hat \rho(\tau)$~\cite{marian2013relative}.
In general, since $\hat{\rho}(\tau)$ is not determined uniquely by its first and second moments, as is the case for Gaussian states, the two states do not coincide, i.e., $\hat{\rho}(\tau)\neq\hat{\rho}_{\mathrm{G}}(\tau)$. 

One way to quantify the difference between two states $\hat{\rho}$ and $\hat{\rho}_{\rm{G}}$ is via the relative entropy $S(\hat{\rho},\hat{\rho}_{\rm{G}})$. It has been shown that the relative entropy $S(\hat{\rho}(\tau),\hat{\rho}_{\mathrm{G}}(\tau))$ is equivalent to the difference between the local von-Neumann entropies of the states~\cite{genoni2008quantifying}.  
The measure of non-Gaussianity $\delta(\tau)$ can therefore be defined as 
\begin{align}\label{measure:of:non:gaussianity}
\delta(\tau):=S(\hat{\rho}_{\mathrm{G}}(\tau))-S(\hat{\rho}(\tau)),
\end{align}
where $S(\hat{\rho})$ is the usual von Neumann entropy of a state $\hat{\rho}$,  defined by $S(\hat{\rho}):=-\textrm{Tr}(\hat{\rho}\,\ln\hat{\rho})$.

Since the reference state $\hat \rho_{\rm{G}}(\tau)$ is Gaussian, it is fully characterised by its first and second moments. We  therefore turn to the continuous variable formalism and consider the covariance matrix $\boldsymbol{\sigma}$ of $\hat \rho_{\rm{G}}(\tau)$. Furthermore, we can define the von Neumann entropy   $S(\boldsymbol{\sigma})$ of the state as given by $S(\boldsymbol{\sigma}) = \sum_j s_V(\nu_j)$, where  $j$ runs over all the modes, $\nu_j$ are the
symplectic eigenvalues of $\boldsymbol{\sigma}$ and $s_V(\nu_j)$ is the binary entropy of the state defined by $s_V(x) = \frac{x + 1}{2} \ln \left( \frac{x+1}{2} \right) - \frac{x - 1}{2} \ln \left( \frac{x - 1}{2} \right)$. The symplectic eigenvalues are defined as $\nu_j = |\lambda_j|$, where $\lambda_j$ are the eigenvalues of the matrix $i \, \boldsymbol{\Omega } \boldsymbol{\sigma}$, where  $\boldsymbol{\Omega}$ is the $4\times4$ symplectic form. Note that, for all physical states, the eigenvalues satisfy $1 \leq \nu_j$. 
It follows from the above that a state is non-Gaussian at time $\tau$ if and only if $\delta(\tau)\neq0$. 

An alternative interpretation of this measure is as a quantification of the impurity of $\hat \rho_{\mathrm{G}}(\tau)$. While the initial state $\hat \rho(\tau)$  remains pure throughout the evolution, such that $S(\hat \rho(0)) = S(\hat \rho(\tau)) = 0$, the constructed Gaussian reference state  $\hat{\rho}_{\mathrm{G}}(\tau)$ does not remain pure. This is not due to external noise, but occurs because we are, loosely speaking, 
`approximating' the actual state with the Gaussian subset of states. 

In this work, we consider unitary dynamics only. If the initial state $\hat \rho(0)$ is pure at $\tau = 0$, it  stays pure throughout its evolution, and the measure thus reduces to 
\begin{equation}
\delta(\tau) = S(\hat \rho_{\mathrm{G}}(\tau)), 
\end{equation}
where $\hat \rho_{\mathrm{G}}(\tau)$ is the Gaussian reference state constructed form the first and second moments of $\hat \rho(\tau)$. Our challenge is therefore to compute the symplectic eigenvalues $\nu_j$ in order to be able to find the expression of $\hat \rho_{\mathrm{G}}$. Using the expression for the decoupled time evolution operator~\eqref{decoupling:form:to:be:used}, we can obtain all of the elements of $\boldsymbol{\sigma}$. These expressions are cumbersome and can be found in~\ref{appendix:CM}. The expression for the symplectic eigenvalues are too involved and we choose not to print them.

Before we proceed, we also consider the effect of mechanical squeezing on the symplectic eigenvalues. In the continuous variable formalism, a squeezing operation can be represented as a symplectic transformation $\boldsymbol{S}$ acting on the covariance matrix $\boldsymbol{\sigma}$ through congruence: $\boldsymbol{S}\,\boldsymbol{\sigma}\,\boldsymbol{S}^\dag$. All symplectic transformations leave the symplectic eigenvalues $\nu_j$ invariant when acted upon in this way. In this work, however, we consider the inclusion of mechanical squeezing as a term in the Hamiltonian, which acts on the fully non-Gaussian state $\hat \rho(\tau)$. The presence of the nonlinearity means that the squeezing term acts non-trivially on the full state and can actually affect the symplectic eigenvalues of the Gaussian reference state. 
 The mechanical squeezing parameter $\tilde{D}_2(\tau)$ affects all $F$-coefficients, meaning that not only the mechanical subsystem but also the optical subsystem will be affected.

\section{Application: Non-Gaussianity for optomechanical systems}\label{sec:applications}

In this section, we demonstrate the applicability of our techniques by computing the non-Gaussianity of an optomechanical system. The solutions allow us to consider both constant and time-dependent light--matter couplings, however, in order to obtain explicit results we choose to set $\tilde{\mathcal{G}}(\tau) \equiv \tilde{g}_0$ constant throughout this work and refer the reader to~\cite{qvarfort2019enhanced} for a thorough analysis of the non-Gaussianity of the optomechanical state given a time-dependent light--matter coupling. 
Furthermore, we set $\tilde{\mathcal{D}}_1 =  0$ throughout the remainder of this work. Since the second moments are not affected by a displacement term, the non-Gaussian character of a state remains unchanged\cite{serafini2017quantum}. 

We consider two cases in this section: one where we assume that the mechanical squeezing parameter is constant, and one where the mechanical squeezing is periodic. Our goal is to derive some general bounds on the non-Gaussianity of the state for each case. First, however, we provide some bounds on the total amount of non-Gaussianity.

\subsection{Bounding the full measure}  \label{sec:bounds}
The exact expression for $\delta(\tau)$ is long and cumbersome due to the complex expressions of the covariance matrix elements~\eqref{eq:CM:elements}. We therefore provide bounds to the measure that can be expressed as simple analytic functions. Since the  full measure $\delta(\tau)$ is an entropy, it can be bounded from above and below by the means of  the Araki--Lieb inequality~\cite{araki2002entropy}, which reads 
\begin{equation} \label{eq:araki:lieb}
|S(\hat{\rho}_{A})-S(\hat{\rho}_{B})|\leq S(\hat{\rho}_{AB})\leq S(\hat{\rho}_{A})+S(\hat{\rho}_{B}) \, ,
\end{equation}
 where $\hat \rho_{AB}$ is the full bipartite state and $\hat \rho_A$ and $\hat \rho_B$ are the traced-out subsystems. This inequality allows us to bound the behaviour of the full measure $\delta(\tau)$ in terms of the subsystem entropies. We therefore proceed to define the lower and upper bounds as $\delta_{\rm{min}}(\tau) :=|S(\hat{\rho}_{A})-S(\hat{\rho}_{B})|$ and $\delta_{\rm{max}}(\tau) := S(\hat{\rho}_{A})+S(\hat{\rho}_{B})$. 

In our case, the subsystems are the traced out optical state $\hat \rho_{\rm{Op}}$ and the traced out mechanical state $\hat \rho_{\rm{Me}}$. To quantify the entropy of the subsystems, we must find the symplectic eigenvalues of the optical and mechanical subsystems, which we call $\nu_{\rm{Op}}$ and $\nu_{\rm{Me}}$ respectively. Lengthy algebra (see~\ref{appendix:CM}), the use of the Bogoliubov identities $|\alpha|^2=1+|\beta|^2$ and $\alpha\,\beta^*=\alpha^*\,\beta$, and observing that $|E_{\hat B_+ \hat B_-}|^2 = e^{- |K_{\hat N_a}|^2}$  (see~\ref{appendix:CM} for a definition of $E_{\hat B_+ \hat B_-} $ and its appearance in the first and second moments) allow us to find
\begin{align}\label{sympelctic:eigenvalue:reduced:state}
\nu_{\rm{Op}}^2 =& 1 +4  \, |\mu_{\mathrm{c}}|^2 \left( 1 - e^{-4|\mu_{\mathrm{c}}|^2 \sin^2{\theta/2}} \,e^{- |K_{\hat N_a}|^2} \right) 
+ 4 \, |\mu_{\mathrm{c}}|^4 \biggl( 1 - 2 \,  e^{-4|\mu_{\mathrm{c}}|^2 \sin^2{\theta/2}} e^{- |K_{\hat N_a}|^2 }\nonumber\\
&  -  e^{- 4 |\mu_{\rm{c}}|^2 \sin^2\theta} \, e^{-4 |K_{\hat N_a}|^2 } 2 \, e^{- 3|K_{\hat N_a}|^2} \,  \Re \left\{  e^{i \theta} \, e^{|\mu_{\mathrm{c}}|^2 ( e^{2i \theta} - 1)} \, e^{2 |\mu_{\rm{c}}|^2 (e^{- i \theta} - 1)} \right\}  \biggr) \, , \nonumber \\
\nu_{\mathrm{Me}}^2=&1+4\,|K_{\hat N_a}|^2 |\mu_{\mathrm{c}}|^2 \, ,
\end{align}
where we recall that  $K_{\hat N_a}:=F_{\hat{N}_a \, \hat{B}_- } +i \, F_{\hat{N}_a \, \hat{B}_+ }$, and where we have defined $\theta(\tau) = 2 \left( F_{\hat N_a^2 }  + F_{\hat N_a \hat B_+} F_{\hat N_a \hat B_-}  \right)$. 

The optical symplectic eigenvalue~\eqref{sympelctic:eigenvalue:reduced:state} is bounded by
\begin{align}
 \nu_{\rm{Op}} < \sqrt{1 + 4|\mu_{\rm{c}}|^2 + 4 |\mu_{\rm{c}}|^4 } \, , 
\end{align}
which can be inferred by noting that $K_{\hat  N_a}$ is generally given by an oscillating function multiplied by the strength of the optomechanical coupling $\tilde{g}_0$. For specific $\tau$ which ensures that $|K_{\hat N_a}|^2 \neq 0$,  and then considering $\tilde{g}_0 \gg 1$, the exponentials in $\nu_{\rm{Op}}$ in~\eqref{sympelctic:eigenvalue:reduced:state} are suppressed, which means we are left with $\nu_{\rm{Op}} \sim \sqrt{1+4\,|\mu_{\mathrm{c}}|^2+4|\mu_{\mathrm{c}}|^4} $.

 When $S(\hat \rho_{\rm{Op}}) \gg S(\hat \rho_{\rm{Me}})$ or $S(\hat \rho_{\rm{Op}}) \ll S(\hat  \rho _{\rm{Me}})$, the bipartite entropy of the Gaussian reference state $S(\hat \rho_{\rm{G}})$ is approximately equal to one of the subsystem entropies. To determine when this is the case, we consider the maximum values of $\nu_{\rm{Op}} $ and $\nu_{\rm{Me}}$.
In general, when  $|\mu_{\rm{c}}|^2 \gg1$, and when $|K_{\hat N_a}|^2 \gg1$, which requires $\tilde{g}_0 \gg1$ and specific values of $\tau$,  the eigenvalues $\nu_{\mathrm{Op}}$ and $\nu_{\mathrm{Me}}$ tend to their maximum values  $\nu_{\rm{Op}, \rm{max}}$ and $\nu_{\rm{Me},\rm{max}}$, which are  
\begin{align}\label{sympelctic:eigenvalue:reduced:state:maximum:value}
\nu_{\mathrm{Op,max}}\sim&1+2\,|\mu_{\mathrm{c}}|^2 \, ,\nonumber\\
\nu_{\mathrm{Me,max}}\sim&2\, |K_{\hat N_a}| \,|\mu_{\mathrm{c}}| \, .
\end{align}
We note that there are three distinct scenarios which arise from the comparison of the coherent state parameter $|\mu_{\rm{c}}|^2$ and the function $K_{\hat N_a}$:
\begin{itemize}
\item[i)]  First, we assume that $1\ll |K_{\hat N_a} |\ll2|\mu_{\mathrm{c}}|$, which implies $\delta(\tau) \sim S(\hat \rho_{\rm{Op}}) = s_V(\nu_{\rm{Op}})$. 
Here, the non-Gaussianity is well-approximated by 
	\begin{align} \label{eq:many:photons}
	S(\hat \rho_{\mathrm{G}}(\tau))\sim s_V(1+2\,|\mu_{\mathrm{c}}|^2);
	\end{align} 
	\item[ii)] Secondly, we assume that $1 \ll 2|\mu_{\rm{c}}|\ll |K_{\hat N_a}| $, which implies that $\delta(\tau) \sim S(\nu_{\rm{Me}}) = s_V(\nu_{\rm{Me}})$. Thus we find that
	\begin{align} \label{eq:nG:large:KNa}
	S(\hat \rho_{\mathrm{G}}(\tau))\sim s_V(2\,|K_{\hat N_a}|\,|\mu_{\mathrm{c}}|);
	\end{align}
	\item[iii)] Finally, when $ |K_{\hat N_a}|\sim2|\mu_{\mathrm{c}}|$ and $|\mu_{\mathrm{c}}|\gg1$, we have $S(\hat{\rho}_{A})\sim S(\hat{\rho}_{B})$. In this case, the Araki--Lieb bound is not very informative since the left-hand-side is zero and must evaluate the non-Gaussianity exactly. 
\end{itemize}
Note that the first two cases might occur only for short periods of time $\tau$ since $K_{\hat N_a}$ is oscillating.
Furthermore, we note that the squeezing parameter $\tilde{\mathcal{D}}_2(\tau)$ affects the peak value of the non-Gaussianity because it enters into $|K_{\hat N_a}|$ through the $F$-coefficients~\eqref{sub:algebra:decoupling:solution}. The dependence is non-trivial, but we will consider the analytic case for constant squeezing below. However, in general, when  $|\mu_{\rm{c}}|\gg |K_{\hat N_a}|$, we see from~\eqref{eq:many:photons} that the non-Gaussianity is independent of $\tilde{\mathcal{D}}_2(\tau)$ and can be accurately modelled by the standard optomechanical Hamiltonian without mechanical squeezing. 

Let us now consider two specific cases where the squeezing term is either constant or modulated.

\subsection{Applications: Constant squeezing parameter}\label{sec:constant:squeezing}
Here we assume that the rescaled squeezing parameter is constant, with $\tilde{\mathcal{D}}_2(\tau) = \tilde{d}_2$. This case is equivalent to the case where the mechanical oscillation frequency $\omega_{\mathrm{m}}$ is shifted by a constant amount and where the initial state is a squeezed coherent state, see~\ref{appendix:CSq}. We begin by deriving analytic expressions for the coefficients in~\eqref{sub:algebra:decoupling:solution} given this choice of parameters. 

\subsubsection{Decoupled dynamics}\label{decoupled:dynamics}

\begin{figure*}[t!]
\subfloat[ \label{fig:quad:mum0:d20}]{%
  \includegraphics[width=.25\linewidth, trim = 00mm 0mm 0mm 0mm]{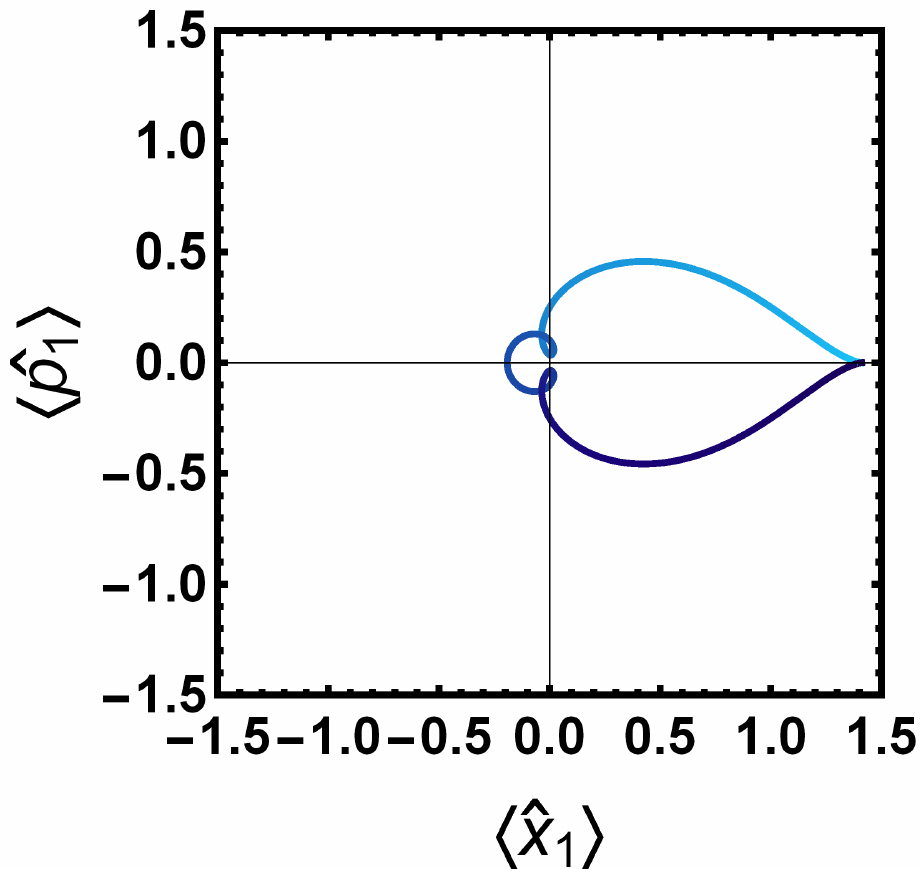}%
}\hfill
\subfloat[ \label{fig:quad:mum0:d201}]{%
  \includegraphics[width=.25\linewidth, trim = 00mm 0mm 0mm 0mm]{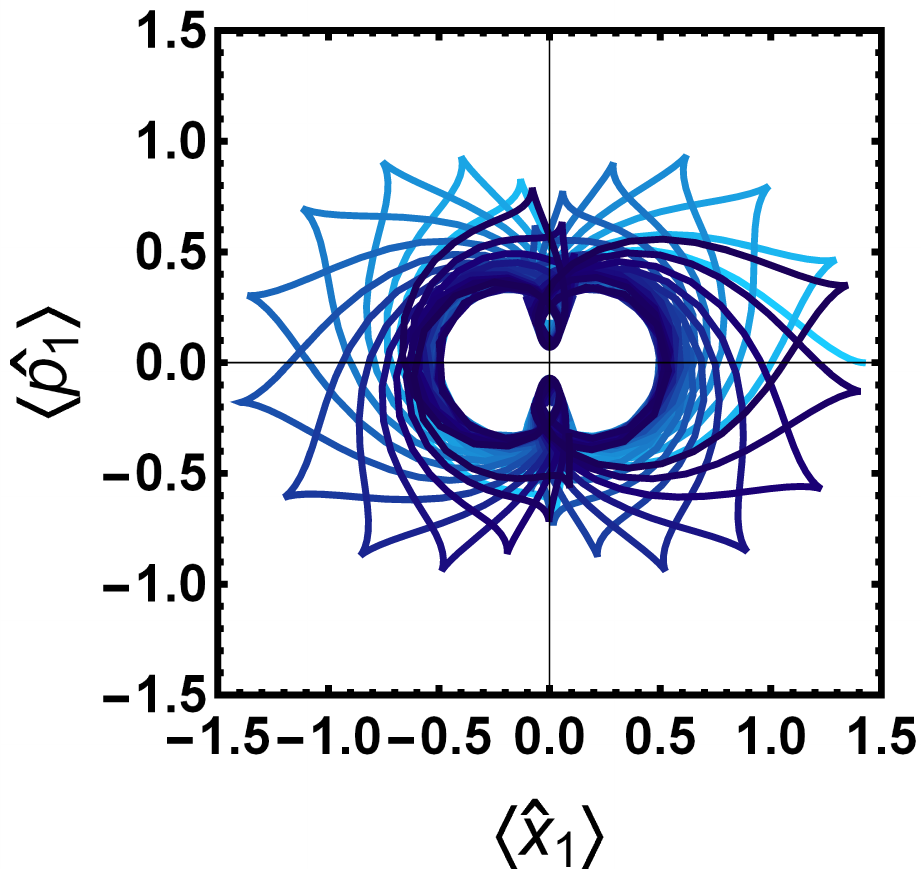}%
}\hfill
\subfloat[
\label{fig:quad:mum0:d205}]{%
  \includegraphics[width=.25\linewidth, trim = 00mm 0mm 0mm 0mm]{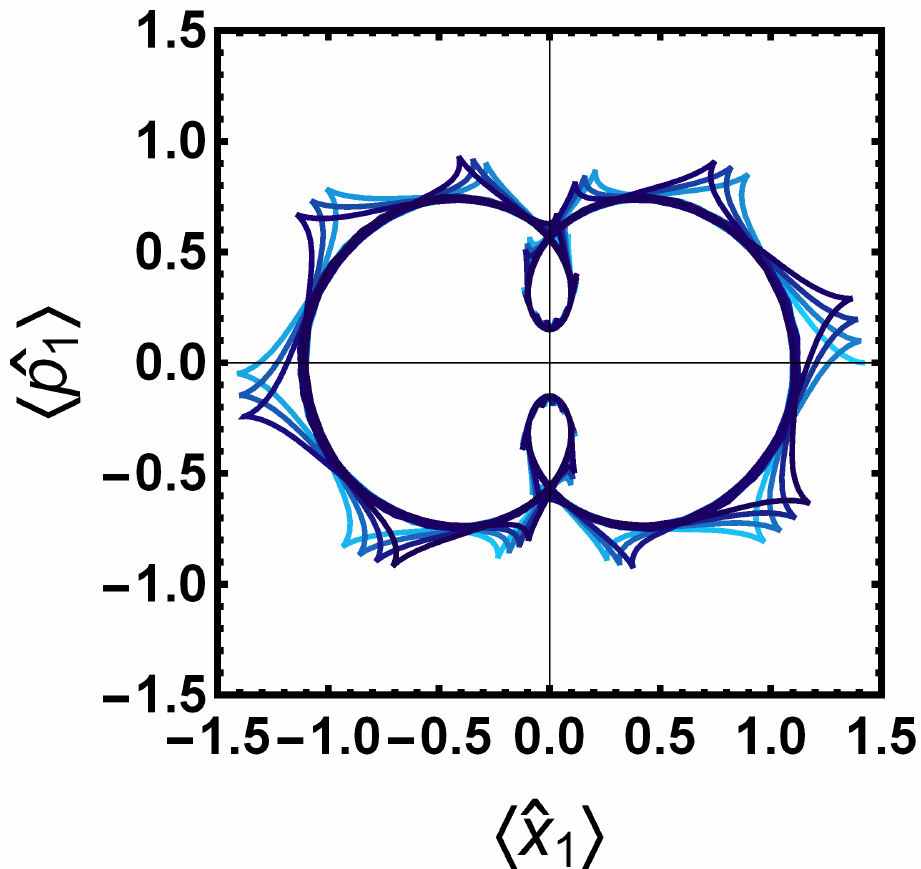}%
}\hfill
\subfloat[
\label{fig:quad:mum0:d21}]{%
  \includegraphics[width=.25\linewidth, trim = 00mm 0mm 0mm 0mm]{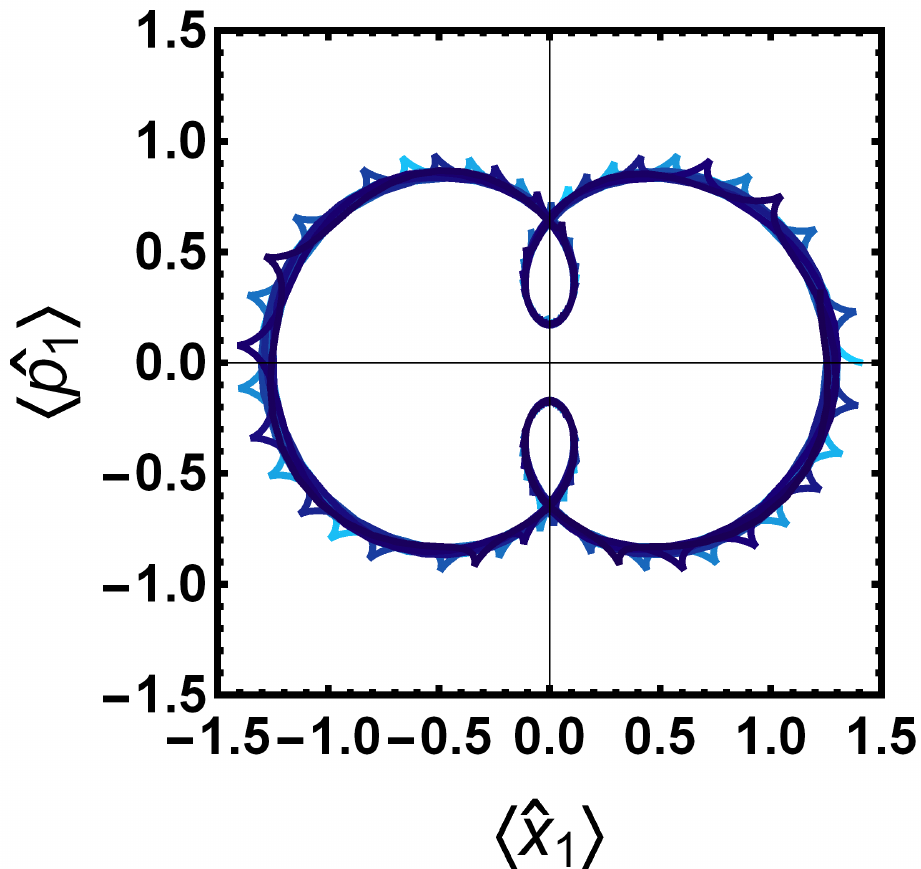}%
}\hfill
\subfloat[
\label{fig:quad:mum1:d20}]{%
  \includegraphics[width=.25\linewidth, trim = 00mm 0mm 0mm 0mm]{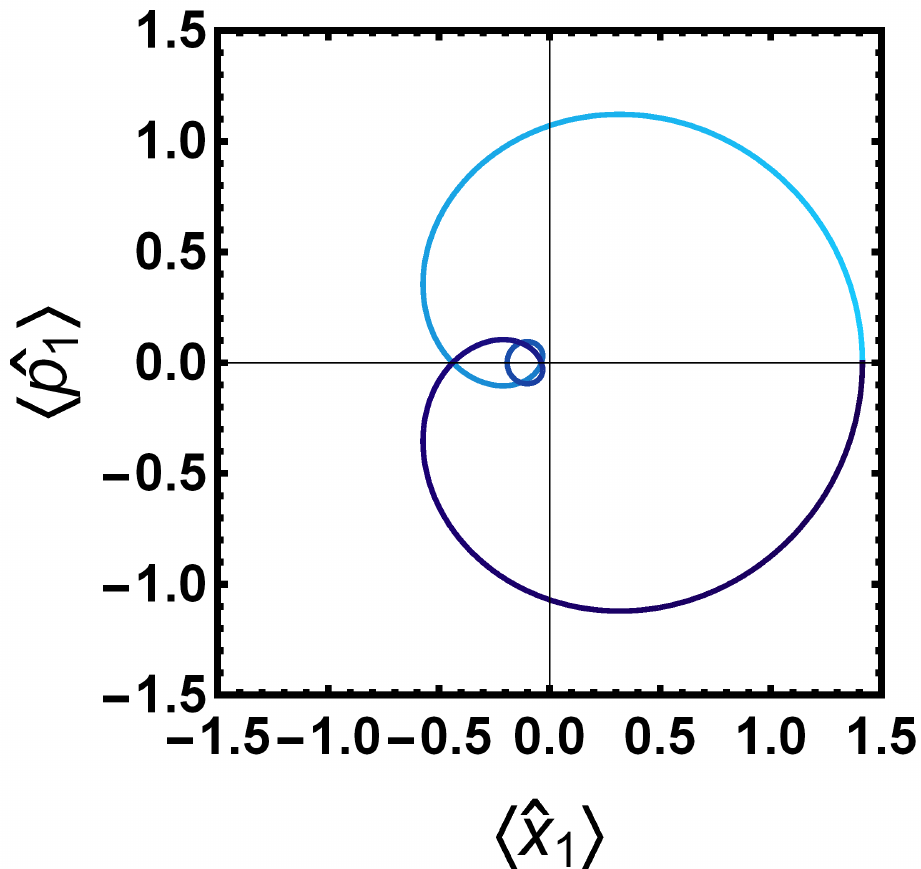}%
}\hfill
\subfloat[
\label{fig:quad:mum1:d201}]{%
  \includegraphics[width=.25\linewidth, trim = 00mm 0mm 0mm 0mm]{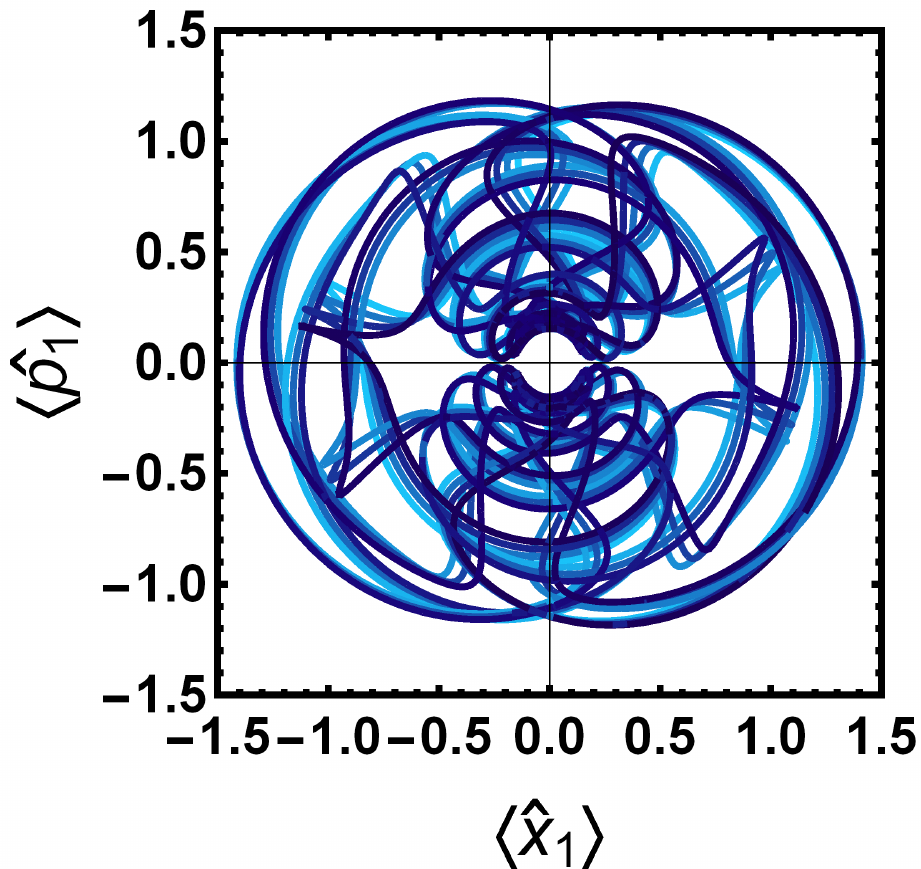}%
}\hfill
\subfloat[
\label{fig:quad:mum1:d205}]{%
  \includegraphics[width=.25\linewidth, trim = 00mm 0mm 0mm 0mm]{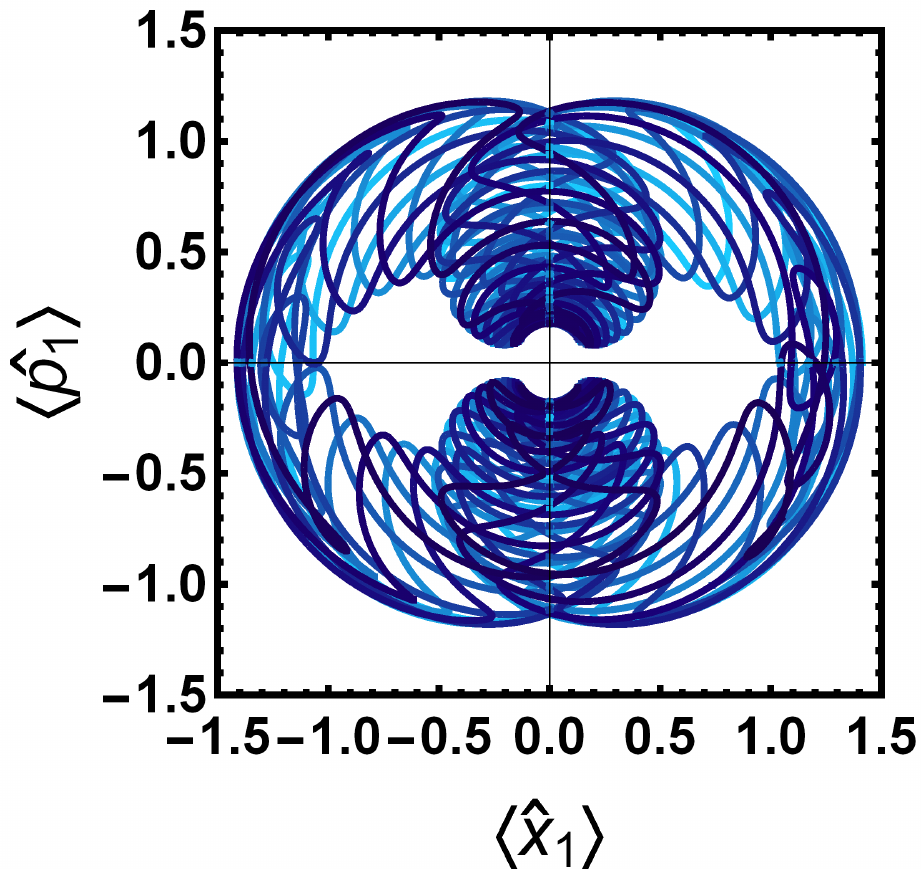}%
}\hfill
\subfloat[
\label{fig:quad:mum1:d21}]{%
  \includegraphics[width=.25\linewidth, trim = 00mm 0mm 0mm 0mm]{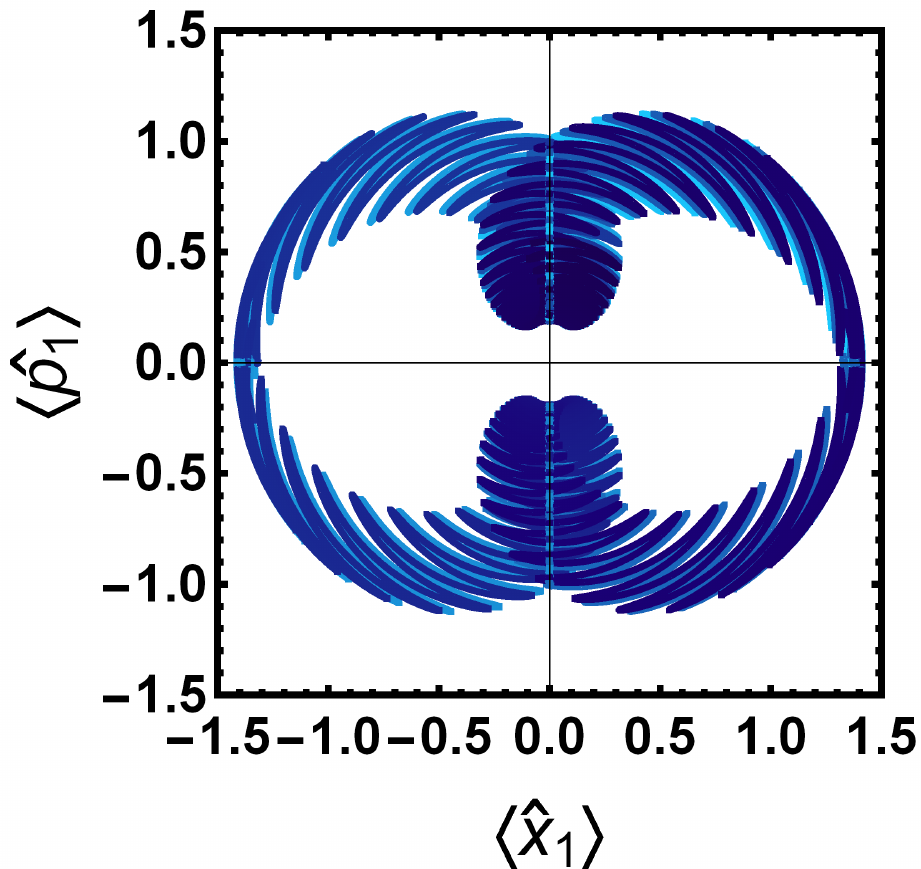}%
}\hfill
\caption{\textbf{Optical quadratures of an optomechanical system with constant mechanical squeezing}. Both rows show the plots of  $\braket{\hat{x}_1} = \braket{\hat a^\dag + \hat a}/\sqrt{2}$ vs.\ $\braket{\hat{p}_1} = i \braket{ \hat a^\dag - \hat a }/\sqrt{2}$. The line starts as light blue at $\tau = 0$ and gradually becomes darker as $\tau$ increase. Plot \textbf{(a)} and \textbf{(e)} show the quadratures for the time range $\tau \in (0, 2\pi)$ and all others have $\tau \in (0, 100\pi)$.  The first row shows the quadratures for the following values: an optical  coherent state parameter $\mu_{\mathrm{c}} = 1$, a mechanical coherent state parameter $\mu_{\mathrm{m}} = 0$, a light--matter coupling strength of $\tilde{g}_0 = 1$, and \textbf{(a)} the squeezing parameter $\tilde{d}_2 = 0$, \textbf{(b)} $\tilde{d}_2 = 0.1$, \textbf{(c)} $\tilde{d}_2 = 0.5$ and \textbf{(d)} $\tilde{d}_2 = 1$. The second row shows the quadratures for values $\mu_{\mathrm{c}} = 1$, $\mu_{\mathrm{m}} = 1$, $\tilde{g}_0 = 1$ and \textbf{(e)} $\tilde{d}_2 = 0$, \textbf{(f)} $\tilde{d}_2 = 0.5$, \textbf{(g)} $\tilde{d}_2 = 1$ and \textbf{(h)} $\tilde{d}_2 = 5$. The increased initial excitation of the mechanical oscillator leads to increased complexity in the quadrature trajectories. A limiting behaviour for large $\tilde{d}_2$ does however appear in which the state is confined to an increasingly narrow trajectory in phase space. Finally, we note that the spikes in \textbf{(b)}, \textbf{(c)}, and \textbf{(d)} appear less pronounced compared with their actual appearance due to restrictions in image resolution. }
\label{fig:constant:squeezing:quadratures}
\end{figure*}

We use  the methods discussed in Section~\ref{tools} to start by solving the differential equations~\eqref{differential:equation:written:in:paper}. We find the solutions $P_{11} =P_{22}= \cos{ \zeta \tau }$, where we define $\zeta := \sqrt{1 + 4 \, \tilde{d}_2}$. This, in turn, yields the following Bogoliubov coefficients (defined in~\eqref{eq:bogoliubov:coefficients}):
\begin{align}
\alpha(\tau) &= \frac{1}{2} \left( 2 \cos{\zeta \tau} - \frac{i}{\zeta} \left( 1 + \zeta^2  \right)\sin{\zeta \tau} \right),  \nonumber \\
\beta(\tau) &= - 2\,i\,\frac{\tilde{d}_2 }{\zeta} \sin{\zeta \tau}.
\end{align}
Furthermore, we find $\xi(\tau) = \cos{\zeta \tau} - \frac{i}{\zeta} \sin{\zeta \tau}$, which in turn can be integrated to obtain the coefficients~\eqref{sub:algebra:decoupling:solution}, which now read
\begin{align}\label{f:functions:constant:squeezing}
&\quad\quad\quad\quad\quad F_{\hat N_a^2} = -\frac{\tilde{g}_0^2}{\zeta^2} \left( 1 - \textrm{sinc}(2\,\zeta\,\tau) \right)\,\tau, \nonumber \\
&F_{\hat N _a \, \hat B_+} = - \frac{\tilde{g}_0}{\zeta} \sin{\zeta \tau}, \quad\quad\quad F_{\hat N _a \, \hat B_-} = \frac{ \tilde{g}_0}{\zeta^2}( \cos{\zeta \tau}-1) \, ,
\end{align}
where $\rm{sinc} (x) = \sin(x) /x$. Since $\tilde{\mathcal{D}}_1 = 0$, all other coefficients are zero. The functions~\eqref{f:functions:constant:squeezing} now fully determine the time evolution through~\eqref{U}.

\subsubsection{Quadratures}\label{subsub:time:independent:quadratures}
To gain intuition about the  evolution of the system, we include plots of the optical quadratures. These can be found in Figure~\ref{fig:constant:squeezing:quadratures}.  The quadratures are the expectation values of $\hat{x}_1 = (\hat{a}^\dag + \hat{a})/\sqrt{2}$ and $\hat{p}_1 = i (\hat{a}^\dag - \hat{a})/\sqrt{2}$ and would correspond to classical trajectories in phase space. The full expression for the expectation values $\braket{\hat x_1}$ and $\braket{\hat p_1}$ can be found in~\eqref{eq:app:expectation:values} in~\ref{appendix:CM}.  

In Figures~\ref{fig:quad:mum0:d20},~\ref{fig:quad:mum0:d201},~\ref{fig:quad:mum0:d205} and~\ref{fig:quad:mum0:d21}, we have plotted the quadratures for $\mu_{\mathrm{c}} = 1$, $\mu_{\mathrm{m}} = 0$, $\tilde{g}_0 = 1$ and increasing values of $\tilde{d}_2$. While it is generally difficult to engineer a coupling of this magnitude, these values are chosen as example values to demonstrate the scaling behaviour of $\delta(\tau)$. Similarly in Figures~\ref{fig:quad:mum1:d20},~\ref{fig:quad:mum1:d201},~\ref{fig:quad:mum1:d205} and~\ref{fig:quad:mum1:d21}, we have plotted the quadratures for $\mu_{\mathrm{c}} = 1$, $\mu_{\mathrm{m}} = 1$, $\tilde{g}_0 = 1$ and again increasing values of $\tilde{d}_2$. To show the directionality of the evolution, the colour of the curve starts as light blue for $\tau = 0$ and becomes increasingly darker as $\tau $ increases. We observe that the addition of mechanical squeezing causes the system to trace out highly complex trajectories, compared with the case when $\tilde{d}_2 = 0$.

\begin{figure*}[t!]
\subfloat[ \label{fig:constant:full:measure:muc1:pi2}]{%
  \includegraphics[width=0.3\linewidth, trim = 0mm 0mm 0mm 0mm]{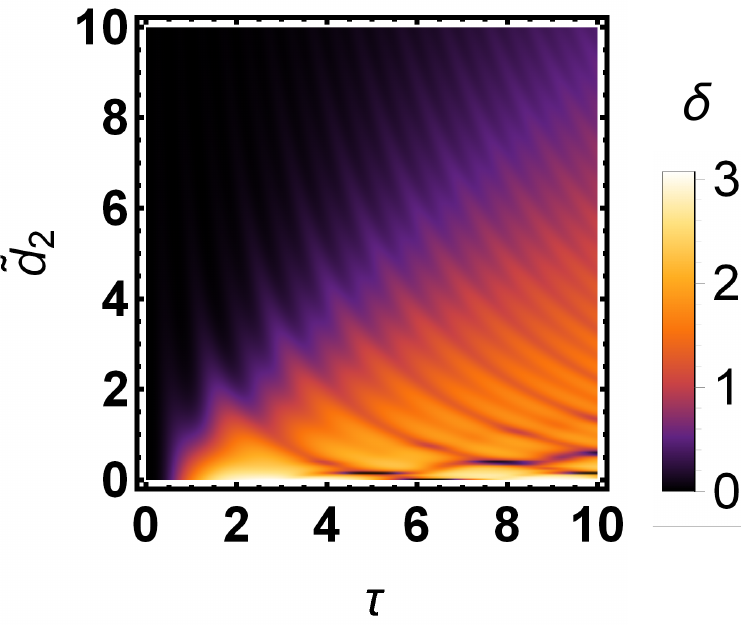}%
}\hfill
\subfloat[ \label{fig:constant:reduced:measure:muc1:pi2}]{
  \includegraphics[width=0.315\linewidth, trim = 0mm 0mm 0mm 0mm]{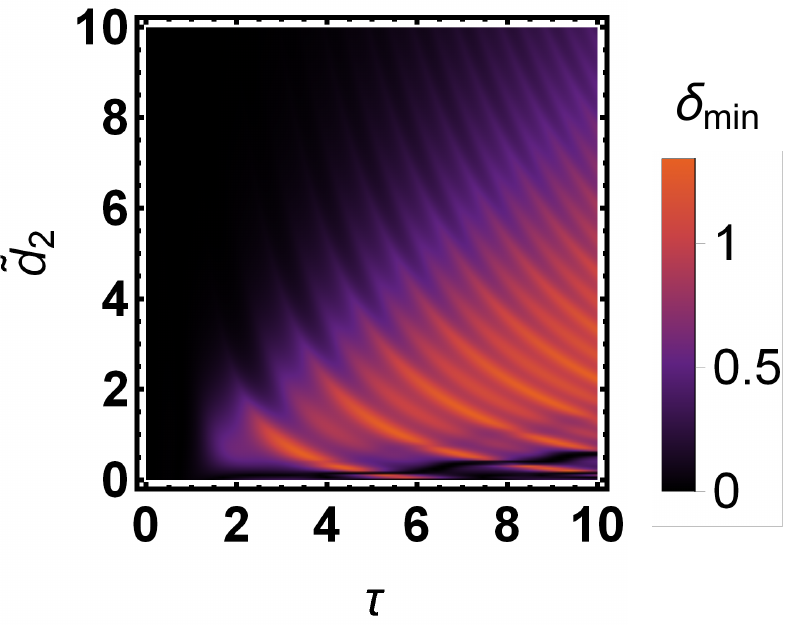}%
  }\hfill
  \subfloat[ \label{fig:constant:approximate:measure:muc1:pi2}]{
  \includegraphics[width=0.3\linewidth, trim = 0mm 0mm 0mm 0mm]{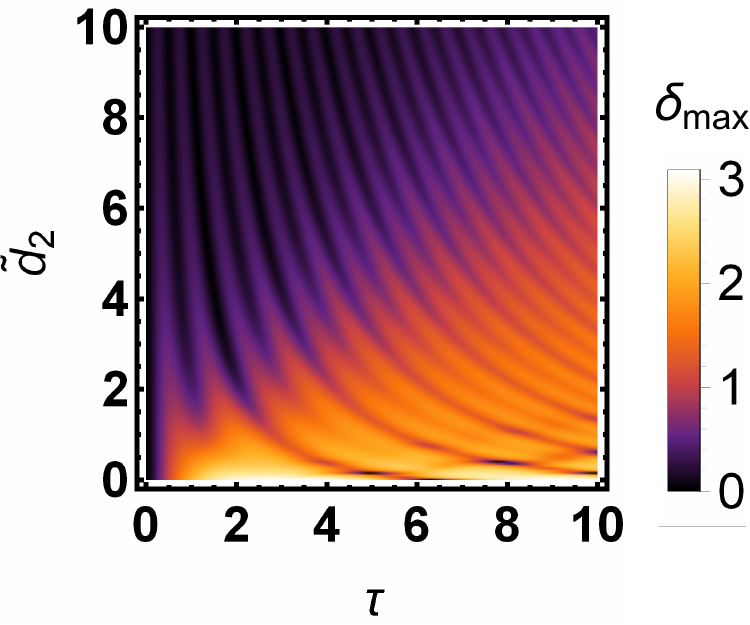}%
}\hfill
\subfloat[ \label{fig:constant:full:measure:muc1:pi}]{%
  \includegraphics[width=0.3\linewidth, trim = 0mm 0mm 0mm 0mm]{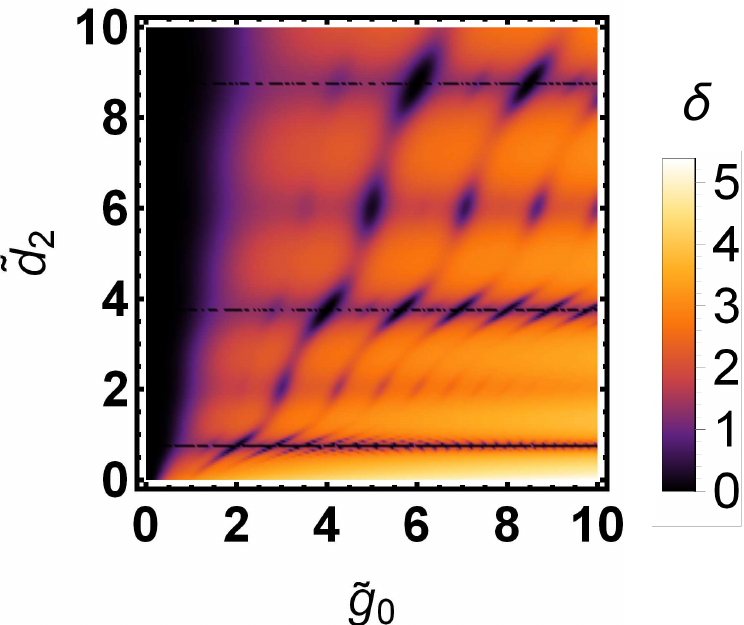}%
}\hfill
\subfloat[ \label{fig:constant:reduced:measure:muc1:pi}]{%
  \includegraphics[width=0.3\linewidth, trim = 0mm 0mm 0mm 0mm]{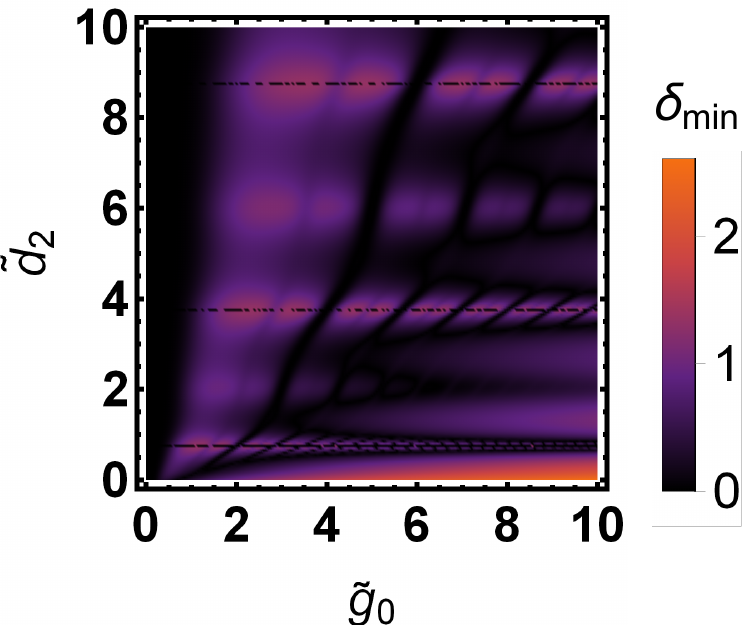}%
}\hfill
\subfloat[ \label{fig:constant:approximate:measure:muc1:pi}]{%
  \includegraphics[width=0.3\linewidth, trim = 0mm 0mm 0mm 0mm]{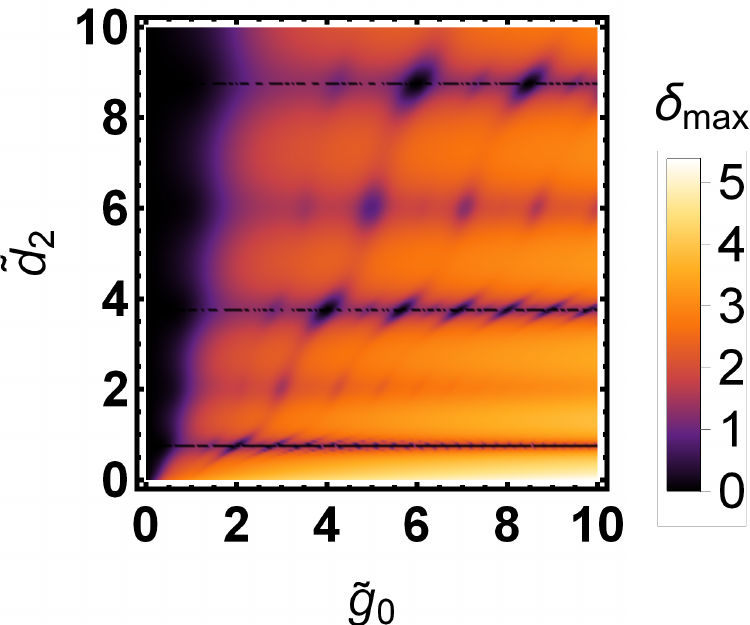}%
}\hfill
\caption{ \textbf{Non-Gaussianity of an optomechanical state with mechanical squeezing}. In each row, the colours have been rescaled to correspond to the same values in the plot. The first row shows the non-Gaussianity as a function of time $\tau$ and the squeezing parameter $\tilde{d}_2$ for the optical coherent state parameter and light--matter coupling $\mu_{\rm{c}} = \tilde{g}_0 = 1$ and mechanical coherent state parameter $\mu_{\rm{m}} = 0$. \textbf{(a)} shows the full measure $\delta(\tau)$, \textbf{(b)} shows the lower bound $\delta_{\rm{min}}(\tau)$, and \textbf{(c)} shows the upper bound $\delta_{\rm{max}}(\tau)$. The non-Gaussianity generally oscillates in time and does slowly increase for increasing time $\tau$. Furthermore, the upper bound $\delta_{\rm{max}}$ approximates the full measure well for these parameters. 
The second row shows the non-Gaussianity $\delta(\tau)$ as a function of the nonlinear coupling $\tilde{g}_0$ and the squeezing parameter $\tilde{d}_2$ for $\mu_{\rm{c}} = 1$ and $\mu_{\rm{m}} = 0$ at time $\tau = \pi$. \textbf{(d)} shows the full measure $\delta(\pi)$, \textbf{(e)} shows the lower bound $\delta_{\rm{min}}(\pi)$ and \textbf{(f)} shows the upper bound $\delta_{\rm{max}}(\pi)$. The non-Gaussianity increases with $\tilde{g}_0$ but decreases with $\tilde{d}_2$. } 
\label{fig:constant:squeezing:measure}
\end{figure*}

\subsubsection{Measure of non-Gaussianity}\label{measure}
We now proceed to compute the non-Gaussianity $\delta(\tau)$, defined in~\eqref{measure:of:non:gaussianity}, of the state evolving at constant squeezing parameter.  A fully analytic expression for $\delta(\tau)$ exists but is too cumbersome to include here. Instead, we plot the measure of non-Gaussianity in Figure~\ref{fig:constant:squeezing:measure}.  In the first row of Figure~\ref{fig:constant:squeezing:measure}, we present a comparison between the full measure $\delta$ (Figure~\ref{fig:constant:full:measure:muc1:pi}) and the lower and upper bounds $\delta_{\rm{min}}$ and $\delta_{\rm{max}}$ provided by the Araki--Lieb inequality in Figures~\ref{fig:constant:reduced:measure:muc1:pi} and~\ref{fig:constant:approximate:measure:muc1:pi}. 

We note that the non-Gaussianity increases for large light--matter coupling $\tilde{g}_0$ and large coherent state parameter $\mu_{\mathrm{c}}$. This feature was also observed for standard optomechanical systems in~\cite{qvarfort2019enhanced}. However, the most striking feature here is that the larger $\tilde{d}_2$ is, the less non-Gaussian the system becomes. To understand why this is the case, we examine the dependence on $\tilde{d}_2$ in the function $|K_{\hat N_a}|$, since this determines the behaviour of the non-Gaussianity in certain regime, as discussed in Section~\ref{sec:bounds}. Using the expression~\eqref{f:functions:constant:squeezing} we find 
\begin{equation}
|K_{\hat N_a}|^2 =  \frac{\tilde{g}_0^2}{\zeta^4} \left[ \left( \zeta^2 + 1 \right) \sin^2(\zeta \, \tau) + \cos( 2\, \zeta\, \tau) - 2 \, \cos(\zeta \, \tau) + 1 \right] \, .
\end{equation}
For large $\tilde{d}_2$, and therefore large $\zeta$, the first term inside the brackets dominates and for $\zeta \tau \neq n \pi$ with integer $n$, we are left with $|K_{\hat N_a}|\sim \tilde{g}_0 \, \sin^2( \zeta \, \tau) / \zeta^2$. In general, we find $\lim_{\tilde{d}_2 \rightarrow \infty} |K_{\hat N_a}|^2 = 0$. The consequences for the non-Gaussianity are difficult to predict given the complexity of the expressions, but we note that the mechanical symplectic eigenvalue $\nu_{\rm{Me}}$ decreases, while the optical symplectic eigenvalue $\nu_{\rm{Op}}$ increases. 

Furthermore, the quantity $\theta(\tau) = 2 \, \left( F_{\hat N_a^2} + F_{\hat N_a \, \hat B_+} \, F_{\hat N_a \, \hat B_-} \right)$ is given by 
\begin{equation}
\theta(\tau)  = 2 \, \frac{\tilde{g}_0^2}{\zeta^3}\left( \sin(\zeta \, \tau)  - \zeta \, \tau \right) \, .
\end{equation} 
We find that $\lim_{\tilde{d}_2 \rightarrow \infty} \theta(\tau) = 0 $.  We then look at the symplectic eigenvalues~\eqref{sympelctic:eigenvalue:reduced:state} in this limit. We find that $\nu_{\rm{Me}}\rightarrow 1$, and $\nu_{\rm{Op}} \rightarrow  1 $, which means that both the upper and the lower bounds of the non-Gaussianity tend to zero, and hence $\delta(\tau) \rightarrow 0$ as $\tilde{d}_2$ increases. We conclude that increasing the amount of constant squeezing in the system reduces the overall non-Gaussianity of the state.

\subsection{Applications: Modulated squeezing parameter}\label{sec:modulated:squeezing}
In this section, we consider a modulated squeezing term. The dimensionless squeezing $\tilde{\mathcal{D}}_2(\tau) = \mathcal{D}_2(t)/\omega_{\rm{m}}$ is time-dependent and of the form 
\begin{equation}  \label{eq:modulated:squeezing}
\tilde{\mathcal{D}}_2(\tau) = \,\tilde{d}_2\,  \cos(\Omega_0\, \tau) \, , 
\end{equation}
where $\tilde{d}_2 = d_2/\omega_{\mathrm{m}}$ is the amplitude of the squeezing and $\Omega_0$ denotes the time-scale of squeezing.\footnote{We remind the reader that our rescaled quantities require us to use $\tilde{d}_2 = d_2/\omega_{\mathrm{m}}$ and we define $\Omega_0 = \omega_0/\omega_{\mathrm{m}}$.} 

The differential equations in~\eqref{differential:equation:written:in:paper}  are not generally analytically solvable for arbitrary choices of $\tilde{\mathcal{D}}_2(\tau)$. However, for the choice of $\tilde{\mathcal{D}}_2(\tau)$ in~\eqref{eq:modulated:squeezing}, both equations have a known form. Consider the differential equation for $P_{11}$, which we reprint here for convenience, 
\begin{equation} \label{eq:reprint:P11}
\ddot{P}_{11} + \left( 1 + 4 \, \tilde{d}_2 \cos(\Omega_0 \, \tau) \right) P_{11} = 0 \, .
\end{equation}
Equation~\eqref{eq:reprint:P11} is that of a parametric oscillator, which is used elsewhere in physics to describe, for example, a driven pendulum. As shown in 
\ref{appendix:Hamiltonian:nonlinear:decoupling}, the equation for the integral of $P_{22}$~\eqref{app:eq:IP22} takes the same form. 

The equation~\eqref{eq:reprint:P11} is known as the Mathieu equation. In its most general form, and using conventional notation,   it reads:
\begin{equation} \label{eq:Mathieu}
\frac{d^2 y}{dx^2} + \left[ a - 2  \, q \, \cos(2 \, x) \right]  y= 0, 
\end{equation}
where $a, q,$ and $x$ are real parameters. The general solutions to this equation are linear combinations of functions known as the Mathieu cosine $C(a,q,x)$ and Mathieu sine $S(a, q, x)$, the exact form of which will be determined by the boundary conditions for $y$. 

To find which values the $a$, $q$ and $x$ parameters correspond to, we note that the cosine-term in $\tilde{\mathcal{D}}_2(\tau)$ has the argument $\Omega_0 \, \tau$, which means that we must rescale time $\tau $ as $\tau' = \Omega_0 \tau /2$. Inserting our expression for $\tilde{\mathcal{D}}_2(\tau)$ and using the chain-rule to change variables from $\tau$ to $\tau'$, we rewrite the equation for $P_{11}$ as
\begin{equation}
\frac{\Omega_0^2}{4} \frac{d^2 P_{11}}{d \tau^{\prime 2}} + \left( 1 + 4 \, \tilde{d}_2  \cos( 2\, \tau' )  \right) P_{11} = 0 \, ,
\end{equation}
where we identify the variables $a = 4 /\Omega_0^2$, and $q = -8 \, \tilde{d}_2 / \Omega_0^2$.  The boundary conditions $P_{11}(0) = 1$ and $\dot{P}_{11}(0) = 0$ will yield the Mathieu cosine $C(a,q,x)$, and for $I_{P_{22}}$ as the solution, and the boundary conditions $I_{P_{22}}(0) = 0$ and $\dot{I}_{P_{22}}(0) = 1$ will yield the Mathieu sine $S(a,q,x)$ as the solution. For our specific choice of $\mathcal{D}_2(\tau)$ in~\eqref{eq:modulated:squeezing}, the system is resonant at $\Omega_0 = 2$ (see~\ref{app:resonance}), which means that $a =1$ and $q = -2\tilde{d}_2$.

\subsection{Approximate solutions at resonance }
The Mathieu equations are notoriously difficult to evaluate numerically. Instead, we use a two-scale method to derive perturbative solutions to $P_{11}$ and $I_{P_{22}}$. The perturbative solutions are valid for  $\tilde{d}_2 \ll 1$ and make use of  specific resonance conditions to ensure that the solutions do not diverge. See~\ref{app:resonance} for the full derivation, where we also show that these approximate solutions correspond exactly to the more physically intuitive rotating-wave approximation (RWA) when $\tau \gg1$. For smaller values of $\tau$, the approximate solutions are still valid, but they cannot be interpreted as equivalent to the RWA. 

The squeezing term is resonant when $\Omega_0 = 2$. We find that the approximate solutions for $P_{11}$ and $I_{P_{22}}$ (the integral of $P_{22}$) are given by, respectively, 
\begin{align}
P_{11} &= \cos (\tau) \,  \cosh (\tilde{d}_2 \, \tau) - \sin (\tau) \,  \sinh ( \tilde{d}_2 \,  \tau) \, \nonumber , \\
I_{P_{22}} &= - \frac{1}{ 1- \tilde{d}_2}\left(\cos (\tau) \,  \sinh (\tilde{d}_2 \, \tau ) - \sin (\tau) \,  \cosh (\tilde{d}_2 \, \tau) \right)  \, .
\end{align}
We then compute $\xi(\tau)$ in~\eqref{app:eq:RWA}. We  assume that $\tilde{d}_2\tau\ll 1$ to find 
\begin{align} \label{eq:approx:xi}
\xi(\tau) \approx & \,  e^{- i \tau}\left( 1 + \frac{ \tilde{d}_2^2 \tau^2}{2}\right) + i \, e^{i \, \tau} \, \tilde{d}_2 \, \tau \, ,
\end{align} 
where we have expanded the hyperbolic functions to second order. By using the relations between $\xi(\tau)$ and the Bogoliubov coefficients~\eqref{app:eq:xi:alpha:beta:relation}, we find that the Bogoliubov condition is approximately satisfied as:
\begin{equation}
|\alpha(\tau)|^2 - |\beta(\tau)|^2 \approx 1  + \mathcal{O}[(\tilde{d}_2 \tau)^4] \, .
\end{equation}
With this expression, we can now compute the non-zero $F$-coefficients~\eqref{sub:algebra:decoupling:solution}:
\begin{align} \label{eq:F:coefficients:modulated:squeezing}
F_{\hat N_a^2} = \, & \tilde{g}_0^2  \tau\, \left( 1 - \tilde{d}_2 \right)  (\mathrm{sinc} (2  \, \tau) - 1 ) + \frac{1}{2} \tilde{g}_0^2 \,  \tilde{d}_2^2 \left(\left(2\,\tau^2 - 3\right) \sin (2 \, \tau) + 2 \, \tau + 4\,\tau \,\cos(2\,\tau)\right)\, ,\nonumber \\
F_{\hat N_a \, \hat B_+} =& - \tilde{g}_0 \, \sin (\tau) - \tilde{g}_0 \, \tilde{d}_2 \, (\tau \,\cos(\tau) - \sin(\tau))  - \frac{1}{2}\,\tilde{g}_0 \, \tilde{d}_2^2 \left[\left(\tau^2 - 2\right) \sin (\tau) + 2\,\tau \,\cos(\tau)\right] ,\nonumber \\
F_{\hat N_a \, \hat B_-} =& - 2\,\tilde{g}_0 \,\sin^2(\tau/2)  + \tilde{g}_0 \, \tilde{d}_2 \, ( \tau\, \sin(\tau) - 2\,\sin^2(\tau/2)  ) \nonumber \\
& + \frac{1}{2} \tilde{g}_0 \,  \tilde{d}_2^2 \left[\left(\tau^2 - 2\right) \cos (\tau) - 2\,\tau\,\sin(\tau) + 2\right]\, ,
\end{align}
where we have discarded terms with $\tilde{d}_2^3$.  With these expressions, we are ready to compute the non-Gaussianity of the system when the squeezing is applied at mechanical resonance.

\begin{figure*}[t!]
\subfloat[ \label{fig:modulated:measure:time}]{%
  \includegraphics[width=0.5\linewidth, trim = 0mm -2mm 0mm 0mm]{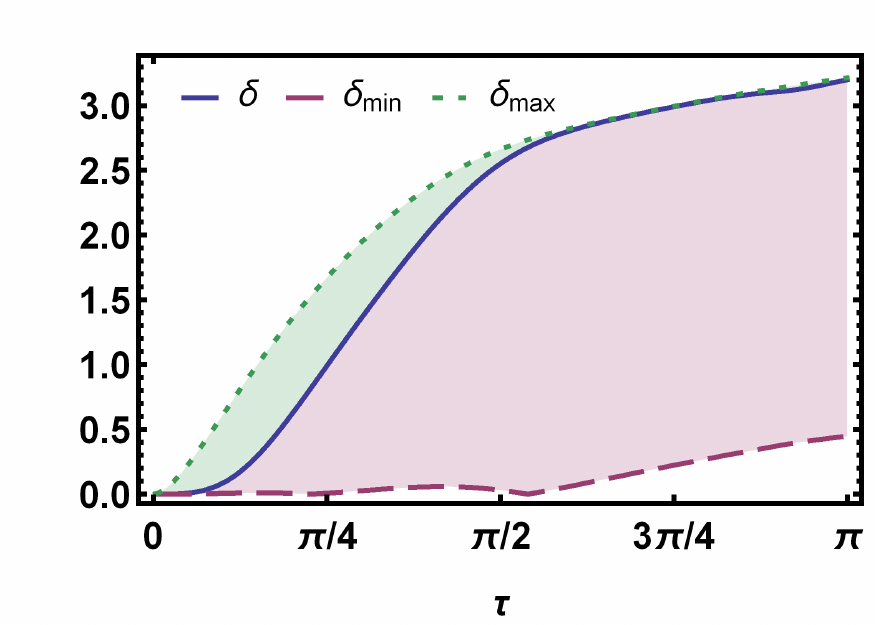}%
}\hfill
\subfloat[ \label{fig:modulated:measure:c1}]{%
  \includegraphics[width=0.49\linewidth, trim = 0mm 0mm 0mm 0mm]{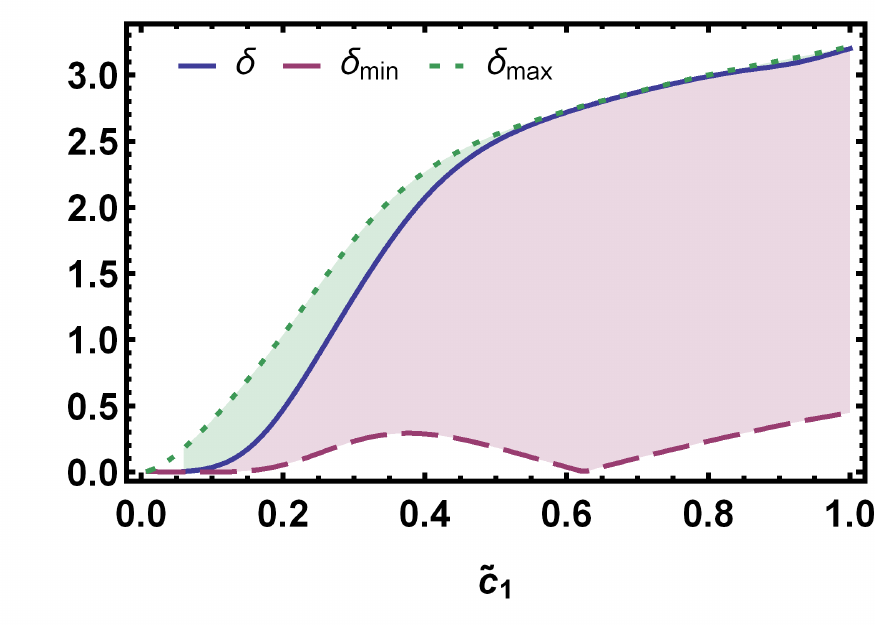}%
}\hfill
\caption{ \textbf{Non-Gaussianity of an optomechanical state with squeezing modulated at mechanical resonance}. The plots show the measure of non-Gaussianity $\delta(\tau)$ together with its lower bound $\delta_{\rm{min}}(\tau)$ and upper bound $\delta_{\rm{max}}(\tau)$. \textbf{(a)} shows the non-Gaussianity as a function of time $\tau$ given the squeezing parameter $\tilde{d}_2 = 0.1$,  the optical coherent state parameter $\mu_{\rm{c}} = 1$,  the mechanical coherent state parameter $\mu_{\rm{m}} = 0$, and the light--matter coupling $\tilde{g}_0 = 1$. \textbf{(b)} shows the non-Gaussianity as a function of $\tilde{g}_0$ at $\tau = \pi$, $\tilde{d}_2 = 0.1$, $\mu_{\rm{c}} = 1$, and $\mu_{\rm{m}} = 0$. The upper bound $\delta_{\rm{max}}(\tau)$ approximates  the full measure $\delta(\tau)$ increasingly well as $\tilde{g}_0$ and $\tau$ become larger.  
}
\label{fig:time:dependent:squeezing}
\end{figure*}

\subsubsection{Measure of non-Gaussianity at resonance}\label{subsec:modulated:measure}
We first compute the full measure of non-Gaussianity $\delta(\tau)$ and plot the results in Figure~\ref{fig:time:dependent:squeezing}.
Figure~\ref{fig:modulated:measure:time} shows the full measure $\delta(\tau)$, the lower bound $\delta_{\rm{min}}(\tau)$ and the upper bound $\delta_{\rm{max}}(\tau)$ as a function of $\tau$ for the parameter  $\tilde{g}_0 = 1$, $\mu_{\rm{c}} = 1$, $\tilde{d}_2 = 0.1$, and $\mu_{\rm{m}} = 0$ as a function of time $\tau$ and the squeezing $\tilde{d}_2$. The second plot in Figure~\ref{fig:modulated:measure:c1} also shows the full measure $\delta(\tau)$, the lower bound $\delta_{\rm{min}}(\tau)$ and the upper bound $\delta_{\rm{max}}(\tau)$ as a function of $\tilde{g}_0$ at $\tau = \pi$, $\tilde{d}_2 = 0.1$, $\mu_{\rm{c}} = 1$, and $\mu_{\rm{m}} = 0$. We find that the non-Gaussianity increases with  $\tilde{g}_0$, as expected.  

 In Figure~\ref{fig:time:dependent:squeezing}, we considered $\tilde{d}_2 = 0.1$; a value consistent with the validity of the approximate solutions to the Mathieu equation. For this value, the  non-Gaussianity is found to increase very slightly with $\tilde{d}_2$. To demonstrate this, we consider the regime where $1 \ll 2 |\mu_{\rm{c}}| \ll |K_{\hat N_a}|$, which occurs when $2|\mu_{\rm{c}}| \ll \tilde{g}_0$ for specific values of $\tau$. In this regime, the non-Gaussianity was approximately given by $s_V \left( 2 \, |K_{\hat N_a}| |\mu_{\rm{c}}|  \right)$~\eqref{eq:nG:large:KNa}. 
Given the functions~\eqref{eq:F:coefficients:modulated:squeezing}, we  find that 
\begin{align} \label{eq:modulated:K}
|K_{\hat N_a}|^2 =&  \, 4 \, \tilde{g}_0^2 \,  \sin ^2(\tau/2) + \tilde{g}_0^2 \,  \tilde{d}_2^2 \left(\tau^2 - 2\left(2 -  \tau^2\right) \sin ^2(\tau/2)\right)  \nonumber \\
&- 2\tilde{g}_0^2  \,  \tilde{d}_2 \,  \left( \tau \,  (\sin (\tau) - \sin (2\,\tau)) + (\cos (\tau) - \cos(2\,\tau)) - 2 \sin ^2(\tau/2) \right),
\end{align}
where  we have again removed terms proportional to $\tilde{d}_2^3$ and $\tilde{d}_2^4$. 
The behaviour of $|K_{\hat N_a}|^2$ is markedly different compared with the constant case. Firstly, while $|K_{\hat N_a}|^2$ still oscillates, it also increases with $\tau$ and $\tilde{d}_2$. If we consider the leading term with $\tau^2$, we find that the non-Gaussianity scales with $\delta \sim \ln(\tau\, \tilde{d}_2\, \tilde{g}_0)$, which confirms that in this specific regime, the non-Gaussianity increases logarithmically with $\tau$, $\tilde{d}_2$, and $\tilde{g}_0$. We conclude that squeezing is not necessarily detrimental to the non-Gaussianity if the squeezing is modulated at resonance, although more work needs to be done to ascertain the full interplay between the two effects. 

\section{Discussion}\label{discussion}

Before  presenting our conclusions, we discuss the applicability and scope of the techniques we developed. 
We also comment on the effect of squeezing on the non-Gaussian character of the system.

\subsection{Advantages over direct numerical simulations}\label{subsec:discussion:numerics}
With our techniques, we have shown that it is possible to analytically solve the dynamics of a nonlinear optomechanical system even when the mechanical squeezing is time-dependent. To emphasise this point, we wish to compare our approach, which relies on numerically solving the differential equations in~\eqref{differential:equation:written:in:paper}, with a general numerical method using a standard higher-order Runge--Kutta solver to evolve a state in a truncated Hilbert space, e.g. using the Python library \textit{QuTiP}~\cite{johansson2013qutip}. 

When the dynamics is solved with a Runge--Kutta method, the continuous variable (pure) states are represented as finite-dimensional vectors in a truncated Hilbert space. When the system is nonlinear, information about the state is quickly distributed across large sectors of the Hilbert space. If the computational Hilbert space is too small, numerical inaccuracies  quickly enter into the evolution, as a result of truncating the vectors. It follows that the dimension of the Hilbert space must be large enough to prevent this, which requires significant amounts of computer memory. It is also very difficult to consider parameters of the magnitude $\tilde{g}_0 = 10$ and $\tilde{d}_2 = 10$, as done in this work, since these cause the system to evolve very rapidly and, consequently, require the evolution of the system to be calculated using smaller and smaller time intervals. 

The methods developed here excel at treating systems numerically for large parameters $\tilde{g}_0, \tilde{d}_2, \mu_{\mathrm{c}}$ and $\mu_{\mathrm{m}}$. However, we note that it becomes increasingly difficult to numerically evaluate the dynamics at longer times $\tau$ when the system of differential equations~\eqref{differential:equation:written:in:paper} is numerically solved for arbitrary functions $\tilde{\mathcal{D}}_2(\tau)$. The difficulty is primarily caused by the double integral that determines the coefficient $F_{\hat N_a^2}$ in~\eqref{sub:algebra:decoupling:solution}, which must  be evaluated numerically. For each value of $\tau$, the integral will be evaluated from $0$ to the final $\tau'$, and then from $0$ to $\tau$. As a result, the integrals take an increasingly long time to evaluate for large $\tau$. We therefore conclude that the key strength in our method lies in evaluating the state of the system at early times $\tau \in (0, 2\pi)$ for large parameters $\mu_{\mathrm{c}}, \tilde{g}_0$, and $\tilde{d}_2 $. We also emphasise that, the computation using our methods is numerically exact, which a naive computation using \textit{QuTiP} or a similar library is not. 

To conclude, our methods allow for the evaluation of the state of the system with 
large parameters, e.g.  $\tilde{g}_0 = 100$ and $\tilde{d}_2 = 10$, which would be nearly impossible to perform with \textit{QuTiP}  or a similar library unless one had access to significantly more computational resources.

\subsection{Competing behaviours of nonlinearity and squeezing}\label{subsec:discussion:squeezing}
We concluded from Figure~\ref{fig:constant:squeezing:measure} that the addition of a constant squeezing term has a detrimental effect on the non-Gaussianity of the system. We also noted that inclusion of a constant squeezing term is equivalent to changing the mechanical trapping frequency $\omega_{\rm{m}}$ to a specific value and starting the computation with a squeezed coherent state (see~\ref{appendix:CSq}). With this interpretation, our results also show that an initially squeezed state evolving under the optomechanical Hamiltonian can be expected to exhibit less non-Gaussianity compared with coherent states.  The reason for this overall behaviour can be found by simple
  inspection of the total Hamiltonian. If a strong squeezing term is
  included in the Hamiltonian~\eqref{main:time:independent:Hamiltonian:to:decouple:dimensionful},
  it dominates over the interaction term, leading to a decrease in the non-Gaussianity. However, such a process is not fully monotonic, since an increase of the squeezing parameter does not always  decrease the non-Gaussianity. This is, however, reasonable, as it  cannot be expected that only the relative weight of the two parts of  the Hamiltonian matter; the precise dynamics is much more complex, and the non-Gaussianity depends on the entire state, which is driven by the full Hamiltonian.
  
The finding that the non-Gaussianity increases with both time $\tau$ and $\tilde{d}_2$ when modulated at mechanical resonance is interesting and warrants further investigation. We leave this to future work.

\section{Conclusion}\label{conclusions}
In this work, we solved the time-evolution of a nonlinear
optomechanical system with a time-dependent mechanical displacement
term and a time-dependent mechanical single-mode squeezing term. We
found analytic expressions for all first and second moments of the quadratures of the nonlinear system and used them to compute the amount of non-Gaussianity of the state. We considered both constant and modulated squeezing parameter, 
and found that a squeezing parameter 
modulated at resonance results in the Mathieu equations, for which we provide approximate solutions equivalent to the rotating-wave approximation.  

In general, we find that the relationship between the squeezing and non-Gaussianity is highly nontrivial. The inclusion of a mechanical squeezing term in the Hamiltonian, which is equivalent to starting with a coherent
squeezed state evolving with the standard optomechanical Hamiltonian with a
shifted mechanical frequency, decreases the overall non-Gaussianity of the state. If the squeezing term is modulated at mechanical resonance, however, we found that the non-Gaussianity increases with both time and the squeezing parameter in specific regimes. These results hold interesting
implications for quantum control of nonlinear optomechanical systems.

Our results also suggest that the combination of non-Gaussian resources and mechanical squeezing may not necessarily be beneficial if the application relies specifically on the non-Gaussian character of the state. However, more work is needed to conclude if this has a significant effect on, for example,  sensing applications. More work is also necessary to properly study the instabilities of the full solutions to the Mathieu equations and how they affect the dynamics. The effect of squeezing the optical rather than mechanical mode is another question we defer to future work. 

The decoupling methods demonstrated here constitute an important step towards fully characterising nonlinear systems with mechanical squeezing and can be used both to model experimental systems and to test numerical methods. The results presented here apply to any system with the same characteristic cubic Hamiltonian interaction term and single mode squeezing term. More broadly, the Lie algebra method can be used to solve any system where a finite set of Lie algebra operators has been identified. Our work can further be extended to more complicated quadratic Hamiltonians of bosonic modes, such as Dicke-like models~\cite{emary2003chaos}, which would allow for applications in other areas of physics to be developed.

\section*{Acknowledgements}
We would like to thank Antonio Pontin, Peter F. Barker, Robert Delaney, Doug Plato, and Ivette Fuentes for useful comments and discussions. 

SQ acknowledges support from the EPSRC Centre for Doctoral Training in Delivering Quantum Technologies and thanks the University of Vienna for its hospitality. DR would like to thank the Humboldt Foundation for supporting his work with their Feodor Lynen Research Fellowship. This work was supported by the European Union's Horizon 2020 research and innovation programme under grant agreement No. 732894 (FET Proactive HOT).  DEB acknowledges support from the CEITEC Nano RI.

\section*{References}

\bibliographystyle{iopart-num}
\bibliography{bibliography}

\providecommand{\newblock}{}
\begin{thebibliography}{10}
\expandafter\ifx\csname url\endcsname\relax
  \def\url#1{{\tt #1}}\fi
\expandafter\ifx\csname urlprefix\endcsname\relax\def\urlprefix{URL }\fi
\providecommand{\eprint}[2][]{\url{#2}}

\bibitem{aspelmeyer2014cavity}
Aspelmeyer M, Kippenberg T~J and Marquardt F 2014 {\em Reviews of Modern
  Physics\/} {\bf 86} 1391
  \urlprefix\url{https://doi.org/10.1103/RevModPhys.86.1391}

\bibitem{bowen2015quantum}
Bowen W~P and Milburn G~J 2015 {\em Quantum optomechanics\/} (CRC Press)

\bibitem{serafini2017quantum}
Serafini A 2017 {\em Quantum Continuous Variables: A Primer of Theoretical
  Methods\/} (CRC Press)

\bibitem{sankey2010strong}
Sankey J~C, Yang C, Zwickl B~M, Jayich A~M and Harris J~G 2010 {\em Nature
  Physics\/} {\bf 6} 707 \urlprefix\url{https://doi.org/10.1038/nphys1707}

\bibitem{leijssen2017nonlinear}
Leijssen R, La~Gala G~R, Freisem L, Muhonen J~T and Verhagen E 2017 {\em Nature
  Communications\/} {\bf 8} ncomms16024
  \urlprefix\url{https://doi.org/10.1038/ncomms16024}

\bibitem{fogliano2019cavity}
Fogliano F, Besga B, Reigue A, Heringlake P, de~L\'{e}pinay L~M, Vaneph C,
  Reichel J, Pigeau B and Arcizet O 2019 {\em arXiv preprint\/}
  (\textit{Preprint} \eprint{1904.01140})

\bibitem{bose1997preparation}
Bose S, Jacobs K and Knight P 1997 {\em Physical Review A\/} {\bf 56} 4175
  \urlprefix\url{https://doi.org/10.1103/PhysRevA.56.4175}

\bibitem{mancini1997ponderomotive}
Mancini S, Man'ko V and Tombesi P 1997 {\em Physical Review A\/} {\bf 55} 3042
  \urlprefix\url{https://doi.org/10.1103/PhysRevA.55.3042}

\bibitem{qvarfort2019enhanced}
Qvarfort S, Serafini A, Xuereb A, R{\"a}tzel D and Bruschi D~E~E 2019 {\em New
  Journal of Physics\/}
  \urlprefix\url{https://doi.org/10.1088/1367-2630/ab1b9e}

\bibitem{zurek2001sub}
Zurek W~H 2001 {\em Nature\/} {\bf 412} 712
  \urlprefix\url{https://doi.org/10.1038/35089017}

\bibitem{toscano2006sub}
Toscano F, Dalvit D~A, Davidovich L and Zurek W~H 2006 {\em Physical Review
  A\/} {\bf 73} 023803
  \urlprefix\url{https://doi.org/10.1103/PhysRevA.73.023803}

\bibitem{howard2018hypercube}
Howard L, Weinhold T, Combes J, Shahandeh F, Vanner M, Ringbauer M and White A
  2018 {\em arXiv preprint\/} (\textit{Preprint} \eprint{1811.03011})

\bibitem{lemonde2016enhanced}
Lemonde M~A, Didier N and Clerk A~A 2016 {\em Nature Communications\/} {\bf 7}
  11338 \urlprefix\url{https://doi.org/10.1038/ncomms11338}

\bibitem{latmiral2016probing}
Latmiral L, Armata F, Genoni M~G, Pikovski I and Kim M 2016 {\em Physical
  Review A\/} {\bf 93} 052306
  \urlprefix\url{https://doi.org/10.1103/PhysRevA.93.052306}

\bibitem{yin2017nonlinear}
Yin T~S, L{\"u} X~Y, Zheng L~L, Wang M, Li S and Wu Y 2017 {\em Physical Review
  A\/} {\bf 95} 053861
  \urlprefix\url{https://doi.org/10.1103/PhysRevA.95.053861}

\bibitem{doolin2014nonlinear}
Doolin C, Hauer B, Kim P, MacDonald A, Ramp H and Davis J 2014 {\em Physical
  Review A\/} {\bf 89} 053838
  \urlprefix\url{https://doi.org/10.1103/PhysRevA.89.053838}

\bibitem{lloyd1999quantum}
Lloyd S and Braunstein S~L 1999 Quantum computation over continuous variables
  {\em Quantum Information with Continuous Variables\/} (Springer) pp 9--17
  \urlprefix\url{https://doi.org/10.1007/978-94-015-1258-9_2}

\bibitem{menicucci2006universal}
Menicucci N~C, van Loock P, Gu M, Weedbrook C, Ralph T~C and Nielsen M~A 2006
  {\em Physical Review Letters\/} {\bf 97} 110501
  \urlprefix\url{https://doi.org/10.1103/PhysRevLett.97.110501}

\bibitem{dell2010teleportation}
Dell'Anno F, De~Siena S, Adesso G and Illuminati F 2010 {\em Physical Review
  A\/} {\bf 82} 062329
  \urlprefix\url{https://doi.org/10.1103/PhysRevA.82.062329}

\bibitem{fiuravsek2002gaussian}
Fiur{\'a}{\v{s}}ek J 2002 {\em Physical Review Letters\/} {\bf 89} 137904
  \urlprefix\url{https://doi.org/10.1103/PhysRevLett.89.137904}

\bibitem{giedke2002characterization}
Giedke G and Cirac J~I 2002 {\em Physical Review A\/} {\bf 66} 032316
  \urlprefix\url{https://doi.org/10.1103/PhysRevA.66.032316}

\bibitem{niset2009no}
Niset J, Fiur{\'a}{\v{s}}ek J and Cerf N~J 2009 {\em Physical Review Letters\/}
  {\bf 102} 120501
  \urlprefix\url{https://doi.org/10.1103/PhysRevLett.102.120501}

\bibitem{zhuang2018resource}
Zhuang Q, Shor P~W and Shapiro J~H 2018 {\em Physical Review A\/} {\bf 97}
  052317 \urlprefix\url{https://doi.org/10.1103/PhysRevA.97.052317}

\bibitem{takagi2018convex}
Takagi R and Zhuang Q 2018 {\em Phys. Rev. A\/} {\bf 97}(6) 062337
  \urlprefix\url{https://doi.org/10.1103/PhysRevA.97.062337}

\bibitem{PhysRevA.98.052350}
Albarelli F, Genoni M~G, Paris M~G~A and Ferraro A 2018 {\em Phys. Rev. A\/}
  {\bf 98}(5) 052350 \urlprefix\url{https://doi.org/10.1103/PhysRevA.98.052350}

\bibitem{aasi2013enhanced}
Aasi J, Abadie J, Abbott B, Abbott R, Abbott T, Abernathy M, Adams C, Adams T,
  Addesso P, Adhikari R {\em et~al.\/} 2013 {\em Nature Photonics\/} {\bf 7}
  613 \urlprefix\url{https://doi.org/10.1038/nphoton.2013.177}

\bibitem{clerk2010introduction}
Clerk A~A, Devoret M~H, Girvin S~M, Marquardt F and Schoelkopf R~J 2010 {\em
  Reviews of Modern Physics\/} {\bf 82} 1155
  \urlprefix\url{https://doi.org/10.1103/RevModPhys.82.1155}

\bibitem{wei1963lie}
Wei J and Norman E 1963 {\em Journal of Mathematical Physics\/} {\bf 4}
  575--581 \urlprefix\url{https://doi.org/10.1063/1.1703993}

\bibitem{wilcox1967exponential}
Wilcox R 1967 {\em Journal of Mathematical Physics\/} {\bf 8} 962--982
  \urlprefix\url{https://doi.org/10.1063/1.1705306}

\bibitem{puri2001mathematical}
Puri R~R 2001 {\em Mathematical Methods of Quantum Optics\/} vol~79
  (Springer-Verlag Berlin Heidelberg)

\bibitem{bruschi2018mechano}
Bruschi D~E and Xuereb A 2018 {\em New Journal of Physics\/}
  \urlprefix\url{https://doi.org/10.1088/1367-2630/aaca27}

\bibitem{bruschi2018time}
Bruschi D~E 2018 {\em arXiv preprint\/} (\textit{Preprint} \eprint{1812.06879})

\bibitem{genoni2008quantifying}
Genoni M~G, Paris M~G and Banaszek K 2008 {\em Physical Review A\/} {\bf 78}
  060303 \urlprefix\url{https://doi.org/10.1103/PhysRevA.78.060303}

\bibitem{marian2013relative}
Marian P and Marian T~A 2013 {\em Physical Review A\/} {\bf 88} 012322
  \urlprefix\url{https://doi.org/10.1103/PhysRevA.88.012322}

\bibitem{favero2009optomechanics}
Favero I and Karrai K 2009 {\em Nature Photonics\/} {\bf 3} 201
  \urlprefix\url{https://doi.org/10.1038/nphoton.2009.42}

\bibitem{jayich2008dispersive}
Jayich A, Sankey J, Zwickl B, Yang C, Thompson J, Girvin S, Clerk A, Marquardt
  F and Harris J 2008 {\em New Journal of Physics\/} {\bf 10} 095008
  \urlprefix\url{https://doi.org/10.1088/1367-2630/10/9/095008}

\bibitem{yin2013large}
Yin Z~q, Li T, Zhang X and Duan L 2013 {\em Physical Review A\/} {\bf 88}
  033614 \urlprefix\url{https://doi.org/10.1103/PhysRevA.88.033614}

\bibitem{eichenfield2009picogram}
Eichenfield M, Camacho R, Chan J, Vahala K~J and Painter O 2009 {\em Nature\/}
  {\bf 459} 550 \urlprefix\url{https://doi.org/10.1038/nature08061}

\bibitem{safavi2014two}
Safavi-Naeini A~H, Hill J~T, Meenehan S, Chan J, Gr{\"o}blacher S and Painter O
  2014 {\em Physical Review Letters\/} {\bf 112} 153603
  \urlprefix\url{https://doi.org/10.1103/PhysRevLett.112.153603}

\bibitem{qvarfort2018gravimetry}
Qvarfort S, Serafini A, Barker P~F and Bose S 2018 {\em Nature
  Communications\/} {\bf 9} 3690
  \urlprefix\url{https://doi.org/10.1038/s41467-018-06037-z}

\bibitem{armata2017quantum}
Armata F, Latmiral L, Plato A and Kim M 2017 {\em Physical Review A\/} {\bf 96}
  043824 \urlprefix\url{https://doi.org/10.1103/PhysRevA.96.043824}

\bibitem{blencowe2004quantum}
Blencowe M 2004 {\em Physics Reports\/} {\bf 395} 159--222
  \urlprefix\url{https://doi.org/10.1016/j.physrep.2003.12.005}

\bibitem{adesso2014continuous}
Adesso G, Ragy S and Lee A~R 2014 {\em Open Systems \& Information Dynamics\/}
  {\bf 21} 1440001 \urlprefix\url{https://doi.org/10.1142/S1230161214400010}

\bibitem{alsing2012observer}
Alsing P~M and Fuentes I 2012 {\em Classical and Quantum Gravity\/} {\bf 29}
  224001 \urlprefix\url{https://doi.org/10.1088/0264-9381/29/22/224001}

\bibitem{birrell1984quantum}
Birrell N~D, Birrell N~D and Davies P 1984 {\em Quantum fields in curved
  space\/} 7 (Cambridge university press)

\bibitem{williamson1936algebraic}
Williamson J 1936 {\em American Journal of Mathematics\/} {\bf 58} 141--163
  \urlprefix\url{https://doi.org/10.2307/2371062}

\bibitem{bruschi2013time}
Bruschi D~E, Lee A~R and Fuentes I 2013 {\em Journal of Physics A: Mathematical
  and Theoretical\/} {\bf 46} 165303
  \urlprefix\url{https://doi.org/10.1088/1751-8113/46/16/165303}

\bibitem{PhysRevD.87.084062}
Brown E~G, Mart\'{\i}n-Mart\'{\i}nez E, Menicucci N~C and Mann R~B 2013 {\em
  Phys. Rev. D\/} {\bf 87}(8) 084062
  \urlprefix\url{https://link.aps.org/doi/10.1103/PhysRevD.87.084062}

\bibitem{moore2016tuneable}
Moore C and Bruschi D~E 2016 {\em arXiv preprint\/} (\textit{Preprint}
  \eprint{quant-phy/1601.01919})

\bibitem{araki2002entropy}
Araki H and Lieb E~H 2002 Entropy inequalities {\em Inequalities\/} (Springer)
  pp 47--57 \urlprefix\url{https://doi.org/10.1007/BF01646092}

\bibitem{johansson2013qutip}
Johansson J~R, Nation P~D and Nori F 2013 {\em Computer Physics
  Communications\/} {\bf 184} 1234--1240
  \urlprefix\url{https://doi.org/10.1016/j.cpc.2012.11.019}

\bibitem{emary2003chaos}
Emary C and Brandes T 2003 {\em Physical Review E\/} {\bf 67} 066203
  \urlprefix\url{https://doi.org/10.1103/PhysRevE.67.066203}

\bibitem{kovacic2018mathieu}
Kovacic I, Rand R and Sah S~M 2018 {\em Applied Mechanics Reviews\/} {\bf 70}
  020802 \urlprefix\url{https://doi.org/10.1115/1.4039144}

\end{thebibliography}

\newpage
\appendix

\section{Decoupling of techniques for time evolution}\label{appendix:decoupling}
In this appendix, we outline the general decoupling techniques that we shall be using throughout this work to find a decoupled time-evolution operator generated by the Hamiltonian in~\eqref{main:time:independent:Hamiltonian:to:decouple}. 

\subsection{Decoupling for arbitrary Hamiltonians}
The time evolution operator $\hat{U}(t)$ induced by a Hamiltonian $\hat{H}(t)$ reads
\begin{equation}
\hat{U}(t)=\overset{\leftarrow}{\mathcal{T}}\,\exp\left[-\frac{i}{\hbar}\int_0^{t} dt'\,\hat{H}(t')\right].
\end{equation}
Any Hamiltonian can be cast in the form $\hat{H}=\sum_n \hbar\, g_n(t)\,\hat{G}_n$, where the $\hat{G}_n$ are time independent, Hermitian operators and the $g_n(t)$ are time-dependent functions. The choice of $\hat{G}_n$ need not be unique.

It has been shown~\cite{wei1963lie, bruschi2013time} that it is always possible to obtain the decoupling
\begin{equation}\label{decoupled:time:evolution:operator:appendix}
\hat{U}(t)=\prod_n\hat{U}_n(t),
\end{equation}
where we have defined $\hat{U}_n:=\exp[-i\,F_n(t)\,\hat{G}_n]$ and the
real, time-dependent functions $F_n(t)$, and the ordering of the
operators is $\hat U_1\hat U_2\ldots $.

The functions $F_n(t)$ are uniquely determined by the coupled, nonlinear, first order differential equations
\begin{align}\label{operator:differential:equations}
\frac{1}{\hbar}\hat{H}=\dot{F}_1\,\hat{G}_1+\dot{F}_2\,\hat{U}_1\,\hat{G}_2\,\hat{U}_1^{\dag}+\dot{F}_3\,\hat{U}_1\,\hat{U}_2\,\hat{G}_3\,\hat{U}_2^{\dag}\,\hat{U}_1^{\dag}+\dot{F}_4\,\hat{U}_1\,\hat{U}_2\,\hat{U}_3\,\hat{G}_4\,\hat{U}_3^{\dag}\,\hat{U}_2^{\dag}\,\hat{U}_1^{\dag}+\dots\,,
\end{align}
 where $\hat{H}$ and all $F_n$ are taken at time $t$.
This is the general method we will be employing in the following sections. 

\subsection{Decoupling for quadratic Hamiltonians}
If the Hamiltonian is quadratic in the mode operators the techniques described in the previous subsection have a more powerful representation. As explained in the main text, the time evolution operator has a symplectic representation $\boldsymbol{S}(t)$ and the decoupled ansatz~\eqref{decoupled:time:evolution:operator:appendix} has the form $\boldsymbol{S}(t)=\prod_n\boldsymbol{S}_n(t)$, where we have introduced  $\boldsymbol{S}_n:=\exp[F_n(t)\,\boldsymbol{\Omega}\,\boldsymbol{G}_n]$ and the real, time-dependent functions $F_n(t)$ are the \emph{same} as those obtained with the technique above.

The real, time-dependent functions $F_n(t)$ can be obtained by solving the following system of coupled nonlinear first order differential equations
\begin{align}\label{matrix:differential:equations}
\frac{1}{\hbar}\boldsymbol{H}=&\dot{F}_1\,\boldsymbol{G}_1+\dot{F}_2\,\boldsymbol{S}_1^{\dag}\,\boldsymbol{G}_2\,\boldsymbol{S}_1+\dot{F}_3\,\boldsymbol{S}_1^{\dag}\,\boldsymbol{S}_2^{\dag}\,\boldsymbol{G}_3\,\boldsymbol{S}_2\,\boldsymbol{S}_1+\dot{F}_4\,\boldsymbol{S}_1^{\dag}\,\boldsymbol{S}_2^{\dag}\,\boldsymbol{S}_3^{\dag}\,\boldsymbol{G}_4\,\boldsymbol{S}_3\,\boldsymbol{S}_2\,\boldsymbol{S}_1+\dots,
\end{align}
where the matrix $\boldsymbol{H}$ can be obtained by $\hat{H}(t)=\frac{1}{2}\mathbb{X}^{\dag}\,\boldsymbol{H}\,\mathbb{X}$ and the summation is over $N\,(2N+1)$ terms~\cite{bruschi2013time}. This is the matrix version of the operator differential equations~\eqref{operator:differential:equations} for quadratic Hamiltonians, which reduces the problem of operator algebra to matrix multiplication.

\section{Decoupling of the nonlinear Hamiltonian}\label{appendix:Hamiltonian:nonlinear:decoupling}
We use the techniques presented in~\ref{appendix:decoupling} to decouple the Hamiltonian~\eqref{main:time:independent:Hamiltonian:to:decouple}. The decoupling below is obtained in the same fashion of previous results in the decoupling of Hamiltonians~\cite{bruschi2013time,moore2016tuneable}. 

\subsection{ Solving the quadratic time-evolution}
We start from the dimensionless Hamiltonian~\eqref{main:time:independent:Hamiltonian:to:decouple}, which we reprint here
\begin{align}
	\hat {H} = \, & \Omega_{\mathrm{c}}\,\hat{a}^\dag\hat{a}+\hat{b}^\dag\hat{b}+ \tilde{\mathcal{D}}_1(\tau)\,\left(\hat b+ \hat b^\dagger\right)+  \tilde{\mathcal{D}}_2(\tau)\left(\hat b+ \hat b^\dagger \right)^2 - \tilde{\mathcal{G}}(\tau)\,\hat a^\dagger\hat a \,\left(\hat b+ \hat b^\dagger\right).
\end{align}
This Hamiltonian can be conveniently re-written as
\begin{align}\label{main:time:independent:Hamiltonian:to:decouple:reprinted}
	\hat {H}= \, & \Omega_{\mathrm{c}} \hat{a}^\dag\hat{a}+ \tilde{\mathcal{D}}_1(\tau) \left(\hat b+ \hat b^\dagger\right)- \tilde{\mathcal{G}}(\tau) \hat a^\dagger\hat a \left(\hat b+ \hat b^\dagger\right) \nonumber \\
&+  2 \left(\frac{1}{2}+\tilde{\mathcal{D}}_2(\tau)\right) \hat{b}^\dag\hat{b} +\tilde{\mathcal{D}}_2(\tau) \left(\hat b^2+ \hat b^{\dagger2} \right).
\end{align}
Our goal now is to separate the Hamiltonian into a linear and a quadratic contribution. We then study the effect of the quadratic Hamiltonian on the nonlinear part, which allows us to focus fully on decoupling the nonlinear dynamics. 

We therefore rewrite the time evolution operator~\eqref{general:time:evolution:operator} in  the alternative form
\begin{align}\label{formal:solution:to:the:decoupling:easier}
\hat{\tilde{U}}(\tau):=e^{-i\,\Omega_\mathrm{c} \hat a^\dagger \hat a\,\tau}\,\hat{U}_{\mathrm{sq}}\,\overset{\leftarrow}{\mathcal{T}}\,\exp\left[-\frac{i}{\hbar}\,\int_0^\tau\,\mathrm{d}\tau'\,\hat{U}_{\mathrm{sq}}^\dag(\tau')\,\hat {H}_1(\tau')\,\hat{U}_{\mathrm{sq}}(\tau')\right],
\end{align}
where we have introduced
\begin{align} \label{app:eq:definition:of:Hsq}
\hat {H}_{\mathrm{sq}}(\tau):=&2\,\left(\frac{1}{2}+\tilde{\mathcal{D}}_2(\tau)\right)\,\hat{b}^\dag\hat{b}+\tilde{\mathcal{D}}_2(\tau)\,\left(\hat b^2+ \hat b^{\dagger2} \right)\, , \nonumber\\
\hat {H}_1(\tau):=&\tilde{\mathcal{D}}_1(\tau)\,\left(\hat b+ \hat b^\dagger\right)- \tilde{\mathcal{G}}(\tau)\,\hat a^\dagger\hat a \,\left(\hat b+ \hat b^\dagger\right)\, , \nonumber\\
\hat{U}_{\mathrm{sq}}(\tau):=&\overset{\leftarrow}{\mathcal{T}}\,\exp\left[-i\,\int_0^\tau\,\mathrm{d}\tau'\,\hat {H}_{\mathrm{sq}}(\tau')\right].
\end{align}
 The transition to the expression~\eqref{formal:solution:to:the:decoupling:easier} is similar to moving from the Heisenberg (or Schr\"odinger) picture to the interaction picture. 

Our first goal is to solve the time-evolution of the quadratic part $\hat U_{\rm{sq}}$. We define the basis vector $\mathbb{X}:=(\hat b, \hat b^\dag)^{\mathrm{T}}$. From standard symplectic (i.e., Bogoliubov) theory we know that 
\begin{align} \label{eq:Bogoliubov:transform}
\mathbb{X}'
=
\hat{U}_{\mathrm{sq}}^\dag
\,\mathbb{X}\,
\hat{U}_{\mathrm{sq}}
=
\begin{pmatrix}
\hat{U}_{\mathrm{sq}}^\dag\,\hat b\,\hat{U}_{\mathrm{sq}} \\
\hat{U}_{\mathrm{sq}}^\dag\,\hat b^\dag\,\hat{U}_{\mathrm{sq}}
\end{pmatrix}
=
\boldsymbol{S}_{\mathrm{sq}}(\tau)\,
\mathbb{X},
\end{align}
where the $2\times2$ symplectic matrix
$\boldsymbol{S}_{\mathrm{sq}}(\tau)$ is the symplectic representation
of $\hat {H}_{\rm{sq}}(\tau)$ 
and satisfies
$\boldsymbol{S}_{\mathrm{sq}}^\dag(\tau)\,\boldsymbol{\Omega}\,\boldsymbol{S}_{\mathrm{sq}}(\tau)=\boldsymbol{\Omega}$. Here
$\boldsymbol{\Omega}=\text{diag}(-i,i)$ is the symplectic form. 

The matrix $\boldsymbol{S}_{\mathrm{sq}}(\tau)$ therefore has the
expression
$\boldsymbol{S}_{\mathrm{sq}}(\tau)=\overset{\leftarrow}{\mathcal{T}}\,\exp[\boldsymbol{ \, \Omega}\,\int_0^\tau\,\mathrm{d}\tau'\,\tilde{\boldsymbol{H}}_{\rm{sq}}(\tau')]$. Here $\hat {H}_{\rm{sq}}=\frac{1}{2}\mathbb{X}^\dag\tilde{\boldsymbol{H}}_{\rm{sq}}
\mathbb{X}$ and one has that 
\begin{align}
\tilde{\boldsymbol{H}}_{\rm{sq}}
=
\begin{pmatrix}
1+2\,\tilde{\mathcal{D}}_2(\tau) & 2\,\tilde{\mathcal{D}}_2(\tau)\\
2\,\tilde{\mathcal{D}}_2(\tau) & 1+2\,\tilde{\mathcal{D}}_2(\tau) 
\end{pmatrix}.
\end{align}
Therefore, we have that 
\begin{align}\label{main:general:expression:time:dependent:squeezing}
\boldsymbol{S}_{\mathrm{sq}}(\tau)=\overset{\leftarrow}{\mathcal{T}}\,\exp\left[-i\,\begin{pmatrix}
1 & 0\\
0 & -1 
\end{pmatrix}\,\int_0^\tau\,\mathrm{d}\tau'\,\begin{pmatrix}
1+2\,\tilde{\mathcal{D}}_2(\tau') & 2\,\tilde{\mathcal{D}}_2(\tau')\\
2\,\tilde{\mathcal{D}}_2(\tau') & 1+2\,\tilde{\mathcal{D}}_2(\tau') 
\end{pmatrix}\right].
\end{align}
 It can be shown that the time independent orthogonal matrix $\boldsymbol{M}_{\text{ort}}$ with expression
\begin{align}
\boldsymbol{M}_{\text{ort}}
=\frac{1}{\sqrt{2}}
\begin{pmatrix}
1 & 1\\
-1 & 1 
\end{pmatrix}\, , 
\end{align}
puts the Hamiltonian matrix $\tilde{\boldsymbol{H}}_s$ in diagonal form through
 \begin{align}
\begin{pmatrix}
1+2\,\tilde{\mathcal{D}}_2(\tau) & 2\,\tilde{\mathcal{D}}_2(\tau)\\
2\,\tilde{\mathcal{D}}_2(\tau) & 1+2\,\tilde{\mathcal{D}}_2(\tau) 
\end{pmatrix}
=
\frac{1}{2}
\begin{pmatrix}
1 & -1\\
1 & 1 
\end{pmatrix}
\begin{pmatrix}
1+4\,\tilde{\mathcal{D}}_2(\tau) & 0\\
0 &1 
\end{pmatrix}
\begin{pmatrix}
1 & 1\\
-1 & 1 
\end{pmatrix}
.
\end{align}
This allows us to manipulate~\eqref{main:general:expression:time:dependent:squeezing} and find
\begin{align}\label{main:general:expression:time:dependent:squeezing:manipulated}
\boldsymbol{S}_{\mathrm{sq}}(\tau)=
\boldsymbol{M}_{\text{ort}}^{\rm{T}}\,\overset{\leftarrow}{\mathcal{T}}\,\exp\left[i\,\int_0^\tau\,\mathrm{d}\tau'\,\begin{pmatrix}
0 & 1 \\
1+4\,\tilde{\mathcal{D}}_2(\tau') & 0 
\end{pmatrix}\right]
\,\boldsymbol{M}_{\text{ort}}.
\end{align}
This means that we have 
\begin{align}
\mathbb{X}'=
\boldsymbol{M}_{\text{ort}}^{\rm{T}}\,\overset{\leftarrow}{\mathcal{T}}\,\exp\left[i\,\int_0^\tau\,\mathrm{d}\tau'\,\begin{pmatrix}
0 & 1 \\
1+4\,\tilde{\mathcal{D}}_2(\tau') & 0 
\end{pmatrix}\right]
\,\boldsymbol{M}_{\text{ort}}\,
\mathbb{X}.
\end{align}
We introduce the new vector $\mathbb{Y}:=\boldsymbol{M}_{\text{ort}}\,\mathbb{X}$, which is just a rotation of the operators. Then we have 
\begin{align}\label{matrix:time:ordered:exponential}
\mathbb{Y}'=
\overset{\leftarrow}{\mathcal{T}}\,\exp\left[i\,\int_0^\tau\,\mathrm{d}\tau'\,\begin{pmatrix}
0 & 1 \\
1+4\,\tilde{\mathcal{D}}_2(\tau') & 0 
\end{pmatrix}\right]
\,\mathbb{Y}.
\end{align}

\subsection{Solving the matrix time-ordered exponential}
Here we seek a formal expression for
\eqref{matrix:time:ordered:exponential}.
 We write
\begin{align}\label{initial:useful:conditions}
\overset{\leftarrow}{\mathcal{T}}\,\exp\left[i\,\int_0^\tau\,\mathrm{d}\tau'\,\begin{pmatrix}
0 & 1 \\
1+4\,\tilde{\mathcal{D}}_2(\tau') & 0 
\end{pmatrix}\right]=\boldsymbol{P}+i\,\int_0^\tau \mathrm{d}\tau'\,\boldsymbol{K}\,\boldsymbol{P},
\end{align}
in terms of the matrix $\boldsymbol{K}$ defined as
\begin{align}\label{matrix:time:ordered:exponential:explicit}
\boldsymbol{K}:=
\begin{pmatrix}
0 & 1 \\
1+4\,\tilde{\mathcal{D}}_2(\tau') & 0 
\end{pmatrix},
\end{align}
and the matrix $\boldsymbol{P}$, which we will determine and which is
diagonal. This follows from the fact that the matrix on the left-hand
side of~\eqref{initial:useful:conditions} is diagonal when squared,
and therefore any even powers in the expansion will be diagonal. We
use the fact that 
\begin{align}
\frac{d}{d\tau}\overset{\leftarrow}{\mathcal{T}}\,\exp\left[i\,\int_0^\tau\,\mathrm{d}\tau'\,\boldsymbol{K}(\tau')\right]
=
i\,\boldsymbol{K}\,
\overset{\leftarrow}{\mathcal{T}}\,\exp\left[i\,\int_0^\tau\,\mathrm{d}\tau'\,\boldsymbol{K}(\tau')\right]\, , 
\end{align}
to find the equation
\begin{align}
-\boldsymbol{K}\,\int_0^\tau \mathrm{d}\tau'\,\boldsymbol{K}\,\boldsymbol{P}=\dot{\boldsymbol{P}}.
\end{align}
Since $\boldsymbol{K}$ is invertible, we manipulate this equation and obtain, after some algebra,
\begin{align}\label{differential:equation:for:matrix}
\ddot{\boldsymbol{P}}- 4\,\frac{\dot{\tilde{\mathcal{D}}}_2(\tau)}{1 \,+4\,\tilde{\mathcal{D}}_2(\tau)}
\begin{pmatrix}
0 &  0\\
0 & 1 
\end{pmatrix}
\dot{\boldsymbol{P}}
+(1+4\,\tilde{\mathcal{D}}_2(\tau))\,\boldsymbol{P}=0.
\end{align}
We can now solve the four differential equations contained in~\eqref{differential:equation:for:matrix} which read
\begin{align}\label{differential:equation:written:down}
\ddot{P}_{11}+(1+4\,\tilde{\mathcal{D}}_2(\tau))\,P_{11}= \, &0\nonumber\\
P_{12}=P_{21}=& \, 0\nonumber\\
\ddot{P}_{22}-4\frac{\dot{\tilde{\mathcal{D}}}_2(\tau)}{1+4\,\tilde{\mathcal{D}}_2(\tau)}\,\dot{P}_{22}+(1+4\,\tilde{\mathcal{D}}_2(\tau))\,P_{22}= & \, 0
\end{align}
The differential equations~\eqref{differential:equation:written:down} must be supplemented by initial conditions. We note that since the left hand side of~\eqref{initial:useful:conditions} is the identity matrix for $\tau=0$ we have that $\boldsymbol{P}(0)=\mathds{1}$ which implies $P_{11}(0)=P_{22}(0)=1$. In addition, taking the time derivative of both sides of~\eqref{initial:useful:conditions} and setting $\tau=0$ implies $\dot{P}_{11}(0)=\dot{P}_{22}(0)=0$.

By introducing the integral $I_{P_{22}} = \int^\tau_0 \mathrm{d} \tau' P_{22} (\tau')$, one can also rewrite the second equation as 
\begin{equation} \label{app:eq:IP22}
\ddot{I}_{P_{22}}  + ( 1 + 4 \, \tilde{\mathcal{D}}_2 (\tau) ) I_{P_{22}} = 0 \, ,
\end{equation}
so that it becomes the same as that for $P_{11}$. The boundary conditions are now $I_{P_{22}}(0) = 0$ and $\dot{I}_{P_{22}} = 1$. 

We were not able to find a general solution to the differential equation for $P_{11}$~\eqref{differential:equation:written:down} and the equation for $I_{P_{22}}$ 
\eqref{app:eq:IP22}, but they can be integrated numerically when an explicit form of $\tilde{\mathcal{D}}_2(\tau)$ is given. For specific choices of $\tilde{\mathcal{D}}_2(\tau)$, which we discuss in the main text of this work, the equations become the well-studied Mathieu equation. 

Proceeding with the solution to the quadratic time-evolution, we have
\begin{align}
\overset{\leftarrow}{\mathcal{T}}\,\exp\left[i\,\int_0^\tau\,\mathrm{d}\tau'\,\begin{pmatrix}
0 & 1 \\
1+4\,\tilde{\mathcal{D}}_2(\tau') & 0 
\end{pmatrix}\right]=&\boldsymbol{P}+i\,\int_0^\tau \mathrm{d}\tau'\,\boldsymbol{K}\,\boldsymbol{P} \nonumber \\
=& 
\begin{pmatrix}
P_{11} &  i\,\int_0^\tau\,\mathrm{d}\tau'\,P_{22}\\
i\,\int_0^\tau\,\mathrm{d}\tau'\,(1+4\,\tilde{\mathcal{D}}_2(\tau'))\,P_{11} & P_{22} 
\end{pmatrix}.
\end{align}
In turn, this allows us to get
\begin{align} 
\boldsymbol{S}_{\mathrm{sq}}(\tau) &=\begin{pmatrix}
\alpha(\tau) & \beta(\tau) \\
\beta^*(\tau) & \alpha^*(\tau) 
\end{pmatrix}
 \nonumber \\
&=\boldsymbol{M}_{\text{ort}}^{\rm{T}}\,
\begin{pmatrix}
P_{11} & i\,\int_0^\tau\,\mathrm{d}\tau'\,P_{22}\\
i\,\int_0^\tau\,\mathrm{d}\tau'\,(1+4\,\tilde{\mathcal{D}}_2(\tau'))\,P_{11} & P_{22} 
\end{pmatrix}\,
\boldsymbol{M}_{\text{ort}},
\end{align}
where we have introduced the Bogoliubov matrix $\boldsymbol{S}_{\mathrm{sq}}(\tau)$ with coefficients $\alpha(\tau)$ and $\beta(\tau)$.

This gives us, after a little algebra,
\begin{align}
\alpha(\tau)=&\frac{1}{2}\,\left[P_{11}(\tau)+P_{22}(\tau)-i\,\int_0^\tau\,\mathrm{d}\tau'\,P_{22}(\tau')-i\,\int_0^\tau\,\mathrm{d}\tau'\,(1+4\,\tilde{\mathcal{D}}_2(\tau'))\,P_{11}(\tau')\right] \, , \nonumber\\
\beta(\tau) =&\frac{1}{2}\,\left[P_{11}(\tau)-P_{22}(\tau)+i\,\int_0^\tau\,\mathrm{d}\tau'\,P_{22}(\tau')-i\,\int_0^\tau\,\mathrm{d}\tau'\,(1+4\,\tilde{\mathcal{D}}_2(\tau'))\,P_{11}(\tau')\right].
\end{align}
These quantities can also be written in terms of $I_{P_{22}}$ as 
\begin{align}
\alpha(\tau) =& \frac{1}{2}\,\left[ P_{11} - i I_{P_{22}} + i\frac{d}{d\tau} ( P_{11}  - i I_{P_{22}} )\right] \, ,\nonumber\\
\beta(\tau) =& \frac{1}{2}\,\left[ P_{11} + i I_{P_{22}} + i\frac{d}{d\tau} ( P_{11}  + i I_{P_{22}} )\right].
\end{align}
This means that the basis vector $\mathbb{X}$~\eqref{eq:Bogoliubov:transform} transforms  as 
\begin{align} \label{eq:Bogoliubov:transform:solved}
\mathbb{X}'=
\hat{U}_{\mathrm{sq}}^\dag
\,\mathbb{X}\,
\hat{U}_{\mathrm{sq}}
=
\begin{pmatrix}
\alpha(\tau) & \beta(\tau) \\
\beta^*(\tau) & \alpha^*(\tau) 
\end{pmatrix}
\,\mathbb{X}.
\end{align}
As a side remark, the Bogoliubov (symplectic) identities $|\alpha(t)|^2-|\beta(t)|^2=1$ read
\begin{align}
P_{11}\,P_{22}+\left(\int_0^\tau\,\mathrm{d}\tau'\,P_{22}\right)\,\left(\int_0^\tau\,\mathrm{d}\tau'\,(1+4\,\tilde{\mathcal{D}}_2(\tau'))\,P_{11}\right)=1.
\end{align}

\subsection{ Decoupling the nonlinear time-evolution}
We can now go back to the time evolution operator
\eqref{formal:solution:to:the:decoupling:easier} which we reprint here
\begin{align}
\hat{\tilde{U}}(\tau)=e^{-i\,\Omega_\textrm{c} \hat a^\dagger \hat a\,\tau}\,\hat{U}_{\mathrm{sq}}(\tau)\,\overset{\leftarrow}{\mathcal{T}}\,\exp\left[-\frac{i}{\hbar}\,\int_0^\tau\,\mathrm{d}\tau'\,\hat{U}_{\mathrm{sq}}^\dag(\tau')\,\hat {H}_1(\tau')\,\hat{U}_{\mathrm{sq}}(\tau')\right].
\end{align}
Our work above allows to obtain
\begin{align}
\hat{\tilde{U}}(\tau)=e^{-i\,\Omega_\textrm{c} \hat a^\dagger \hat a\,\tau}\,\hat{U}_{\mathrm{sq}}(\tau)\,\overset{\leftarrow}{\mathcal{T}}\,&\exp\biggl[-i\,\int_0^\tau\,\mathrm{d}\tau'\,\bigl(\tilde{\mathcal{D}}_1(\tau') \bigl(\xi(\tau')\,\hat b+ \xi^*(\tau')\,\hat b{}^\dagger\bigr) \nonumber \\
&\quad\quad\quad\quad\quad\quad\quad\quad-\tilde{\mathcal{G}}(\tau') \,  \hat a^\dagger\hat a\,\bigl(\xi(\tau')\,\hat b+ \xi^*(\tau')\,\hat b{}^\dagger\bigr)\bigr)\biggr],
\end{align}
which can be conveniently rewritten as 
{\small\begin{align}\label{almost:final:time:evolution:operator}
\hat{\tilde{U}}(t)&=e^{-i\,\Omega_\textrm{c} \hat a^\dagger \hat a\,\tau}\,\hat{U}_{\mathrm{sq}}\,\overset{\leftarrow}{\mathcal{T}}\,\exp\biggl[-i\,\int_0^\tau\,\mathrm{d}\tau'\,\biggl(\tilde{\mathcal{D}}_1(\tau') \Re\xi(\tau')\,\hat B_+-i \, \tilde{\mathcal{D}}_1(\tau')\,\Im\xi(\tau')\,\hat B_- \nonumber \\
&\quad\quad\quad\quad\quad\quad\quad\quad\quad\quad\quad-\tilde{\mathcal{G}}(\tau') \,\Re\xi(\tau')\,\hat a^\dagger\hat a\,\hat B_++i \, \tilde{\mathcal{G}}(\tau') \,\Im\xi(\tau')\,\hat a^\dagger\hat a\,\hat B_-\biggr)\biggr].
\end{align}}
Here we have introduced 
\begin{align} \label{app:eq:simplified:xi}
\xi(\tau):=\alpha(\tau)+\beta^*(\tau)=P_{11}-i\,\int_0^\tau\,\mathrm{d}\tau'\,P_{22}.
\end{align}
for conveniency of presentation, where $\Re \xi(\tau)$ and $\Im \xi(\tau)$ are the real and imaginary parts of $\xi(\tau)$. Our definition of $\xi(\tau)$ also implies that  
\begin{align} \label{app:eq:xi:alpha:beta:relation}
\alpha(\tau) &= \frac{1}{2}(\xi(\tau) + i \dot{\xi}(\tau)) &\mbox{and} &&\beta(\tau) = \frac{1}{2}(\xi^*(\tau) + i \dot{\xi}^*(\tau)) .
\end{align}
We now note that the operators $\hat N_a,\hat N_a^2, \hat B_+,\hat B_-,\hat N_a\,\hat B_+,\hat N_a\,\hat B_-$ form a closed Lie sub-algebra of the full algebra of our operators. Therefore, we can apply the decoupling techniques described above.

We  proceed to decouple the remaining part of the operator
\eqref{almost:final:time:evolution:operator}, which reads 
\begin{align}
\hat{U}_\text{remain}(t):=\overset{\leftarrow}{\mathcal{T}}\,&\exp\biggl[-i\,\int_0^\tau\,\mathrm{d}\tau'\,\biggl(\tilde{\mathcal{D}}_1(\tau') \Re\xi(\tau')\,\hat B_+-\tilde{\mathcal{D}}_1(\tau')\,\Im\xi(\tau')\,\hat B_- \nonumber \\
&\quad\quad\quad\quad\quad\quad\quad\quad\quad\quad-\tilde{\mathcal{G}}(\tau') \,\Re\xi(\tau')\,\hat a^\dagger\hat a\,\hat B_++\tilde{\mathcal{G}}(\tau') \,\Im\xi(\tau')\,\hat a^\dagger\hat a\,\hat B_-\biggr)\biggr].
\end{align}
We make the ansatz
\begin{align}\label{decoupling:ansatz}
\hat{U}_\text{remain}(t)=&e^{-i\,F_{\hat{N}_a}\,\hat{N}_a}\,e^{-i\,F_{\hat{N}^2_a}\,\hat{N}^2_a}\,e^{-i\,F_{\hat{B}_+}\,\hat{B}_+}\,e^{-i\,F_{\hat{N}_a\,\hat{B}_+}\,\hat{N}_a\,\hat{B}_+}\,e^{-i\,F_{\hat{B}_-}\,\hat{B}_-}\,e^{-i\,F_{\hat{N}_a\,\hat{B}_-}\,\hat{N}_a\,\hat{B}_-}.
\end{align}
Taking the time derivative on both sides and arranging in a similar fashion to~\eqref{operator:differential:equations} we find
\begin{align}\label{main:docupling:Hamiltonian:technique:simplified}
&\tilde{\mathcal{D}}_1(\tau) \Re\xi(\tau)\,\hat B_+-\tilde{\mathcal{D}}_1(\tau')\,\Im\xi(\tau)\,\hat B_--\tilde{\mathcal{G}}(\tau) \,\Re\xi(\tau)\,\hat a^\dagger\hat a\,\hat B_++\tilde{\mathcal{G}}(\tau) \,\Im\xi(\tau)\,\hat a^\dagger\hat a\,\hat B_-\nonumber\\
=\, &\dot F_{\hat{N}_a}\,\hat{N}_a
+\dot F_{\hat{N}^2_a}\,\hat{N}^2_a
+ \dot F_{\hat{B}_+}\,\hat{B}_+
+ \dot F_{\hat{N}_a\,\hat{B}_+}\,\hat{N}_a\,\hat{B}_+\nonumber\\
&+ \dot F_{\hat{B}_-}\,e^{-i\,F_{\hat{B}_+}\,\hat{B}_+}\,e^{-i\,F_{\hat{N}_a\,\hat{B}_+}\,\hat{N}_a\,\hat{B}_+}\,
\hat{B}_-
\,e^{i\,F_{\hat{N}_a\,\hat{B}_+}\,\hat{N}_a\,\hat{B}_+}\,e^{i\,F_{\hat{B}_+}\,\hat{B}_+}\nonumber\\
&+ \dot F_{\hat{N}_a\,\hat{B}_-}\,e^{-i\,F_{\hat{B}_+}\,\hat{B}_+}\,e^{-i\,F_{\hat{N}_a\,\hat{B}_+}\,\hat{N}_a\,\hat{B}_+}\,
\hat{N}_a\,\hat{B}_-
\,e^{i\,F_{\hat{N}_a\,\hat{B}_+}\,\hat{N}_a\,\hat{B}_+}\,e^{i\,F_{\hat{B}_+}\,\hat{B}_+}\nonumber\\
=\, &\dot F_{\hat{N}_a}\,\hat{N}_a
+\dot F_{\hat{N}^2_a}\,\hat{N}^2_a
+ \dot F_{\hat{B}_+}\,\hat{B}_+
+ \dot F_{\hat{N}_a\,\hat{B}_+}\,\hat{N}_a\,\hat{B}_+\nonumber\\
&+ (\dot F_{\hat{B}_-}+\dot F_{\hat{N}_a\,\hat{B}_-}\,\hat{N}_a)\,e^{-i\,F_{\hat{B}_+}\,\hat{B}_+}\,e^{-i\,F_{\hat{N}_a\,\hat{B}_+}\,\hat{N}_a\,\hat{B}_+}\,
\hat{B}_-
\,e^{i\,F_{\hat{N}_a\,\hat{B}_+}\,\hat{N}_a\,\hat{B}_+}\,e^{i\,F_{\hat{B}_+}\,\hat{B}_+}\nonumber\\
=\, &\dot F_{\hat{N}_a}\,\hat{N}_a
+\dot F_{\hat{N}^2_a}\,\hat{N}^2_a+ \dot F_{\hat{B}_+}\,\hat{B}_+
+ \dot F_{\hat{N}_a\,\hat{B}_+}\,\hat{N}_a\,\hat{B}_+\nonumber \\
&+ (\dot F_{\hat{B}_-} +\dot F_{\hat{N}_a\,\hat{B}_-}\,\hat{N}_a)\,(\hat{B}_-+2\, F_{\hat{B}_+}+2\, F_{\hat{N}_a\,\hat{B}_+}\,\hat{N}_a).
\end{align}
Therefore our main differential equations can be obtained by equating the coefficient of the different, linearly independent operators of the Lie algebra from the equation below
\begin{align}\label{main:docupling:Hamiltonian:technique:simplified}
&\tilde{\mathcal{D}}_1(\tau) \Re\xi(\tau)\,\hat B_+-\tilde{\mathcal{D}}_1(\tau')\,\Im\xi(\tau)\,\hat B_--\tilde{\mathcal{G}}(\tau) \,\Re\xi(\tau)\,\hat a^\dagger\hat a\,\hat B_++\tilde{\mathcal{G}}(\tau) \,\Im\xi(\tau)\,\hat a^\dagger\hat a\,\hat B_-\nonumber\\
=\, &\dot F_{\hat{N}_a}\,\hat{N}_a
+\dot F_{\hat{N}^2_a}\,\hat{N}^2_a
+ \dot F_{\hat{B}_+}\,\hat{B}_+
+ \dot F_{\hat{N}_a\,\hat{B}_+}\,\hat{N}_a\,\hat{B}_+\nonumber \\
&+ (\dot F_{\hat{B}_-}+\dot F_{\hat{N}_a\,\hat{B}_-}\,\hat{N}_a)\,(\hat{B}_-+2\, F_{\hat{B}_+}+2\, F_{\hat{N}_a\,\hat{B}_+}\,\hat{N}_a).
\end{align}
The solutions with operators proportional to $\hat B_\pm$ can be independently solved. However, the solutions for  $F_{\hat N_a}$, $F_{\hat N_a^2} $, and $F_{\hat N_a \, \hat B_\pm}$ are less straight-forward. We identify the following four coupled differential equations:
\begin{align}\label{eq:diff:eqs:explicit}
\dot{F}_{\hat N_a}  &= - 2  \, \dot{F}_{\hat B_-} \, F_{\hat N_a \, \hat B_+} - 2 \, F_{\hat B_+} \, \dot{F}_{\hat N_a \, \hat B_-} \, , \nonumber \\
\dot{F}_{\hat N_a^2} &= - 2 \, \dot{F}_{\hat N_a \, \hat B_-} \, F_{\hat N_a \, \hat B_+} \, ,  \nonumber \\
\dot{F}_{\hat N_a \, \hat B_+} &=- \tilde{\mathcal{G}} ( \tau) \, \Re \xi(\tau) \, , \nonumber \\
\dot{F}_{\hat N_a \, \hat B_-} &= \tilde{\mathcal{G}} \,  \Im \xi(\tau) \, .
\end{align}
By first solving the equations for $\dot{F}_{\hat N_a \, \hat B_\pm}$ and $\dot{F}_{\hat B_\pm}$, it is then possible to insert the solutions into the expressions for $\dot{F}_{\hat N_a} $ and $\dot{F}_{\hat N_a^2}$. We find the following key expression for this work
\begin{align}\label{appendix:sub:algebra:decoupling:solution}
  F_{\hat{N}_a}(\tau)&= -2 \,\int_0^\tau\,\mathrm{d}\tau'\,\tilde{\mathcal{D}}_1(\tau')\,\Im\xi(\tau')\int_0^{\tau'}\mathrm{d}\tau''\, \tilde{\mathcal{G}}(\tau'')\,\Re\xi(\tau'')\, \nonumber \\
&-2 \, \int^\tau_0\,\mathrm{d}\tau' \,\tilde{\mathcal{G}}(\tau')\, \Im \xi(\tau') \, \int^{\tau'}_0 \,\mathrm{d}\tau''\, \tilde{\mathcal{D}}_1(\tau'') \, \Re \xi(\tau'') \, ,  \nonumber\\
F_{\hat{N}^2_a}(\tau)&=  2 \,\int_0^\tau\,\mathrm{d}\tau'\,\tilde{\mathcal{G}}(\tau')\,\Im\xi(\tau')\int_0^{\tau'}\mathrm{d}\tau''\,\tilde{\mathcal{G}}(\tau'')\,\Re\xi(\tau'')\, , \nonumber\\
F_{\hat{B}_+}(\tau)&= \int_0^\tau\,\mathrm{d}\tau'\,\tilde{\mathcal{\mathcal{D}}}_1(\tau')\,\Re\xi(\tau')\, , \nonumber\\
F_{\hat{B}_-}(\tau)&=- \int_0^\tau\,\mathrm{d}\tau'\,\tilde{\mathcal{D}}_1(\tau')\,\Im\xi(\tau')\, , \nonumber\\
F_{\hat{N}_a\,\hat{B}_+}(\tau)&=- \int_0^\tau\,\mathrm{d}\tau'\,\tilde{\mathcal{G}}(\tau')\,\Re\xi(\tau')\, , \nonumber\\
F_{\hat{N}_a\,\hat{B}_-}(\tau)&=\int_0^\tau\,\mathrm{d}\tau'\,\tilde{\mathcal{G}}(\tau')\,\Im\xi(\tau') \, , 
\end{align}
where $\Re\xi$ denotes the real part of $\xi$ and $\Im\xi$ denotes the imaginary part of $\xi$. 
This result concludes the decoupling part of our work. The expressions~\eqref{appendix:sub:algebra:decoupling:solution}, together with the decoupling form~\eqref{decoupling:ansatz}, can be used in the  expression for $\hat{U}(\tau)$~\eqref{formal:solution:to:the:decoupling:easier} to obtain an explicit (up to a formal solution for $\xi(\tau)$) time-evolved expression for the observables of the system. 

Finally, let us once more write down the final expression of the time-evolution operator
\begin{align} \label{eq:app:evolution:operator}
\hat U(\tau)=& \, e^{-i\,\Omega_\mathrm{c} \hat a^\dagger \hat a\,\tau}\,\hat{U}_{\mathrm{sq}}(\tau)\,e^{-i\,F_{\hat{N}_a}\,\hat{N}_a}\,e^{-i\,F_{\hat{N}^2_a}\,\hat{N}^2_a}\,e^{-i\,F_{\hat{B}_+}\,\hat{B}_+}\,e^{-i\,F_{\hat{N}_a\,\hat{B}_+}\,\hat{N}_a\,\hat{B}_+}\,\nonumber \\
&\times e^{-i\,F_{\hat{B}_-}\,\hat{B}_-}\,e^{-i\,F_{\hat{N}_a\,\hat{B}_-}\,\hat{N}_a\,\hat{B}_-}, 
\end{align}
to be complemented with the functions listed in~\eqref{appendix:sub:algebra:decoupling:solution}.

\section{First and second moments and covariance matrix elements}\label{appendix:CM}
We can employ the results summarised in Section~\ref{tools} to compute all the relevant quantities needed in this work. They include the first and second moments of the state of the system at any moment in time, which we use to compute the measure of non-Gaussianity, see Section~\ref{sec:measure}. To do this, we use the explicit form of the time-evolution operator, written out above in~\eqref{eq:app:evolution:operator}. 
We further assume that the initial state is a separable coherent state $\ket{\mu_{\rm{c}}}\otimes \ket{\mu_{\rm{m}}}$, defined in~\eqref{initial:state:two}. 

In order to compute the second moments that constitute the covariance matrix $\boldsymbol{\sigma}$ for the state, we must calculate the expectation values of $\langle \hat a \rangle, \langle \hat b \rangle, \langle \hat a ^\dag a \rangle, \langle \hat b ^\dag \hat b \rangle, \langle \hat a ^2 \rangle, \langle \hat b^2 \rangle, \langle \hat a \hat b \rangle$, and $\langle \hat a\hat b^\dag \rangle$. 
To achieve this goal, we start by defining the following quantities for ease of notation:
\begin{align}
\varphi(\tau) :=& \, F_{\hat N_a} + F_{\hat N_a^2}  + 2  \, F_{\hat N_a \hat B _+} F_{\hat B_-} \, , \nonumber\\
\theta(\tau) :=& \, 2 \left( F_{\hat N_a^2 }  + F_{\hat N_a \hat B_+} F_{\hat N_a \hat B_-}  \right) \, , \nonumber\\
\Gamma(\tau) :=\, & (\alpha(\tau) + \beta(\tau)) F_{\hat{B}_-} - i ( \alpha(\tau) - \beta(\tau) ) \, F_{\hat{B}_+}\, , \nonumber\\
\Delta(\tau) :=\, & (\alpha(\tau) + \beta(\tau)) F_{\hat{N}_a \, \hat{B}_- } - i ( \alpha(\tau) - \beta(\tau)) F_{\hat{N}_a \, \hat{B}_+ }\, , \nonumber\\
E_{\hat B _+ \hat B_-} :=& \, \left\langle e^{- i \, F_{\hat N_a \, \hat B_+}\, \hat{B}_+} \, e^{- i \, F_{\hat N_a \, \hat B_-} \, \hat B_-} \right\rangle \, , \nonumber\\
E_{\hat B_+ \hat B_- \hat B_+ \hat B_- } :=&\,  \left\langle  e^{- i \, F_{\hat N_a \hat B_+}\, \hat{B}_+} \, e^{- i \, F_{\hat N_a \, \hat B_-} \, \hat B_-}  \, e^{- i \, F_{\hat N_a \, \hat B_+}\, \hat{B}_+} \, e^{-i \,  F_{\hat N_a \, \hat B_-} \, \hat B_-}\right\rangle, 
\end{align}
and where we also introduce 
\begin{align}
K &= F_{\hat B_-} + i F_{\hat B_+}\, ,  \nonumber \\
K_{\hat N_a} &= F_{\hat N_a \, \hat B_-} + i F_{\hat N_a \, \hat B_+}\, , 
\end{align}
which we use when deriving the state~\eqref{non:linear:state:evolution}. The last two quantities can be computed using the expression for the Weyl displacement operator $\hat D_\alpha = e^{\alpha \, \hat b^\dag - \alpha^* \hat b}$ and the combination $\hat D_\alpha \, \hat D_\beta = e^{\left( \alpha \beta^* - \alpha^* \beta \right)/2} \hat D_{\alpha + \beta}$, and read
\begin{align}
E_{\hat B _+ \hat B_-} &=  \mathrm{exp} \biggl[ \frac{1}{2} \biggl( - |K_{\hat N_a}|^2 - 2 \, i \, F_{\hat N_a \hat B_-} F_{\hat N_a \hat B_+} - 2 \, \mu_{\mathrm{m}}K_{\hat N_a} + 2 \,  \mu_{\mathrm{m}}^* K^*_{\hat N_a} \biggr) \biggr]\nonumber\\
E_{\hat B_+ \hat B_- \hat B_+ \hat B_- } &= \mathrm{exp} \biggl[ - 2\,  \biggl( |K_{\hat N_a}|^2 + i \, F_{\hat N_a \hat B_+} F_{\hat N_a \hat B_-} + \mu_{\mathrm{m}} \, K_{\hat N_a} -\mu_{\mathrm{m}}^* \, K_{\hat N_a}^*  \biggr) \biggr] \nonumber \\
&= e^{-|K_{\hat N_a}|^2} \,  E_{\hat B_+ \hat B_-}^2 \, .
\end{align}
The time-evolution of the operators $\hat{a}$ and $\hat{b}$ in the Heisenberg picture is therefore 
\begin{align}
\hat a(\tau) =& \,  e^{- i \, \Omega_{\rm{c}} \, \tau} e^{- i \,  \varphi(\tau) } \, e^{-  i \, \theta(\tau) \, \hat{N}_a} e^{- i F_{\hat{N}_a \, \hat{B}_+} \, \hat{B}_+ } \, e^{- i F_{\hat{N}_a \, \hat{B}_- } \, \hat{B}_- } \, \hat{a} , \nonumber \\
\hat b(\tau)=& \, \alpha(\tau ) \, \hat  b + \beta(\tau) \,  \hat b^\dagger + \Gamma(\tau) + \Delta(\tau) \hat N_a.
\end{align}
These expressions allow us to compute the expectation values of the
first and second moments given our initial state. We have here
transformed into a frame rotating with 
$e^{- i \, \Omega_{\rm{c}} \, \tau} $, which read
\begin{align} \label{eq:app:expectation:values}
\braket{\hat a (\tau) } =\, & e^{- i \varphi } \, e^{|\mu_{\mathrm{c}}|^2  \, (e^{- i \theta}-1)}  \, E_{\hat B_+ \hat B_- } \mu_{\mathrm{c}} \, , \nonumber \\
\braket{\hat b(\tau) } =\, & \alpha(\tau) \,  \mu_{\mathrm{m}} + \beta(\tau) \,  \mu_{\mathrm{m}}^*  + \Gamma(\tau) + \Delta(\tau)  \, |\mu_{\mathrm{c}}|^2 \, ,\nonumber \\ 
\braket{\hat{a}^2} = \, & e^{- 2i \varphi} \, \mu_{\mathrm{c}}^2  \, e^{- i \theta} \,  e^{|\mu_{\mathrm{c}}|^2 \,  (e^{- 2 i \theta} - 1)} \, e^{-|K_{\hat N_a}|^2} \,  E_{\hat B_+ \hat B_-}^2\, , \nonumber  \\
 \braket{\hat b^{2} } =&\,  \alpha^2(\tau) \,  \mu_{\mathrm{m}}^2  + \alpha(\tau) \,  \beta(\tau) \,(2 \, |\mu_{\mathrm{m}}|^2 + 1) + \beta^2(\tau) \, \mu_{\mathrm{m}}^{*2}  \nonumber \\
 &+ 2 \, (\alpha(\tau)\,\mu_{\mathrm{m}}+\beta(\tau)\,\mu_{\mathrm{m}}^*) \,  \left[\Gamma(\tau) + \Delta(\tau)\,  |\mu_{\mathrm{c}}|^2 \right] \nonumber \\
 &+ \Gamma^2 (\tau) + 2 \, \Gamma(\tau) \Delta(\tau) \, |\mu_{\mathrm{c}}|^2 + \Delta^2(\tau) \, |\mu_{\mathrm{c}}|^2 \, ( 1 + |\mu_{\mathrm{c}}|^2)\, , \nonumber  \\
 \braket{\hat a^\dag \hat a } = \, & |\mu_{\rm{c}}|^2\, , \nonumber \\
 \braket{\hat b ^\dag \hat b}  = \, &  (|\alpha(\tau)|^2 + |\beta(\tau)|^2) | \, \mu_{\mathrm{m}}|^2 + \alpha^*(\tau) \,  \beta (\tau) \,  (\mu_{\mathrm{m}}^*)^2 + \alpha(\tau) \, \beta^*(\tau) \,  \mu_{\mathrm{m}}^2 \nonumber \\
 &+ (\alpha^*(\tau)\,\mu_{\mathrm{m}}^* +\beta^*(\tau)\,\mu_{\mathrm{m}}) \, \left( \Gamma(\tau) + \Delta(\tau)  \, |\mu_{\mathrm{c}}|^2 \right)\nonumber \\
 &+ (\alpha(\tau)\,\mu_{\mathrm{m}}+\beta(\tau)\,\mu_{\mathrm{m}}^*) \, \left( \Gamma(\tau) + \Delta(\tau) | \, \mu_{\mathrm{c}}|^2 \right)^*\nonumber \\
 &+ (\Gamma^*(\tau) \, \Delta(\tau) + \Gamma(\tau) \,  \Delta^*(\tau)) | \, \mu_{\mathrm{c}}|^2 + |\Delta(\tau)|^2 |\mu_{\mathrm{c}}|^2 \, ( 1 + |\mu_{\mathrm{c}}|^2)\nonumber\\
 &+ |\beta(\tau)|^2  + |\Gamma(\tau)|^2 \, , \nonumber \\
\braket{\hat a(\tau) \hat b(\tau) } =& \, e^{- i \varphi} \,e^{ |\mu_{\mathrm{c}}|^2 \, ( e^{- i \theta}  - 1)} \, \mu_{\mathrm{c}} \, E_{\hat B_+ \hat B_-} \,  \bigl[ \alpha(\tau)\, \mu_{\mathrm{m}} +\beta(\tau)\,(\mu_{\mathrm{m}}^*- K_{\hat N_a}) \nonumber \\
&+ 
\Gamma(\tau) +\left( |\mu_{\mathrm{c}}|^2 \,  e^{- i \theta} + 1\right) \, 
\Delta (\tau)  \bigr]\, ,\nonumber   \\
\braket{\hat a (\tau) \, \hat b^\dag(\tau) }= \, & e^{- i \varphi} \, e^{ |\mu_{\mathrm{c}}|^2 \, ( e^{- i \theta}  - 1)} \, \mu_{\mathrm{c}} \,E_{\hat B_+ \hat B_-}\,  \bigl[  \alpha^*(\tau)\, (\mu_{\mathrm{m}}^* - K_{\hat N_a}) \nonumber \\
&+\beta^*(\tau)\,\mu_{\mathrm{m}} +
\Gamma^*(\tau)  + \left( |\mu_{\mathrm{c}}|^2 \, e^{- i \theta} + 1\right) \, \Delta^* (\tau)  \bigr]  \, .
\end{align}
The two-mode covariance matrix is fully determined by its elements $\sigma_{nm}$, defined in this basis as 
\begin{align}\label{general:CM:coefficients}
\sigma_{11} &= \sigma_{33} = 1 + 2\braket{\hat a^\dag \hat a} - 2 \braket{\hat a^\dag } \braket{\hat a } \, , \nonumber \\
\sigma_{31} &=2 \braket{\hat{a}^2} - 2 \braket{\hat{a}}^2 \, ,\nonumber \\
\sigma_{22} &= \sigma_{44} = 1 + 2 \braket{\hat{b}^\dag \hat{b}} - 2\braket{\hat{b}^\dag} \braket{\hat{b}} \, , \nonumber \\
\sigma_{42}  &= 2 \braket{\hat{b}^2} - 2 \braket{\hat{b}}^2\, ,  \nonumber \\
\sigma_{21} &= \sigma_{34} = 2 \braket{\hat{a} \hat{b}^\dag} - 2 \braket{\hat{a}} \braket{\hat{b}^\dag} \, ,  \nonumber \\
\sigma_{41} &= \sigma_{32} = 2 \braket{\hat{a} \hat{b}} - 2 \braket{\hat{a} }\braket{\hat{b}} \, ,
\end{align}
where all the other elements follow from the symmetries of $\boldsymbol{\sigma}$, imposed by the requirement that $\boldsymbol{\sigma}^\dagger = \boldsymbol{\sigma}$. Given all of the above, we can compute an explicit expression for elements~\eqref{general:CM:coefficients} of the covariance matrix, which reads
\begin{align} \label{eq:CM:elements}
\sigma_{11} &=1 + 2\, |\mu_{\mathrm{c}}|^2 \left( 1 - e^{-4|\mu_{\mathrm{c}}|^2 \sin^2{\theta/2}} \, |E_{\hat B_+ \hat B_-}|^2 \right)\, , \nonumber \\
\sigma_{31} &=  2\, \mu_{\mathrm{c}}^2 \, e^{- 2i \varphi} \left(  e^{- i \theta} \, e^{|\mu_{\mathrm{c}}|^2 ( e^{-  2i \theta} - 1)} e^{-|K_{\hat N_a}|^2} \,   - e^{2|\mu_{\mathrm{c}}|^2 ( e^{- i \theta} - 1)} \right) \, E_{\hat B_+ \hat B_-}^2\, , \nonumber \\
\sigma_{21} &=  2 \,e^{- i \varphi(\tau)} \, e^{ |\mu_{\mathrm{c}}|^2 ( e^{- i \theta(\tau)}  - 1)}  \,E_{\hat B_+ \hat B_-} \, \mu_{\mathrm{c}}\, \left[\Delta^*(\tau)\left( |\mu_{\mathrm{c}}|^2 (e^{- i \theta(\tau)} - 1) + 1  \right) - \alpha^*(\tau) \, K_{\hat N_a}\right]\, , \nonumber \\
\sigma_{41} &= 2\, e^{- i \varphi(\tau) } \, e^{|\mu_{\mathrm{c}}|^2 (e^{- i \theta(\tau)}-1)} E_{\hat B_+ \hat B_- } \mu_{\mathrm{c}} \, \left[\Delta(\tau) \left( |\mu_{\mathrm{c}}|^2 (e^{- i \theta(\tau)} - 1) + 1\right)- \beta(\tau) K_{\hat N_a}  \right]\, ,\nonumber\\
\sigma_{22} &= 1 + 2\, |\beta(\tau)|^2  + 2\, |\Delta(\tau)|^2 \, |\mu_{\mathrm{c}}|^2\, , \nonumber \\
\sigma_{42}  &=2\,  \alpha(\tau) \, \beta(\tau) +2\,  \Delta^2(\tau) \, |\mu_{\mathrm{c}}|^2.
\end{align}

\subsection{Symplectic eigenvalues of the optical and mechanical subsystems}\label{appendix:subsystems}
In this Appendix, we use the covariance matrix elements to compute the symplectic eigenvalues of the optical and mechanical subsystems. 

\subsubsection{The optical symplectic eigenvalue}
We wish to compute the symplectic eigenvalue of the optical subsystem. It is given by 
\begin{equation}
\nu_{Op}^2 = \sigma_{11}^2 - |\sigma_{13}|^2  \, .
\end{equation}
From the covariance matrix elements in~\eqref{eq:CM:elements} we find
\begin{align}
\sigma_{11}^2 &= 1 + 4\, |\mu_{\mathrm{c}}|^2 \left( 1 - e^{-4|\mu_{\mathrm{c}}|^2 \sin^2{\theta/2}} \,e^{- |K_{\hat N_a}|^2} \right)+ 4 \, |\mu_{\mathrm{c}}|^4 \left( 1 - e^{-4|\mu_{\mathrm{c}}|^2 \sin^2{\theta/2}} \,e^{- |K_{\hat N_a}|^2} \right) ^2  \, ,\nonumber \\
|\sigma_{31}|^2 &= 4\, |\mu_{\mathrm{c}}|^4 \, |E_{\hat B_+ \hat B_-}|^4 \left(  e^{i \theta} \, e^{|\mu_{\mathrm{c}}|^2 ( e^{2i \theta} - 1)} e^{-|K_{\hat N_a}|^2} \,   - e^{2|\mu_{\mathrm{c}}|^2 ( e^{i \theta} - 1)} \right) \, \nonumber \\
&\quad\quad\quad\quad\quad\quad\quad\quad\times \left(  e^{- i \theta} \, e^{|\mu_{\mathrm{c}}|^2 ( e^{-  2i \theta} - 1)} e^{-|K_{\hat N_a}|^2} \,   - e^{2|\mu_{\mathrm{c}}|^2 ( e^{- i \theta} - 1)} \right)\nonumber \\
&= 4\, |\mu_{\mathrm{c}}|^4 \, |E_{\hat B_+ \hat B_-}|^4 \biggl( e^{-2|K_{\hat N_a}|^2}e^{|\mu_c|^2 \left( e^{2i \theta} + e^{- 2 i \theta} -2\right)}    + e^{2|\mu_{\mathrm{c}}|^2 ( e^{i \theta} + e^{- i \theta} -  2)}  \nonumber \\
&\quad\quad\quad\quad\quad\quad\quad\quad\quad- 2 \, \Re \left[  e^{i \theta} \, e^{|\mu_{\mathrm{c}}|^2 ( e^{2i \theta} - 1)} e^{-|K_{\hat N_a}|^2} e^{2 |\mu_c|^2 (e^{- i \theta} - 1)} \right] \biggr)  \,. 
\end{align}
By using the following trigonometric identities, 
\begin{align}
e^{2i \theta} + e^{-2 i \theta } - 2  &= 2 \cos{2\theta} - 2= - 4 \sin^2{\theta}  \, , \nonumber \\
e^{i \theta} + e^{- i \theta } - 2 &= 2( \cos\theta - 1) = - 4 \sin^2{\theta/2}   \, ,
\end{align}
 and from the fact that $|E_{\hat B_+ \hat B_-}|^2 = e^{- |K_{\hat N_a}|^2 }$, we obtain
\begin{align}
\sigma_{11}^2 &= 1 + 4\, |\mu_{\mathrm{c}}|^2 \left( 1 - e^{ -4|\mu_{\mathrm{c}}|^2 \sin^2{\theta/2}} \,e^{- |K_{\hat N_a}|^2} \right)+ 4 \, |\mu_{\mathrm{c}}|^4 \left( 1 - e^{ -4|\mu_{\mathrm{c}}|^2 \sin^2{\theta/2}} \,e^{- |K_{\hat N_a}|^2} \right) ^2 \, ,\nonumber \\
|\sigma_{31}|^2 &= 4\, |\mu_{\mathrm{c}}|^4 \, e^{-2 |K_{\hat N_a}|^2 } \biggl( e^{-2|K_{\hat N_a}|^2}e^{ - 4|\mu_c|^2  \sin^2{\theta}}    + e^{-8 |\mu_{\mathrm{c}}|^2  \sin^2\theta/2} \nonumber \\
&\quad\quad\quad\quad\quad\quad\quad\quad\quad- 2 \, \Re \left[  e^{i \theta} \, e^{|\mu_{\mathrm{c}}|^2 ( e^{2i \theta} - 1)} e^{-|K_{\hat N_a}|^2} e^{2 |\mu_c|^2 (e^{- i \theta} - 1)} \right] \biggr) \, .
\end{align}
Putting them together, we find 
\begin{align}
\nu_{Op}^2 &= 1 +4  \, |\mu_{\mathrm{c}}|^2 \left( 1 - e^{-4|\mu_{\mathrm{c}}|^2 \sin^2{\theta/2}} \,e^{- |K_{\hat N_a}|^2} \right) \nonumber \\
&+ 4 \,  |\mu_c|^4 \biggl( 1 - 2 \,  e^{-4|\mu_{\mathrm{c}}|^2 \sin^2{\theta/2}} e^{- |K_{\hat N_a}|^2 } + e^{ -8|\mu_{\mathrm{c}}|^2 \sin^2{\theta/2}} e^{- 2|K_{\hat N_a}|^2 } \nonumber \\
&\quad\quad\quad\quad\quad- e^{-4 |K_{\hat N_a}|^2 } e^{- 4 |\mu_c|^2 \sin^2\theta} - e^{- 2 |K_{\hat N_a}|^2 } e^{- 8 |\mu_{\rm{c}}|^2 \sin^2\theta/2} \nonumber \\
&\quad\quad\quad\quad\quad+ 2 \, e^{- 3|K_{\hat N_a}|^2} \,  \Re \left[  e^{i \theta} \, e^{|\mu_{\mathrm{c}}|^2 ( e^{2i \theta} - 1)}  e^{2 |\mu_c|^2 (e^{- i \theta} - 1)} \right]  \biggr) \, ,
\end{align}
which can  be simplified into 
\begin{align}
\nu_{Op}^2 &= 1 +4  \, |\mu_{\mathrm{c}}|^2 \left( 1 - e^{-4|\mu_{\mathrm{c}}|^2 \sin^2{\theta/2}} \,e^{- |K_{\hat N_a}|^2} \right) \nonumber \\
&+ 4 |\mu_c|^4 \biggl( 1 - 2 e^{-4|\mu_{\mathrm{c}}|^2 \sin^2{\theta/2}} e^{- |K_{\hat N_a}|^2 }  - e^{-4 |K_{\hat N_a}|^2 } e^{- 4 |\mu_c|^2 \sin^2\theta} \nonumber \\
&\quad\quad\quad\quad\quad\quad+ 2 \, e^{- 3|K_{\hat N_a}|^2} \,  \Re \left[  e^{i \theta} \, e^{|\mu_{\mathrm{c}}|^2 ( e^{2i \theta} - 1)}  \, e^{2 |\mu_c|^2 (e^{- i \theta} - 1)} \right]  \biggr) \, .
\end{align}
Next, we compute the mechanical symplectic eigenvalue.

\subsubsection{The mechanical symplectic eigenvalue}
We first recall the Bogoliubov identities, which are $|\alpha(\tau)|^2 - |\beta(\tau)|^2 = 1$ and $\alpha(\tau) \beta^*(\tau) - \alpha^*(\tau) \beta(\tau) = 0$. 

The mechanical eigenvalue is given by 
\begin{equation}
\nu_{\rm{Me}}^2 = \sigma_{22}^2 - |\sigma_{42}|^2 \, .
\end{equation} 
Given the covariance matrix elements in~\eqref{eq:CM:elements}, we find 
\begin{align}
\sigma_{22}^2 &= 1 + 4 |\beta|^2 + 4 |\beta|^4+ 4 |\Delta|^2 |\mu_{\rm{c}}|^2 + 8 |\beta|^2 \, |\Delta|^2 |\mu_{\rm{c}}|^2   + 4 |\Delta|^4 |\mu_{\rm{c}}|^4 \, ,\nonumber \\
|\sigma_{42}|^2 &= 4 |\alpha|^2 |\beta|^2 + 4 \alpha^* \beta^* \Delta^2 |\mu_{\rm{c}}|^2 + 4 \alpha \beta \Delta^{*2} |\mu_{\rm{c}}|^2 + 4 |\Delta|^4 |\mu_{\rm{c}}|^4 \, .
\end{align}
This allows us to write 
\begin{equation}
\nu_{\rm{Me}}^2 = \sigma_{22}^2 - |\sigma_{42}|^2 = 1 + 4\left[ \left( 1 + 2 |\beta|^2 \right) |\Delta|^2 - 2 \, \Re \left( \alpha \, \beta \Delta^{*2} \right) \right] |\mu_{\rm{c}}|^2 \, ,
\end{equation}
where we have suppressed the dependence of $\tau$ for notational clarity. 

We wish to simplify this expression by examining each term in turn and using the Bogoliubov conditions. We recall that 
\begin{equation}
\Delta = \left( \alpha + \beta \right) F_{\hat N_a \, \hat B_-} - i \, \left( \alpha - \beta \right) F_{\hat N_a \,\hat B_+} \, , 
\end{equation}
We can now use the Bogoliubov identities to show that 
\begin{equation}
|\Delta|^2 = \left( 1 +  2|\beta|^2 \right) \, |K_{\hat N_a}|^2 - ( \alpha \beta^* + \alpha^*  \beta ) \left( F_{\hat N_a \, \hat B_+}^2 - F_{\hat N_a \, \hat B_-}^2 \right) \, ,
\end{equation}
and
\begin{align}
\Delta^{*2} = \, & ( \alpha^*   + \beta^*)^2 F_{\hat N_a \, \hat B_-}^2  - ( \alpha^* - \beta^*)^2 F_{\hat N_a \, \hat B_+}^2 + 2 \, i \, ( \alpha^{*2} - \beta^{*2}) \, F_{\hat N_a \, \hat B_+} \, F_{\hat N_a \, \hat B_-} \nonumber \\
= \, & ( \alpha^{*2}  + \beta^{*2}   + 2 \alpha^* \beta^*) F_{\hat N_a \, \hat B_-}^2 - ( \alpha^{*2}  + \beta^{*2}   - 2 \, \alpha^* \beta^*) \, F_{\hat N_a \, \hat B_+}^2  \nonumber \\
&+ 2 \, i \, ( \alpha^{*2} - \beta^{*2} ) F_{\hat N_a \, \hat B_+} F_{\hat N_a \, \hat B_-} \nonumber\\
= \, & \left( \alpha^{*2} + \beta^{*2}  \right) \left( F_{\hat N_a \, \hat B_-}^2 - F_{\hat N_a \, \hat B_+}^2 \right) + 2\, \alpha^* \beta^* |K_{\hat N_a}|^2 \nonumber \\ 
&+ 2 i \left( \alpha^{*2} 
- \beta^{*2}  \right) F_{\hat N_a \, \hat B_+} \, F_{\hat N_a \, \hat B_-}  \, ,
\end{align}
where we recall that $|K_{\hat N_a}|^2 = F_{\hat N_a \, \hat B_+}^2 + F_{\hat N_a \, \hat B_-}^2 $. Furthermore, we find
\begin{align}
\alpha \beta \, \Delta^{*2} = \, & \left( |\alpha|^2 \alpha^* \beta + \alpha \beta^* |\beta|^2 \right) \left( F_{\hat N_a \, \hat B_-}^2 - F_{\hat N_a \, \hat B_+} ^2\right) + 2\, |\alpha|^2 |\beta|^2 |K_{\hat N_a}|^2 \nonumber \\
&+ 2 i \left( |\alpha|^2 \alpha^* \beta - \alpha \beta^* |\beta|^2 \right) F_{\hat N_a \,\hat B_+} F_{\hat N_a \, \hat B_-} \nonumber \\
= \, & \alpha^* \beta \left( F_{\hat N_a \, \hat B_-}^2 - F_{\hat N_a \, \hat B_+}^2 \right) + |\beta|^2 \left( \alpha^* \beta + \alpha \beta^* \right) \left( F_{\hat N_a \, \hat B_-}^2 - F_{\hat N_a \, \hat B_+}^2 \right)  \nonumber \\
&+ 2 \left( 1 + |\beta|^2 \right) |\beta|^2 |K_{\hat N_a}|^2 + 2 i \alpha^* \beta F_{\hat N_a \, \hat B_+} F_{\hat N_a \, \hat B_-} \nonumber \\
&+ 2 i |\beta|^2 \left( \alpha^* \beta - \alpha \beta^* \right) F_{\hat N_a \, \hat B_+} F_{\hat N_a \, \hat B_-} \, ,
\end{align} 
where we used $|\alpha|^2 = |\beta|^2 + 1$ everywhere. We now note that the last term disappears because $\alpha^* \beta - \alpha \beta^* = 0$. We also note that $ \alpha^* \beta = \frac{1}{2} ( \alpha ^* \beta + \alpha \beta^* ) $ is real, which follows from the Bogoliubov condition $\alpha^* \beta = \alpha \beta^*$. When we take the real part, the second-to-last term disappears as well because it has an additional $i$, meaning that we are left with 
\begin{align}
\Re \left( \alpha \beta \Delta^{*2} \right) =& \frac{1}{2} \left( \alpha^* \beta + \alpha \beta^* \right) \left( F_{\hat N_a \, \hat B_-}^2 - F_{\hat N_a \, \hat B_+}^2 \right) + |\beta|^2 \left( \alpha^* \beta + \alpha \beta^* \right) \left( F_{\hat N_a \, \hat B_-}^2 - F_{\hat N_a \, \hat B_+}^2 \right)   \nonumber \\
&+ 2\left( 1 + |\beta|^2 \right) |\beta|^2 |K_{\hat N_a}|^2 \, .
\end{align}
We turn again to the symplectic eigenvalue, which we can now simplify as
\begin{align}
\nu_{\rm{Me}}^2 =&  1 + 4\left[ \left( 1 + 2 |\beta|^2 \right) |\Delta|^2 - 2 \, \Re \left( \alpha \, \beta \Delta^{*2} \right) \right] |\mu_{\rm{c}}|^2 \nonumber \\
=& 1 + 4 \biggl[ ( 1 + 2 |\beta|^2 )^2 |K_{\hat N_a}|^2  - ( 1 + 2 |\beta|^2) \left( \alpha \beta^* + \alpha^* \beta\right) \left( F_{\hat N_a \, \hat B_+}^2 - F_{\hat N_a \, \hat B_-}^2 \right) \nonumber \\
&\quad\quad\quad- \left( \alpha^* \beta + \alpha \beta^* \right) \left( F_{\hat N_a \, \hat B_-}^2 - F_{\hat N_a \, \hat B_+}^2 \right) - 2 |\beta|^2 (\alpha^* \beta + \alpha \beta^*) \left( F_{\hat N_a \, \hat B_-} ^2 - F_{\hat N_a \, \hat B_+}^2 \right)  \nonumber \\
&\quad\quad\quad- 4 (1 + |\beta|^2 ) |\beta|^2 |K_{\hat N_a}|^2 \biggr] |\mu_{\rm{c}}|^2 \nonumber \\
=& 1 + 4 \left[ \left( 1 + 4 |\beta|^2 + 4|\beta|^4 \right) |K_{\hat N_a}|^2 - 4 \left( |\beta|^2 + |\beta|^4 \right) |K_{\hat N_a}|^2 \right] |\mu_{\rm{c}}|^2 \nonumber \\
=& 1 + 4 |K_{\hat N_a}|^2 |\mu_{\rm{c}}|^2 \, .
\end{align}

\section{Considerations of the scenario with constant squeezing}\label{appendix:CSq}
In this Appendix, we show how a constant squeezing can be interpreted as a shift in the mechanical oscillation frequency $\omega_{\mathrm{m}}$. The initial Hamiltonian~\eqref{main:time:independent:Hamiltonian:to:decouple:dimensionful} can be written as
\begin{align}
	\hat {H} &=  \hat{H}'_0 - \hbar\left(\mathcal{G}(t) \hat a^\dagger\hat a - \mathcal{D}_1(t)\right) \left(\hat b+ \hat b{}^\dagger\right),
\end{align}
where the quadratic part, which we call $\hat{H}'_0$, reads
\begin{align}
	\hat{H}'_0 &:=\hbar\,\omega_\mathrm{c} \hat a^\dagger \hat a + \hbar\,\omega_\mathrm{m} \,\hat b{}^\dagger \hat b +  \hbar \, \mathcal{D}_2(t)\left(\hat b+ \hat b{}^\dagger \right)^2. 
\end{align}
To show how the squeezing affects the mechanics, we rewrite this as 
\begin{align}
\hat{H}_0'	&= \hbar\,\omega_\mathrm{c} \hat a^\dagger \hat a + \hbar\,\omega_\mathrm{m} \,\hat b{}^\dagger \hat b +  \hbar \, \mathcal{D}_2(t)\left(\hat b+ \hat b{}^\dagger \right)^2 \nonumber \\
&= \hbar\,\omega_\mathrm{c} \hat a^\dagger \hat a + \frac{m\omega_\mathrm{m}^2}{2} \,\left(\hat x_\mathrm{m} - \frac{i}{m\omega_\mathrm{m}}\hat p_\mathrm{m}\right)\left(\hat x_\mathrm{m} + \frac{i}{m\omega_\mathrm{m}}\hat p_\mathrm{m}\right) +  2 m\omega_\mathrm{m} \mathcal{D}_2(t) \hat x_\mathrm{m}^2 \nonumber \\
	&= \hbar\,\omega_\mathrm{c} \hat a^\dagger \hat a + \frac{1}{2m} \hat p_\mathrm{m}^2 +  \frac{m\omega_\mathrm{m}^2}{2} \,\left(1 + \frac{4\mathcal{D}_2(t)}{\omega_\mathrm{m}}\right) \hat x_\mathrm{m}^2 - \frac{\hbar \omega_\mathrm{m}}{2} \,.
\end{align}
where 
\begin{align}
&\hat{x}_{\mathrm{m}} = \sqrt{\frac{\hbar}{2\omega_{\mathrm{m}} m }} ( \hat{b}^\dag + \hat{b} ), &&\hat{p}_{\mathrm{m}} =i  \sqrt{\frac{\hbar m \omega_{\mathrm{m}}}{2}}(\hat{b}^\dag - \hat{b}) 
\end{align}
This shows that $\mathcal{D}_2(t)$ can be understood and implemented as a possibly time-dependent modulation of the frequency $\omega_\mathrm{m}$ of the mechanical oscillator. For the case of constant squeezing $\mathcal{D}_2$, the Hamiltonian $\hat{H}'_0$ becomes time-independent and we can define $\hat b'$ and $\hat b'{}^\dagger$ with respect to $\omega_\mathrm{m}':= \omega_\mathrm{m} \sqrt{1+4\mathcal{D}_2/\omega_\mathrm{m}}$ such that
\begin{align}
	\hat{H}'_0 &:=\hbar\,\omega_\mathrm{c} \hat a^\dagger \hat a + \hbar\,\omega_\mathrm{m}' \,\hat b'{}^\dagger \hat b' + \frac{\hbar}{2}(\omega'_\mathrm{m} - \omega_\mathrm{m}) \,.
\end{align}
This transformation can be implemented by the squeezing operation $ \hat U^\dagger_{\mathrm{sq}} \hat{H}'_0 \hat U_{\mathrm{sq}}$, where
\begin{align}
	\hat U_{\rm{sq}} &:=\exp\left(\frac{r}{2}( \hat b^{\dagger 2} - {\hat b}^2)\right)  \,,
\end{align}
which induces the mapping
\begin{align}
	\hat U^\dagger_{\rm{sq}} \, \hat b \,  \hat U_{\rm{sq}} &= \cosh(r) \, \hat b + \sinh(r) \, \hat b^{\dagger} \,.
\end{align}
When we apply this to the quadratic Hamiltonian, we obtain 
\begin{align}
\hat U^\dagger_{\rm{sq}} \hat{H}'_0 \hat U_{\rm{sq}} =& \hbar\,\omega_{\mathrm{c}} \hat a^\dagger \hat a + \hbar\,\omega_{\mathrm{m}} \,(\cosh(r)\hat b^\dagger + \sinh(r)\hat b) (\cosh(r)\hat b + \sinh(r)\hat b^\dagger)\\
	& +  \hbar \, \mathcal{D}_2\left(\cosh(r)\hat b + \sinh(r)\hat b^\dagger + \cosh(r)\hat b^\dagger + \sinh(r)\hat b \right)^2  \nonumber \\
=& \hbar\,\omega_{\mathrm{c}} \hat a^\dagger \hat a + \hbar\,\omega_{\mathrm{m}} \,\left((1+2\sinh^2(r)) \hat b^\dagger \hat b  +  \cosh(r)\sinh(r)\left(\hat b^{\dagger2} + \hat b^2\right) + \sinh^2(r)  \right) \nonumber \\
& +  \hbar \, \mathcal{D}_2 e^{2r} \left(2 \hat b{}^\dagger\hat  b + {\hat b{}^\dagger}^2 + \hat b^2 + 1\right) \,.
\end{align}
To cancel the term proportional to $\hat{b}^{\dagger2} + \hat b^2$, we have to fix $\mathcal{D}_2 e^{2r} = -\omega_{\mathrm{m}} \cosh(r)\sinh(r) = -\omega_{\mathrm{m}}(e^{2r} - e^{-2r})/4$, and therefore, $e^{-2r} = \sqrt{ 1 + 4\mathcal{D}_2/\omega_{\mathrm{m}}} = \omega'_{\mathrm{m}}/\omega_{\mathrm{m}}$. With $\mathcal{D}_2 = \omega_{\mathrm{m}}((\omega'_{\mathrm{m}}/\omega_{\mathrm{m}})^2-1)/4$, we obtain
\begin{align}
	 \hat{H}''_0 :=& \hat U^\dagger_{\rm{sq}} \hat{H}'_0 \hat U_{\rm{sq}} \nonumber \\
=& \hbar\,\omega_\mathrm{c} \hat a^\dagger \hat a + \hbar\,\omega_\mathrm{m} \,\left((1+2\sinh^2(r)) \hat b{}^\dagger \hat b  + \sinh^2(r)  \right) +  \hbar \, \mathcal{D}_2 e^{2r} \left(2 \hat  b{}^\dagger \hat b + 1\right) \nonumber \\
=& \hbar\,\omega_\mathrm{c} \hat a^\dagger \hat a + \hbar\,\omega_\mathrm{m} \,\left(\frac{1}{2}\left(\frac{\omega'_\mathrm{m}}{\omega_\mathrm{m}} + \frac{\omega_\mathrm{m}}{\omega'_\mathrm{m}}\right) \hat b^\dagger \hat b  + \frac{1}{4}\left(\frac{\omega'_\mathrm{m}}{\omega_\mathrm{m}} - 2 + \frac{\omega_\mathrm{m}}{\omega'_\mathrm{m}}\right)  \right) \nonumber \\
	& + \frac{\hbar}{4} \, \omega_\mathrm{m}\left(\frac{\omega'_\mathrm{m}}{\omega_\mathrm{m}} - \frac{\omega_\mathrm{m}}{\omega'_\mathrm{m}}\right) \left(2 \hat b^\dagger \hat b + 1\right) \nonumber \\
	=& \hbar\,\omega_\mathrm{c} \hat a^\dagger \hat a + \hbar\,\omega'_\mathrm{m} \, \hat b^\dagger \hat b  + \frac{\hbar}{2}(\omega'_\mathrm{m} - \omega_\mathrm{m})  \,.
\end{align}
In particular,  $\hat{H}_0 \rightarrow \hat{H}''_0 - \hbar(\omega'_\mathrm{m} - \omega_\mathrm{m})/2$ under the replacement $\omega_\mathrm{m} \rightarrow  \omega'_\mathrm{m}$. Furthermore, we find that $\hat b+ \hat b{}^\dagger$ transforms as
\begin{eqnarray}
	\hat U^\dagger_{\rm{sq}}\left(\hat b + \hat b{}^\dagger\right)\hat U_{\rm{sq}} = \sqrt{\frac{\omega_\mathrm{m}}{\omega'_\mathrm{m}}}\left(\hat b+ \hat b{}^\dagger\right)\,.
\end{eqnarray}
When applying the same transformation to the nonlinear part of the Hamiltonian, we find
\begin{align}
	\hat {H}'':= \hat U^\dagger_{\rm{sq}} \hat {H} \hat U_{\rm{sq}} &=  \hat{H}''_0 - \hbar\sqrt{\frac{\omega_\mathrm{m}}{\omega'_\mathrm{m}}}\left(\mathcal{G}(t) \hat a^\dagger\hat a - \mathcal{D}_1(t)\right) \left(\hat b+ \hat b{}^\dagger\right)\,.
\end{align}
If $\mathcal{G}(t)\propto 1/\sqrt{\omega_\mathrm{m}}$ and $\mathcal{D}_1(t)\propto 1/\sqrt{\omega_\mathrm{m}}$, which is indeed fulfilled for the interesting cases, we find that $\hat{H}_\mathrm{opt} \rightarrow \hat{H}'' - \hbar(\omega'_\mathrm{m} - \omega_\mathrm{m})/2$ under the replacement $\omega_\mathrm{m} \rightarrow  \omega'_\mathrm{m}$, where
\begin{equation}
	\hat {H}_\mathrm{opt} =  \hat{H}_0 - \hbar\left(\mathcal{G}(t) \hat a^\dagger\hat a - \mathcal{D}_1(t)\right) \left(\hat b+ \hat b{}^\dagger\right)\,.
\end{equation} 
We define $\hat {H}_\mathrm{opt}^{\omega'_\mathrm{m}}:=\hat {H}_\mathrm{opt}[\omega_\mathrm{m}\rightarrow \omega'_\mathrm{m}]$, and we obtain for the full time evolution
\begin{eqnarray}
	\hat{U}(t)&=&\overset{\leftarrow}{\mathcal{T}}\,\exp\left[-\frac{i}{\hbar}\int_0^{t} dt'\,\hat{H}(t')\right] \nonumber \\
	&=& \hat U_{\rm{sq}}\overset{\leftarrow}{\mathcal{T}}\,\exp\left[-\frac{i}{\hbar}\int_0^{t} dt'\,\hat{H}''(t')\right] \hat U^\dagger_{\rm{sq}} \nonumber \\
	&=& e^{-\frac{i}{2}(\omega'_\mathrm{m} - \omega_\mathrm{m})t} \hat U_{\rm{sq}}\overset{\leftarrow}{\mathcal{T}}\,\exp\left[-\frac{i}{\hbar}\int_0^{t} dt'\,\hat {H}_\mathrm{opt}^{\omega'_\mathrm{m}}(t')\right] \hat U^\dagger_{\rm{sq}} \nonumber  \\
	&=& e^{-\frac{i}{2}(\omega'_\mathrm{m} - \omega_\mathrm{m})t} \hat U_{\rm{sq}} \hat{U}^{\omega'_\mathrm{m}}_\mathrm{opt}(t) \hat U^\dagger_{\rm{sq}}\,.
\end{eqnarray}
For the expectation values of quadrature of a state $\hat \rho$, this leads to 
\begin{eqnarray}
	\nonumber \langle \hat{x}_\mathrm{m}(t) \rangle &=& \mathrm{Tr}(\hat{x}_\mathrm{m} \hat{U}(t) \hat \rho \hat{U}^\dagger(t)) \\
	\nonumber &=&  \mathrm{Tr}(\hat U^\dagger_{\rm{sq}} \hat{x}_\mathrm{m} \hat U_{\rm{sq}} \hat{U}^{\omega'_\mathrm{m}}_\mathrm{opt}(t) \,  \hat \rho^\mathrm{sq} \,  \hat{U}^{\omega'_\mathrm{m}\dagger}_\mathrm{opt}(t)) \nonumber \\
	&=&  \mathrm{Tr}\left(\sqrt{\frac{\hbar }{2m\omega'_\mathrm{m}}}\left(\hat b^\dagger + \hat b\right) \hat{U}^{\omega'_\mathrm{m}}_\mathrm{opt}(t)  \, \hat \rho^\mathrm{sq} \, \hat{U}^{\omega'_\mathrm{m}\dagger}_\mathrm{opt}(t)\right) \, ,\nonumber \\
	\langle \hat{p}_\mathrm{m}(t) \rangle &=& \mathrm{Tr}\left(i\sqrt{\frac{\hbar m\omega'_\mathrm{m}}{2}}\left(\hat b^\dagger - \hat b\right) \hat{U}^{\omega'_\mathrm{m}}_\mathrm{opt}(t) \,  \hat \rho^\mathrm{sq} \,  \hat{U}^{\omega'_\mathrm{m}\dagger}_\mathrm{opt}(t)\right)\, , \nonumber \\
		\langle \hat{x}_\mathrm{c}(t) \rangle &=& \mathrm{Tr}\left(\hat{x}_\mathrm{c} \hat{U}^{\omega'_\mathrm{m}}_\mathrm{opt}(t)  \, \hat \rho^\mathrm{sq}  \, \hat{U}^{\omega'_\mathrm{m}\dagger }_\mathrm{opt}(t)\right)\, , \nonumber \\
	\langle \hat{p}_\mathrm{c}(t) \rangle &=& \mathrm{Tr}\left(\hat{p}_\mathrm{c} \hat{U}^{\omega'_\mathrm{m}}_\mathrm{opt}(t)  \, \hat \rho^\mathrm{sq}  \, \hat{U}^{\omega'_\mathrm{m}\dagger}_\mathrm{opt}(t)\right) \, ,
\end{eqnarray}
where $\hat{\rho}^\mathrm{sq} := \hat U^\dagger_{\rm{sq}} \hat{\rho} \hat U_{\rm{sq}}$. For the initial separable coherent state $|\mu_\mathrm{c}\rangle|\mu_\mathrm{m}\rangle$, we find that the time evolution of the quadratures induced by the full Hamiltonian $\hat{H}$ with constant $\mathcal{D}_2$ can be obtained by calculating the corresponding time evolution of the quadratures induced by $\hat{H}$ with vanishing $\mathcal{D}_2$ by replacing $\omega_\mathrm{m}$ with $\omega'_\mathrm{m}$ and considering the squeezed coherent initial state $\hat U^\dagger_{\rm{sq}}|\mu_\mathrm{c}\rangle|\mu_\mathrm{m}\rangle =|\mu_\mathrm{c}\rangle|\mu'_\mathrm{m},r\rangle $, where $\mu'_\mathrm{m}= \mu_\mathrm{m}\cosh(r) + \mu_\mathrm{m}^*\sinh(r)$.

As a result, the techniques we have developed here can also be utilised to model all the expectation values for an optomechanical system for an initially squeezed states $\ket{\mathcal{D}_2(z)} $.

%
\section{Solutions to the differential equations at resonance} \label{app:resonance}
%
In this Appendix, we first derive perturbative solutions to the Mathieu equations and then compare them with dynamics that we obtain from performing the rotating-wave approximation. We present two approaches which both amount to the same approximation while starting from different assumptions. 

\subsection{Approximate two-scale solution}
Our goal is to obtain approximate solutions to the differential equations~\eqref{differential:equation:written:down} and~\eqref{app:eq:IP22}. We will do so by following the derivation in Ref~\cite{kovacic2018mathieu} with some modifications. 

The general form of Mathieu's equation is given by 
\begin{equation} \label{app:Mathieu}
\frac{d^2 y}{dx^2} + \left[ a - 2  \, q \, \cos(2 \, x) \right] \, y= 0 \, .
\end{equation}
We will use the general notation in this Appendix and then compare with the system in the main text. 

We begin by defining a slow time scale $X= q x$. We then assume that the solutions $y$ depend on both scales, such that $y(x, X)$. This means that we can treat $x$ and $X$ as independent variables and the absolute derivative $d/dx$ in~\eqref{app:Mathieu} can be split in two:
\begin{equation}
\frac{d}{dx} = \partial_x +  q \,  \partial_X \, .
\end{equation}
Mathieu's equation~\eqref{app:Mathieu} therefore becomes
\begin{equation}
\left( \partial_x +  q \,  \partial_X \right)^2 y(x, X) + ( a - 2 q \cos(2 x) ) \, y(x, X) = 0 \, .
\end{equation}
We then expand the solution $y(x, X)$ for small $q$ as $y(x,X) = y_0(x,X) +  q\, y_1 (x,X) + \mathcal{O}(q^2)$ and insert this into the differential equation above. Our goal is to obtain a solution for $y_0$ which incorporates a number of restrictions from the differential equation for $y_1$. 

To zeroth order, we recover the regular harmonic differential equation for $y_0$, which is the limiting case as $q \rightarrow 0$:
\begin{equation}
\partial^2_x y_0 + a \, y_0 = 0 \, ,
\end{equation}
where we know that the solutions are sinusoidal, while the coefficients must depend on $X$.  We choose the following trial solution:
\begin{equation}
y_0(x,X) = A(X) \, e^{i \sqrt{a} \, x} +A^*(X) \, e^{- i \sqrt{a} \, x} \, .
\end{equation}
Our goal is now to find explicit solutions to the complex function $A(X)$. We continue with the equation for $y_1$. We discard all terms of order $q^2$ to find
\begin{equation}
 q \, \partial^2_x y_1 + 2\, q \,  \partial_x \partial _X y_0 +   a \, q \, y_1 -2 \, q \cos(2 x) y_0 = 0 \, .
\end{equation}
We divide by $ q$ and insert our solution for $y_0$ to find 
\begin{align}
 &\partial^2_x y_1 + a \, y_1 + 2 \, i \, \sqrt{a} \, \left(\frac{\partial A(X)}{\partial X} \, e^{i \sqrt{a} x} - \frac{\partial A^*(X)}{\partial X} \, e^{ -i \sqrt{a} x}    \right)  \nonumber \\
 &\quad- 2  \, \cos(2 x)  \left( A(X) \, e^{i \sqrt{a} x}  + A^*(X) \, e^{- i \sqrt{a} x} \right) = 0 \, .
\end{align}
At this point, we specialise to $a = 1$, which corresponds to setting $\Omega_0 = 2$ in the main text. We combine the exponentials to find
\begin{align}
&\partial^2_x \, y_1 + a\, y_1 + \left( 2  i \frac{\partial A(X)}{\partial X} -  \, A^*(X) \right) \, e^{i x} + \left( 2 i \frac{\partial A^*(X)}{\partial X} + \, A(X) \right) \, e^{- i x} \nonumber \\
&\quad  -  A(X) \, e^{3 i x} - A^*(X) \, e^{- 3 i x} = 0 \, .
\end{align}
In order for the solution to be stable, we require that secular terms such as resonant terms $e^{i x}$ vanish. If these do not vanish, the solution will grow exponentially~\cite{kovacic2018mathieu}.  We also neglect terms that oscillate much faster, such as $e^{3 i x}$.  This leaves us with the condition that
\begin{equation} \label{app:resonance:condition}
 \left( 2 i \frac{\partial A^*(X)}{\partial X} + \, A(X) \right)  = 0 \,  , 
\end{equation}
which can be differentiated again and solved with the trial solution $A(X) = (c_1 - i \, c_2) \, e^{X/2} + (c_3- i\,  c_4) \, e^{- X/2}$ for the  parameters $c_1,c_2,c_3$ and $c_4$. From the requirement in~\eqref{app:resonance:condition}, it is now possible to fix two of the coefficients in~\eqref{app:y0:almost:final:solution}. We differentiate $A(X)$ and use~\eqref{app:resonance:condition} to find that the conditions $g_0 =  c_2 $ and $c_3 = -c_4$ must always be fulfilled. 

We then recall that $X =  q x$ and after combining some exponentials, we obtain the full trial solution for the zeroth order term $y_0$:
\begin{align} \label{app:y0:almost:final:solution}
y_0(x) &=  A(qx) \, e^{i  \, x} +A^*(qx) \, e^{ -i  x} \, \nonumber \\
&= 2 \,  \left( c_1 \, e^{qx /2} + c_3 \, e^{- qx/2} \right) \, \cos(x) +  2 \, \left( c_1 \, e^{qx/2} - c_3 \, e^{- qx/2} \right) \sin(x) \, .
\end{align}
We now proceed to compare this solution with the parameters and initial conditions given for $P_{11}$ in~\eqref{differential:equation:written:down} and $I_{P_{22}}$ in~\eqref{app:eq:IP22} in the main text, which are both solved by the Mathieu equation. 

First, we note that $q = - 2\,  \tilde{d}_2$ and that $x = \tau$. Then we consider the boundary conditions for $P_{11}$, which are $P_{11}(0) = 1$ and $\dot{P}_{11} (0) = 0$. From these conditions, we find that  $c_1 = c_3 = 1/4 $, and the the approximate solution to $P_{11}$ is given by 
\begin{equation} \label{app:eq:P11:approx}
P_{11}(\tau) =\cos (\tau) \,  \cosh (\tilde{d}_2 \, \tau) - \sin (\tau) \,  \sinh ( \tilde{d}_2 \,  \tau) \, .
\end{equation}
The equation for $I_{P_{22}}$ has the opposite initial conditions $I_{P_{22}}(0) = 0$ and $\dot{I}_{P_{22}} = 1$. For this case, we find that $c_1 = -c_3 =  1/(4 (1 - \tilde{d}_2))$. The full solution to $I_{P_{22}}$ is therefore
\begin{equation} \label{app:eq:IP22:approx}
I_{P_{22}}(\tau) = - \frac{1}{ 1- \tilde{d}_2}\left(\cos (\tau) \,  \sinh (\tilde{d}_2 \, \tau ) - \sin (\tau) \,  \cosh (\tilde{d}_2 \, \tau) \right)  \, , 
\end{equation}
and thus
\begin{equation}
P_{22} = \cos (\tau) \cosh (\tilde{d}_2 \,  \tau)-\frac{\tilde{d}_2 + 1}{\tilde{d}_2-1} \sin (\tau) \sinh (\tilde{d}_2 \,  \tau) \, .
\end{equation}
Both solutions reduce to the correct solutions for the zero-squeezing case as $\tilde{d}_2 \rightarrow 0$. 

The validity of the perturbative approach can be determined as follows. By inserting the trial solution for $P_{11}(\tau)$ into the Mathieu equation, we are left with terms that are multiplied by $\tilde{d}_2$. The leading term is in fact $\tilde{d}_2 \, \cosh(\tilde{d}_2 \, \tau)$. These terms must be approximately zero to solve the Mathieu equation, which means that we require $\tilde{d}_2 \, \cosh(\tilde{d}_2 \, \tau) \ll1$. This means that while $\tilde{d}_2 \ll1$, we can allow $\tilde{d}_2 \, \tau \sim 1$, which means that $\tau$ can be large provided that $\tilde{d}_2$ is sufficiently small.

From the expression for $\xi(\tau)$ in~\eqref{app:eq:simplified:xi} we then find
\begin{align}
\xi(\tau) =&\cos (\tau) \,  \cosh (\tilde{d}_2 \,  \tau) - \sin (\tau)  \, \sinh (\tilde{d}_2 \, \tau) \nonumber \\
&- \frac{i}{1- \tilde{d}_2} \,  \left( \sin (\tau) \cosh (\tilde{d}_2 \, \tau) -\cos (\tau) \sinh (\tilde{d}_2 \, \tau)\right) \, . 
\end{align} 
For very small $\tilde{d}_2 \ll 1$, which was the condition for deriving the approximate solutions in the first place, we can approximate the fraction as unity and we find the compact expression
\begin{align} \label{app:eq:RWA}
\xi(\tau) =& \,  e^{- i \, \tau} \, \cosh (\tilde{d}_2 \, \tau) + i \, e^{i \, \tau} \,  \sinh (\tilde{d}_2 \, \tau) \, .
\end{align} 
To better understand what this approximation entails physically, we compare it with the rotating-wave approximation, which has a well-known physical interpretation. 

\subsection{Alternative solution}
There is another solution which more explicitly demonstrates how the resonance conditions helps constrain the solution. 
We write the solution to the differential equation for $P_{11}$ as
\begin{equation}
	P_{11}(\tau) = Q_{c}(\tau) \cos(\tau + \pi/4 ) + Q_{s}(\tau) \sin(\tau + \pi/4 )\,.
\end{equation}
Then, the differential equation $\ddot P_{11} + (1+f(\tau)) P_{11} = 0$ is solved approximately,
considering only terms of first order in $\tilde{d}_2$ and neglecting terms rotating with
frequency $3$ off resonantly, by the following set of differential equations 
\begin{eqnarray}
	\dot Q_c(\tau) =  \tilde{d}_2 Q_c(\tau) \quad\mathrm{and}\quad	\dot Q_s(\tau) =  - \tilde{d}_2 Q_s(\tau)\, .
\end{eqnarray}
These can be solved as
\begin{eqnarray}
	Q_c(\tau) =  e^{\tilde{d}_2 \tau} Q_c(0) \quad\mathrm{and}\quad Q_s(\tau) = e^{- \tilde{d}_2 \tau} Q_s(0)\,.
\end{eqnarray}
As the initial conditions for $P_{11}$ are $P_{11}(0)=1$ and $\dot P_{11}(0)=0$ ,
we find 
\begin{equation}
	Q_{c}(0) + Q_{s}(0) = \sqrt{2} \, , \quad\rm{and}\quad  \left(1 - \tilde{d}_2\right) \left( Q_{c}(0) - Q_{s}(0) \right) = 0  \, ,
\end{equation}
which implies $Q_{c}(0) = Q_{s}(0) =  1/\sqrt{2}$, and
\begin{equation}
	P_{11}(\tau) = \frac{1}{\sqrt{2}}\left( e^{\tilde{d}_2\tau}  \cos(\tau + \pi/4 ) + e^{-\tilde{d}_2\tau} \sin(\tau + \pi/4 )\right)\,.
\end{equation}
The same steps as above can be applied to find an approximate solution for $I_{P_{22}}$:
\begin{equation}
	I_{P_{22}}(\tau) = \bar Q_{c}(\tau) \cos(\tau + \pi/4 ) + \bar Q_{s}(\tau) \sin(\tau + \pi/4 )\,, 
\end{equation}
with
\begin{eqnarray}
	\bar Q_c(\tau) =  e^{\tilde{d}_2\tau} \bar Q_c(0) \quad\mathrm{and}\quad \bar Q_s(\tau) = e^{-\tilde{d}_2\tau} \bar Q_s(0)\,.
\end{eqnarray}
and
\begin{equation}
	\bar Q_{c}(0) + \bar Q_{s}(0) =  0 \, , \quad\rm{and}\quad  - \left(1 - \tilde d_2\right) \left( \bar Q_{c}(0) - \bar Q_{s}(0) \right) =  \sqrt{2} \, ,
\end{equation}
which leads to 
\begin{align}
	\bar Q_c(0)= & -\frac{1}{\sqrt{2}\left(1 - \tilde d_2\right)} \, ,\quad \mathrm{and} \quad \bar Q_s(0)=  \frac{1}{\sqrt{2}\left(1 - \tilde d_2\right)} \, ,
\end{align}
and
\begin{equation}
I_{P_{22}}(\tau) = -\frac{1}{\sqrt{2}\left(1 - \tilde d_2\right)}\left( e^{\tilde{d}_2\tau}  \cos\left(\tau + \frac{\pi}{4}\right) - e^{-\tilde{d}_2\tau} \sin\left(\tau + \frac{\pi}{4}\right)\right)\,.
\end{equation}
We find that
\begin{equation}
	\xi(\tau) = \frac{1}{\sqrt{2}}\left(\left( 1 + i\frac{1}{\left(1 - \tilde d_2\right)}\right) e^{\tilde{d}_2\tau}  \cos\left(\tau + \frac{\pi}{4}\right) + \left( 1 - i\frac{1}{\left(1 - \tilde d_2\right)}\right) e^{-\tilde{d}_2\tau} \sin\left(\tau + \frac{\pi}{4}\right)\right) \,  \, ,
\end{equation}
which exactly coincides with~\eqref{app:eq:RWA}. 

\subsection{Comparison with the rotating-wave approximation}
Here we compare the approximate resonance solution for $P_{11}$ in~\eqref{app:eq:P11:approx} and $I_{P_{22}}$  in~\eqref{app:eq:IP22:approx} with the rotating-wave approximation, which is obtained as an approximation to the Hamiltonian in~\eqref{main:time:independent:Hamiltonian:to:decouple:dimensionful} when $\tau \gg 1$. 
In the main text, we separated the Hamiltonian~\eqref{main:time:independent:Hamiltonian:to:decouple:dimensionful} into a free term and a squeezing term~\eqref{app:eq:definition:of:Hsq}, which for our specific choice of $\tilde{\mathcal{D}}_2(\tau) = \tilde{d}_2 \, \cos(\Omega_0 \, \tau)$ becomes
\begin{align}
\hat{{H}}_{\rm{sq}} &= \hat b^\dag \hat b + \tilde{d}_2   \cos(\Omega_0 \tau)  \left( \hat b^\dag + \hat b  \right)^2 \, .
\end{align}
We now define the free evolution Hamiltonian $\hat{H}_0 =  \hat b^\dag \hat b$ and the squeezing term $\hat{H}'_{\rm{sq}} = \tilde{d}_2  \cos(\Omega_0 \, \tau)  ( \hat b^{\dag} + \hat b  )^2$ as separate operators. 
We transform into a frame rotating with $\exp[-i \,  \tau \, \hat b^\dag \hat b ]$, which means that the squeezing term transforms into 
\begin{align} \label{eq:rotating:frame:more:exponentials}
e^{i \hat{H}_0 \, \tau} \hat{H}_{\rm{sq}}' e^{ - i \hat{H}_0 , \tau} =& \, \tilde{d}_2 \, \cos(\Omega_0 \, \tau)  \left( e^{2 \, i \, \tau} \, b^{2\dag} +  e^{- 2 \, i \, \tau} \, \hat b^2 + 2 \, \hat b^\dag \hat b + 1\right) \, .
\end{align}
For this specific case, the system becomes resonant when $\Omega_0 = \omega_0/\omega_{\mathrm{m}} = 2$. We can see this by expanding the cosine in terms of exponentials to obtain
\begin{align}
e^{i \hat{H}_0 \, \tau} \hat{H}_{\rm{sq}}' e^{ - i \hat{H}_0 , \tau} =& \frac{1}{2} \tilde{d}_2 \left[ \left( e^{i \,(2 + \Omega_0 ) \, \tau} + e^{i \, (2 - \Omega_0 )\, \tau} \right) \hat b^{2\dag} + \left( e^{  -i \, (2 + \Omega_0 ) \, \tau} + e^{- i \, ( 2 - \Omega_0 )\, \tau} \right) \hat b^2 \right] \, \nonumber \\
&+ \tilde{d}_2 \, \cos(\Omega_0 \, \tau) \left( 2 \, \hat b^\dag \hat b  + 1 \right) \, .
\end{align}
When $\Omega_0 = 2$, two of the time-dependent terms will cancel. We then perform the rotating-wave approximation, i.e.~we neglect all remaining time-dependent terms. This approximation is only valid for $\tau \gg1$. 
In the interaction frame, we find
\begin{align} \label{app:rotating:frame:approx}
\hat H_{\rm{sq}, I} = e^{i \hat H_0\, \tau} \hat{H}_{\rm{sq}}' e^{ - i \hat H_0 \tau} \approx  \frac{1}{2} \tilde{d}_2    \, (  \hat b^{\dag 2}  +  \hat b^2 )  \, .
\end{align}
In the symplectic basis $\hat \vec{r} = (\hat b, \hat b ^\dag )^{\rm{T}}$ the corresponding symplectic operator, given by $\boldsymbol{S}_{\rm{sq}} = e^{\boldsymbol{\Omega} \, \boldsymbol{H}_{\rm{sq}}}$, where $\boldsymbol{\Omega} = i \, \mathrm{diag}(-1, 1)$ and $\boldsymbol{H}_{\rm{sq}}$ is given by 
\begin{align}
\boldsymbol{H}_{\rm{sq}} =  \tilde{d}_2  \, \begin{pmatrix} 0 & 1 \\ 1  & 0 \end{pmatrix}  \, .
\end{align}
The symplectic representation of the squeezing operator~\eqref{main:general:expression:time:dependent:squeezing} in the lab frame  reads 
\begin{align}\label{app:lab:frame:evolution}
\boldsymbol{S}_{\mathrm{sq}}(\tau)=&\boldsymbol{S}_0(\tau) \,\begin{pmatrix}
\cosh (\tilde{d}_2 \,\tau) & -i\,\sinh(\tilde{d}_2 \,\tau) \\
i\,\sinh(\tilde{d}_2 \,\tau)  & \cosh (\tilde{d}_2 \,\tau) 
\end{pmatrix},
\end{align}
where $\boldsymbol{S}_0 = e^{-  i \, \tau}$ encodes the evolution from the Hamiltonian $\hat{H}_0$. We therefore find the Bogoliubov coefficients 
\begin{align}
\alpha(\tau)=&  \, e^{-  i \, \tau}\,\cosh (\tilde{d}_2 \,\tau) \, ,\nonumber\\
\beta(\tau)=&-i\, e^{-  i \, \tau}\,\sinh (\tilde{d}_2 \,\tau) \, ,
\end{align}
which 
evidently satisfy the Bogoliubov conditions, and obtain 
\begin{equation}
\xi = \alpha(\tau) + \beta^*(\tau) = e^{- i \, \tau} \, \cosh (\tilde{d}_2 \, \tau) + i \, e^{i \, \tau} \,  \sinh (\tilde{d}_2 \, \tau) \, .
\end{equation}
This expression exactly matches the one we derived as a perturbative solution to the Mathieu equations in~\eqref{app:eq:RWA}. However, the requirement for the validity of the RWA is that $\tau \gg1$, while the approximate solutions are still valid for small $\tau$. We conclude that the approximate solutions only coincide with the RWA for large $\tau$, while this interpretation cannot be used when $\tau \sim1$. 

\end{document}